\documentclass[11pt]{article}

\usepackage[utf8]{inputenc}
\usepackage{amsmath,amssymb,amsthm, bm, bbm}
\usepackage{graphicx, xcolor}
\usepackage{natbib}
\usepackage{enumerate}
\usepackage{comment}
\usepackage{subcaption}
\usepackage{stackrel}

\usepackage{setspace}
\usepackage{multirow}
\usepackage{booktabs}
\usepackage{hyperref}
\usepackage{setspace}
\usepackage[top=1in, bottom=1in, left=1in, right=1in]{geometry}
\hypersetup{colorlinks,citecolor=blue,urlcolor=blue,linkcolor=blue}

\usepackage{stefan_tex}

\newtheorem{theorem}{Theorem}
\newtheorem{proposition}[theorem]{Proposition}

\newtheorem{lemma}[theorem]{Lemma}
\newtheorem{assumption}{Assumption}
\newtheorem{design}{Design}
\theoremstyle{definition}

\graphicspath{{../../figures}}

\newcommand\blfootnote[1]{%
  \begingroup
  \renewcommand\thefootnote{}\footnote{#1}%
  \addtocounter{footnote}{-1}%
  \endgroup
}

\begin{document}

\title{Treatment Effects in Market Equilibrium\blfootnote{
We thank
Isaiah Andrews,
Joshua Angrist,
Dmitry Arkhangelsky,
PM Aronow,
Susan Athey,
Lanier Benkard,
Han Hong,
Guido Imbens,
Michael Kosorok,
Casey Mulligan,
Whitney Newey,
Fredrik S\"avje,
Paulo Somaini,
Edward Vytlacil,
Ali Yurukoglu,
and seminar participants at
Columbia,
Cornell,
Google,
Harvard,
Michigan State University,
Princeton,
Stanford,
UC San Diego,
UT Austin,
University of Chicago,
University of Montreal,
Yale,
the Bay Area Tech Economics Seminar,
the Conference on Digital Experimentation,
the International Conference on Statistics and Data Science,
the Joint Statistical Meetings,
the National Association for Business Economics,
and the Online Causal Inference Seminar
for helpful comments and discussions. This research is partially supported by a gift from Meta and NSF SES-2242876. Code for the simulations in this paper is available at \texttt{https://github.com/evanmunro/market-interference}. Xu Kuang published under a different full name in earlier versions of this manuscript. Please use ``E. Munro, X. Kuang and S. Wager" when citing this paper.}
\vspace{2mm}
}

\author{
Evan Munro
\and
Xu Kuang
\and
Stefan Wager}
\date{\today \\
 \vspace{2mm}
Stanford Graduate School of Business}
  \maketitle

\bigskip
\begin{abstract}
Policy-relevant treatment effect estimation in a marketplace setting requires taking into account
both the direct benefit of the treatment and any spillovers induced by changes to the market
equilibrium. The standard way to address these challenges is to evaluate interventions via
cluster-randomized experiments, where each cluster corresponds to an isolated market.
This approach, however, cannot be used when we only have access to a single market (or a small number
of markets). Here, we show how to identify and estimate treatment effects using
a unit-level randomized trial run within a single large market. A standard Bernoulli-randomized
trial allows consistent estimation of direct effects, and of treatment heterogeneity measures
that can be used for welfare-improving targeting. Estimating spillovers---as well as providing
confidence intervals for the direct effect---requires estimates of price elasticities, which we
provide using an augmented experimental design. Our results rely on
all spillovers being mediated via the (observed) prices of a finite number of traded goods, and
the market power of any single unit decaying as the market gets large. We illustrate our
results using a simulation calibrated to a conditional cash transfer experiment in the Philippines. \\ 

\end{abstract}


\onehalfspacing
\newpage
\section{Introduction}

The standard analysis of treatment effects in a randomized trial relies on an assumption that the treatment status of one individual does not affect the outcomes of other individuals \citep{fisher1935design, banerjee2011poor,imbens2015causal}. When an intervention affects supply or demand and market prices affect the outcome of interest, then general equilibrium effects lead to spillover effects, and typical estimators using data from randomized trials do not correspond to a meaningful measure of policy impact \citep{heckman1998general}.

In practice, the most common approach to address interference of this type is clustering units at a higher-level at which there are no spillover effects \citep{baird2018optimal, hudgens2008toward}.
Examples include the evaluation of general equilibrium impacts of cash transfers in \citet{banerjee2021food},  \citet{egger2022general} and \citet{filmer2023}, and incentives for suppliers or consumers in online two-sided markets
\citep{brennan2022cluster, holtz2024reducing}.
However, if a market is tightly integrated, splitting the market into effectively isolated
sub-markets may not be feasible. This highlights the need for spillover-aware methods
for treatment effect estimation that can be used on data from an experiment run on a
single interconnected market.

In this paper, we introduce a framework for treatment effect estimation in
a market equilibrium setting without relying on parametric, distributional,
or homogeneity assumptions.  Rather than relying on market splitting, we leverage
a structural assumption whereby all spillovers are mediated through the (observed) equilibrium
prices of a finite number of traded goods. Our main finding is that integrating such price
equilibrium effects into the potential-outcomes model for causal inference enables
us to estimate spillover-aware treatment effects using unit-level experiments. In particular,
we show how our approach yields asymptotic characterizations of policy-relevant average and
conditional average treatment effects, and motivates the design of new forms of randomized trials
and targeting rules.

We frame our model in terms of a potential outcomes specification that allows for cross-unit interference  \citep{hudgens2008toward, manski2013identification, aronow2017estimating}. When equilibrium effects are removed, the model reduces to the standard Neyman-Rubin potential outcomes framework  \citep{imbens2015causal}. In a market with $n$ participants, an intervention is assigned randomly conditional on pre-treatment covariates. Each individual's production
and consumption choices are determined by latent supply and demand functions that may be shifted by the intervention
\citep{angrist2000interpretation,heckman2005structural}. Units interact via a market modeled as a multiple goods economy in general equilibrium, where individuals' production and consumption decisions depend only on their own type, their own treatment, and their expectation of the market price \citep{walras1900elements, mas1995microeconomic}. In this setup, equilibrium prices induce an exposure mapping in the sense of \citet{manski2013identification} and \citet{aronow2017estimating}: One unit's treatment impacts another's outcomes only through the treatment's impact on the equilibrium price, which rules out peer effects or other forms of network-type interference.  The price ensures that average net demand (demand minus supply) is equal to zero. The validity of this restriction on spillovers is context-specific, and depends both on the nature of the intervention and the type of market.


Our analysis begins by considering the marginal effect of increasing treatment probabilities
for all market participants; following \citet{carneiro2010evaluating}, we refer to this quantity
as a marginal policy effect. As shown in \citet{hu2021average}, the marginal policy effect can be
decomposed into a direct and indirect treatment effect, where the indirect effect is a general measure
of spillover effects. We then analyze the sample average direct and indirect effects of
a binary treatment in equilibrium \citep{halloran1995causal, hu2021average, savje2021average}.  In
finite samples, these estimands are difficult to characterize statistically because the price
equilibrium induces a complex form of data dependence. We show, however, that under an assumption
that the $n$ participants in the market are independently sampled from a population distribution,
we can use techniques based on empirical-process theory to derive simple large-sample limiting expressions
for the direct and indirect effects. These limiting expressions can then be used to guide estimation
and inference for both sample and population treatment effects.

In our setting, the average direct effect can be consistently estimated via a 
simple difference in means estimator that ignores spillovers. Consistent estimation
of the average indirect effect---and construction of confidence intervals for either
estimand---also requires estimates of price elasticities. We consider augmenting
 standard Bernoulli-randomized experiments with small, random price perturbations
that do not disrupt the overall market equilibrium. The use of such random price
perturbations is in line with the documented practice of online marketplaces using
randomly assigned coupons, discounts or bonuses to learn about price elasticities
\citep{castillo2023}. We then show how such augmented experiments can be used to provide
valid large-sample inference for both the average direct and indirect effects using
data from a single market.

Our model allows for unrestricted heterogeneity across agents,
and thus also enables us to estimate how treatment effects vary with pre-treatment
characteristics---and to exploit this heterogeneity to learn improved policies that
can be deployed in a market similar to the one observed. We propose versions of the
sample direct effect and indirect effect that are conditional on pre-treatment covariates.
These estimands depend both on the population distribution and the size of the market.
They are related to the large literature on targeting without spillovers
\citep{manski2004statistical, athey2016recursive, wager2018estimation, vanderweele2019selecting},
in that the conditional average direct effect is equal to the conditional average treatment
effect when interference is removed. We show that as the market size grows large,
the conditional estimands converge to population estimands that are simple combinations
of price elasticities and conditional expectations of the direct effect of the
treatment on outcomes and net demand. Furthermore, we show that it is possible to use
data from a standard randomized experiment (without price perturbations) to estimate
the outcome-maximizing treatment rule within the class of all treatment rules that
do not move the observed market equilibrium. This optimal rule takes the form of a
linear thresholding rule in the space of conditional average direct effects on outcomes and net demand.

\citet{filmer2023} use a village-level cluster randomized experiment to demonstrate that the Pantawid conditional cash transfer program in the Philippines affects the equilibrium price of perishable protein sources, such as eggs, which leads to spillover effects on health outcomes for children in non-eligible families.
In Section \ref{sec:sim}, we argue that this application fits the assumptions of our paper. We then use the data from this setting to estimate an equilibrium model of children's health outcomes and the market for eggs in a small village in the Philippines, and demonstrate the consistency and coverage properties of our treatment effect and variance estimators based on an augmented individual-level randomized experiment generated by simulating data from this model.  We also augment the model to include heterogeneous responses to the cash transfer and describe the properties of the equilibrium-stable targeting rule on samples drawn from this model.

\subsection{Related Work}


Much of the existing work on treatment effect estimation with spillovers focuses on setting
where the Stable Unit Treatment Value Assumption (SUTVA) holds for higher-level clusters of units
\citep{baird2018optimal, basse2019randomization, hudgens2008toward, karrer2021network, liu2014large, tchetgen2012causal}.
Another related approach is the network interference model, which posits that interference operates
along a network and the connections between units are sparse
\citep{athey2018exact,leung2020treatment,savje2021average}.
The sparsity of connections then enables randomization-based methods for studying cluster-randomized experiments
to be extended to this setting.
In our setting, however, the interference pattern produced by marketplace price effects is dense and simultaneously affects all units, so neither cluster nor sparsity-based methods are applicable.

Our use of a random sampling model presents a departure from the literature cited above, where
typically the sample is held fixed and inference is only driven by random treatment assignment.
This random sampling model, however, plays a key role in enabling our analysis, as it allows us
to leverage tools from empirical process theory \citep[e.g,][]{van1996weak} to address difficulties
around potentially unstable and non-unique equilibrium formation in finite samples.\footnote{As
discussed in detail in the following section, we assume that there is a unique market-clearing
price in the population; however, we allow each market participant to respond discontinuously to
market prices and this may lead to non-uniqueness of possible equilibria in finite samples.}
In doing so, our paper adds to a handful of recent papers that use a stochastic model in a
causal inference setting.  \citet{johari2020experimental} use a mean-field model to examine
the benefits of a two-sided randomization design in two-sided market platforms and quantify
the bias due to interference between market participants.
\citet{li2022random} study large sample treatment effect estimation under network interference,
where the exposure graph is randomly generated from a graphon model. Finally, \citet{wager2019experimenting} consider a model where exogenous demand is matched with endogenous supply, with interference via supply cannibalization. They propose using zero-mean perturbations to estimate a revenue gradient and optimize a continuous decision variable dynamically. Our work also uses zero-mean perturbations as part of a unit-level experiment, but is otherwise distinct, given our focus is on treatment effects of a binary intervention in a general price equilibrium.

Our model combines a potential-outcomes model with some limited structural assumptions on
the price equilibrium. The insight that structural modeling can be used to address
endogeneity issues in causal inference has a long tradition in economics
\citep{heckman1979sample,heckman1998general,roy1951some}, and there is a large literature that uses
structural modeling to develop econometric methods for causal inference in models
with endogenous treatment selection  
\citep{angrist2000interpretation,carneiro2011estimating,heckman2005structural}.
To date, however, there has been less emphasis on methods for estimating equilibrium effects in systems with randomized trials, where treatment selection is not an issue.

We do note a growing literature that combines data from a randomized trials with parametric structural models to estimate both partial and general equilibrium policy effects. For example,  \citet{allende2019approximating} uses a randomized control trial to estimate the direct effect of information provision on school choice but simulates a fully parametric structural model to evaluate equilibrium effects of the treatment. In contrast, our approach uses a more complex unit-level experiment and the structure of the price equilibrium to analyze equilibrium effects without imposing a parametric model. To accomplish this, we rely on a representation of population estimands as a combination of price elasticities and expectations of the outcome and demand distribution. This is related to the sufficient statistics approach, as described in \citet{chetty2009sufficient}, which also expresses policy effects in terms of combinations of estimable elasticities without imposing a specific parametric model, although this is usually in a partial equilibrium setting. In the macroeconomics literature, \citet{wolf2019missing} uses a semi-structural approach to decompose the effect of economic shocks into partial equilibrium effects and general equilibrium effects (e.g., price effects), but uses time series methods to estimate the price effects, in contrast to the unit-level approach presented in this paper.

The literature on targeting treatments under interference is quite limited. \citet{viviano2019policy} estimates treatment allocation rules in a network interference model, where his results depend on sparsity of the network.  \citet{sahoo2022policy} study policy learning with strategic agents, where the policymaker has a budget constraint, which leads to a type of interference. Both approaches rely on a form of direct empirical welfare maximization and do not define spillover-aware measures of treatment heterogeneity as we do in this paper.

Finally, the price perturbations that we use to estimate average price elasticities can be thought of as generating an ideal instrument for our setting. By construction, everyone in the sample is a complier, and the size of the perturbations shrink so that local elasticities can be estimated without imposing parametric assumptions on demand or supply. In empirical work with observational data, instrumental variables approaches have been used in both parametric and non-parametric settings to estimate price elasticities  or a weighted average of elasticities \citep{angrist1996identification, berry2021foundations}. \cite{angrist2001instrumental} reviews IV approaches and their relation to causal inference for observational data. In settings where it is not possible to randomize individual-level fees, under additional structural assumptions, a traditional IV-based approach could be used to estimate relevant  elasticities from observational data.


\section{Model and Estimands }
\label{sec:method}

We first introduce a potential-outcomes model of a market with $n$ participants, where each participant is drawn independently from a population. A binary intervention affects individual supply, demand and outcomes in the market.
We use this model to define the estimands
of interest for our paper, both at the sample level and at the population level. We use ``sample" to
refer to a finite-sized market with $J$ goods and $n$ participants.
Throughout this section, we will refer to two example interventions: The first is a supplier-level subsidy in an online market setting; the second is a household-level cash transfer that affects economic choices in a village economy.

We assume that for all $n$ market
participants,\footnote{In Section \ref{sec:estimation}, our inference approach will
also allow for settings where we can only observe outcomes for a sub-sample of the $n$-sized market
(but equilibrium effects  are still determined by the full size-$n$ sample).}
we observe an outcome $Y_i \in \RR$, treatment $W_i \in \cb{0, \, 1}$,
net demand (i.e., demand minus supply) for the $J$ traded goods $Z_i \in \RR^J$, along
with covariates $X_i \in \mathcal X$.
Following the potential-outcomes model with spillovers, we write $Y_i(\bm w)$ and $Z_i(\bm w)$ for
the outcome and net demand we would have obtained for $i$-th unit under treatment
market-level treatment assignment $\bm w \in \cb{0, \, 1}^n$.\footnote{We
use bold-face notation to denote a vector-valued quantity collected across all market participants.}
We assume that spillovers are mediated by endogenous prices $P_n \in \RR^J$ that match supply and
demand for the $J$ traded goods, as detailed below.
We also allow for an exogenous individual-level price shifter $U_i \in \mathbb R^J$. For our analysis of classical randomized trials, $U_i = 0$,
but we will also consider a class of experiments that randomize discounts or fees in a market, where $U_i \neq 0$ and is also exogenous.


\begin{assumption} \label{as:sampling}
Each market participant is characterized by a latent
outcome function $Y_{i}(w_i, \, p)$ and a net demand function $Z_{i}(w_i, \ p)$,
such that given treatment $W_i \in \cb{0, \, 1}$, market prices $P_n \in \RR^J$ and (optional) price perturbations $U_i \in \RR^J$,
we have $Y_i = Y_i(W_i, \, P_n + U_i)$ and $Z_i = Z_i(W_i, \, P_n + U_i)$.
The quantities $Y_{i}(w_i, \, p)$, $Z_{i}(w_i, \ p)$ and $X_i$ for $i = 1, \, \ldots, \, n$ are drawn independently from a population distribution. 
\end{assumption}




\begin{assumption}
\label{as:interference}
The market prices satisfy $P_n = P_n(\wvec)$, where $P_n(\wvec)$
endogenously sets net demand to approximately 0 with high probability in the following sense.
There exists a sequence $a_n$ with $\lim \limits_{n \to \infty} a_n \, \sqrt{n} = 0$ and
constants $b, \, c_1 > 0$ such that, for every
$\wvec \in \cb{0,\, 1}^n$ and for $U_i$ drawn IID from any distribution on $[-b, +b]^J$, 
\begin{equation}
\label{eq:approxzero}
\set_{\wvec} = \cb{p \in \RR^J : \Norm{\frac{1}{n} \sum \limits_{i=1}^n Z_{i}(w_i, p + U_i)}_2 \leq a_n }
\end{equation}
is non-empty with probability at least $1 - e^{-c_1n}$ for all $n$. On the event where this set is non-empty, the market price
is in this set, $P_n(\wvec) \in \set_{\wvec}$. 
\end{assumption}

In Section \ref{sec:estimation}, we will use observations of net demand, outcomes, treatments and price-shifters in a single market of size $n$ to estimate policy-relevant treatment effects. Assumption \ref{as:interference} is the main structural assumption in the paper that makes this possible. Treatment and market prices directly impact outcomes and net demand for each individual. Prices are endogenous, and are determined by a sample equilibrium condition on net demand. Because of this endogeneity, there are spillovers from individual $i$ to individual $j$ through their effect on the equilibrium that the market reaches. Although our model allows for equilibrium spillovers, it rules out other types of spillovers, such as through peer effects or some other network mechanism. The validity of this restriction on interference primarily depends on the nature of the intervention and the context of the market. A subsidy allocated to suppliers in an online market may, by design, be impossible to share with those in the control group, so spillovers occur only through market prices. On the other hand, an intervention that affects consumption in a village economy may have spillovers both through social networks in the village and through the market equilibrium.

The second part of the assumption characterizes how prices form in finite samples.
The finite market price does not need to exactly clear the market; instead, we impose a restriction on the rate at which the error in the market-clearing condition shrinks with the sample size. Furthermore, it does not need to be the unique price meeting this condition, although we will require uniqueness at the population level, i.e., in the limit as the sample size grows large
(Assumption \ref{as:market}).\footnote{Proposition \ref{prop:clear} in Appendix \ref{app:add} provides explicit sufficient conditions for Assumption \ref{as:interference} to hold in a single good market without price shifters and with discontinuous net demand functions.}


We will start by considering results that hold under a standard Bernoulli-randomized RCT (Design \ref{def:rct}).
In Section \ref{sec:estimation}, we will show how augmenting Bernoulli-randomized trial with
random individual-level price shifters (Design \ref{def:aug}) enables further estimation and inferential results.
Design \ref{def:aug} will discussed in more detail in Section \ref{sec:aug}. 


\begin{design}\textbf{Bernoulli Randomized Trial.}
\label{def:rct}
Each unit has a known randomization probability $\pi_i$ satisfying
$\eta \leq \pi_i \leq 1 - \eta$ for some $\eta > 0$, and treatment is generated as
$W_i \sim \text{Bernoulli}(\pi_i)$. The $\pi_i$ may be constant, functions of $X_i$, or random.\footnote{In
Design \ref{def:rct} there are no price perturbations, i.e., $U_i = 0$ for all $i \in \{1, \ldots, n \}$.}
\end{design}

\begin{design} \label{def:aug} \textbf{Augmented Randomized Trial.} Treatment $W_i$ is randomized according to Design \ref{def:rct}. In addition, the experimenter generates random price perturbations $U_i \in \RR^J$, such that for
$i \in \{ 1, \, \ldots, n \} $ and $j \in \{1, \, \ldots, J \}$, $ U_{ij}$ is drawn independently and uniformly at random from $\{-h_n  , +h_n \}$
where $h_n = c n^{-\alpha}$ with $\frac{1}{4} < \alpha < \frac{1}{2}$ and $c > 0$ is a constant. 
\end{design}

Given this causal model, there are many treatment effects one can seek to study. The total
treatment effect is the average difference in outcomes from treating everyone versus no one
in the market,
\begin{equation} \label{eq:tot}
\tau_{\text{TOT}} = \frac{1}{n} \sum_{i = 1}^n \p{Y_{i}(1, \, P_n(\mathbf{1}_n) + U_i ) - Y_{i}(0, \, P_n(\mathbf{0}_n)+ U_i )},
\end{equation}where $\mathbf{1}_n$ and $\mathbf{0}_n$ are $n$-length vectors of 1s and 0s. Total treatment effects can readily be estimated across disconnected markets using cluster-randomized
experiments clustered at the market level \citep{baird2018optimal}.
However, without further assumptions, estimating $\tau_{\text{TOT}}$ from a single (connected) market
is impossible, as it requires predicting behavior at prices that are far from anything seen in the experiment.
A more tractable---and still fruitful---approach is to focus on treatment effects that,
in the sense of \citet{carneiro2010evaluating}, are marginal to the treatment-assignment policy used to collect the data.
For any $n$-length vector of Bernoulli randomization probabilities $\pivec$, write
\begin{equation}
V_n(\pivec) = \EE[\pivec]{\frac{1}{n} \sum_{i = 1}^n Y_i},
\end{equation}
where we use the notation $\EE[\pivec]{\cdot} = \EE{\cdot \cond \cb{Y_i(w, \, p), \, Z_i(w, \, p), U_i}_{i=1}^n}$
to denote expectations over the Bernoulli treatment assignment while holding the sample (i.e., the potential outcomes)
fixed. The marginal policy effect in the sample, $\tau_{\text{MPE}}$, is a local approximation to the total treatment effect defined in \eqref{eq:tot}. It measures the average effect of increasing treatment probabilities
for everyone:
\begin{equation}
\tau_{\text{MPE}} = \frac{d}{d\varepsilon} V(\pivec + \varepsilon \mathbf{1}_n)|_{\varepsilon=0}.
\end{equation}
\citet{hu2021average} show that the marginal policy effect can be decomposed into the sum of two sample average treatment effects, which are the first set of target estimands in our paper. Here, the notation $(W_i = a ; \bm W_{-i}) $ represents the random vector we obtain by first sampling $ \bm W $, and then setting its $ i $-th coordinate to a fixed value $ a $:
\begin{equation}
\label{eqn:ade_aie}
\begin{split}
& \tau_{\text{MPE}}  = \tau_{\text{ADE}}  + \tau_{\text{AIE}}, \\
& \tau_{\text{ADE}} =   \frac{1}{n}  \sum \limits_{i=1}^n \mathbb E_{\pi}[ Y_i(W_i = 1, P_n( W_i = 1 ; \bm W_{-i}) + U_i )- Y_i(W_i = 0, P_n(W_i = 0; \bm W_{-i}) + U_i )  ],
\\  & \tau_{\text{AIE}} =  \frac{1}{n}  \sum \limits_{i=1}^n \sum \limits_{j \neq i}  \mathbb E_{\pi}[ Y_j(W_j, P_n(W_i = 1; \bm W_{-i}) + U_i ) - Y_j(W_j, P_n(W_i = 0; \bm W_{-i}) + U_i)  ].
\end{split}
\end{equation}
$\tau_{\text{ADE}}$ is the average effect on an individual's outcome of changing their own treatment in a market with $n$ participants, and is a standard quantity in the causal inference literature \citep[e.g.,][]{halloran1995causal,savje2021average}. It is a measure of the direct effect of the treatment while holding the equilibrium fixed. When SUTVA holds, then $\tau_{\text{ADE}}$ is equal to the sample Average Treatment Effect (SATE) and $\tau_{\text{AIE}}$ is 0. The average indirect effect in the sample is the average effect on everyone else's outcomes of changing an individual's treatment. It is a general measure of spillover effects.  



The estimands
$\tau_{\text{ADE}}$ and $\tau_{\text{AIE}}$ are defined at the sample level. We also find it  useful to introduce population versions of our target estimands, and define the following whenever the limits below exist. $\mathbb E[\cdot]$ reflects expectations over all sources of randomness in our model: the sampling from the population distribution in Assumption \ref{as:sampling}, the price-shifter assignment process, and the treatment assignment process. 

\begin{equation}
\begin{split}
& \tau^*_{\text{ADE}} = \lim \limits_{n \rightarrow \infty} \mathbb E[\tau_{\text{ADE}}] = \lim \limits_{n \rightarrow \infty} \mathbb E[ Y_i (W_i = 1; \bm W_{-i})  - Y_i(W_i =0; \bm W_{-i})],   \\
& \tau^*_{\text{AIE}} = \lim \limits_{n \rightarrow \infty} \mathbb E[\tau_{\text{AIE}} ]  = \lim \limits_{n \rightarrow \infty} (n-1) \cdot \mathbb E[ Y_{j} (W_i = 1; \bm W_{-i}) - Y_{j}( W_i = 0; \bm W_{-i}) ].
\end{split}
\end{equation}
In the next section, we will show that under our model and an additional set of regularity conditions,
these limits in fact exist and the population estimands have a tractable representations.
These representations will then be useful for designing estimators that are consistent for both sample and population-level targets. In addition, just as in the literature on causal inference under SUTVA, in some settings the population estimand is the more relevant target. For example, we may have data from a small market and are interested in doing inference on a similar but larger market.


\subsection{Large Sample Characterization}

Our asymptotic characterization of the average direct and indirect effect requires some additional regularity conditions. It is first helpful to introduce some additional notation for moments of the population distribution. Let $y(p) = \mathbb E[Y_i(W_i, p)]$ and $z(p) = \mathbb E[Z_i(W_i, p)]$. For $w \in \{0, 1\}$, let $y(w, p) = \mathbb E [ Y_i(w,  p)]$ and $ z(w, p) =  \mathbb E[  Z_i(w, p)]$. Define $ p^*_{\pi}$ to be the price that clears the population market when the treatment is allocated according to $\pi(\cdot)$, i.e., $p^*_{\pi} = \{ p: z( p) = 0 \}$.


We start by making regularity assumptions on the expected net demand, and in particular assume that there is a unique equilibrium price in the population.
Market prices will be in a compact set as long as all net demand functions are weakly negative for prices above an upper bound and weakly positive for prices smaller than some lower bound. In a single good market, uniqueness of $p^*_{\pi}$ only requires continuity and strict monotonicity of $z(p)$ in $p$. In a market with multiple goods, uniqueness and existence of the equilibrium price can be shown by ensuring the expected net demand function is a contraction, as is common in the literature on models of strategic behavior \citep{cornes1999equilibrium, van2000existence}. Other approaches that rely on more primitive assumptions are also possible; for example, under a zero-degree homogeneity assumption in net demand a gross-substitutes condition ensures uniqueness \citep{arrow1971general}.

\begin{assumption}
\label{as:market}
Market prices take values in a compact set $\set \subset \RR^J$ almost surely.
Given any randomization policy $\pi$, there is a unique population market-clearing price
$p^*_\pi \in \set$ that satisfies
$z(p^*_{\pi}) = 0.$
The Jacobian $\xi_z =  \nabla_{ p} z(p^*_{\pi}) $ (i.e., the $J \times J$ matrix with $j$-th row $\nabla^{\top}_p z_j(p^*_{\pi})$) is full rank.
\end{assumption}

Next, we make regularity assumptions on the unit-level net demand and outcome functions.
To ensure that a variety of random processes studied in the paper concentrate, Assumption \ref{as:monotone} and Assumption \ref{as:lips} impose some constraints on outcomes and net demand at an individual level.
These restrictions have a simple economic interpretation and are general enough to encompass a wide variety of possible data-generating processes that underlie market behavior.
Assumption \ref{as:monotone} is our main assumption on the net demand, and is a generalization
of the familiar assumption in a single-good setting that net demand be non-increasing in price.
In the multiple good case, we also require that every unit $i$ will decrease their net demand
of good $j$ more in response to a decrease in the price of good $j$ than in response to a
relatively small change in the price of other goods. This assumption restricts the magnitudes
of cross-price elasticities compared to own-price elasticities.

\begin{assumption}
\label{as:monotone}
For each market participant $i$, net demand for the $j$-th good is approximately monotone
decreasing in the price of the $j$-th good.  There exists a constant $C > 0$ such that the
following holds almost surely for all units $i$, goods $j$, treatment levels $w \in \cb{0, \, 1}$,
and prices $p \in \mathcal{S}$, and for any $0 < \varepsilon \leq 1$:
\begin{equation}
Z_{ij}(w, \, p - \varepsilon e_j) \geq Z_{ij}(w, \, p + \delta) \geq Z_{ij}(w, \, p + \varepsilon e_j), \ \text{ for all } \ \Norm{\delta}_2 \leq C\varepsilon,
\end{equation}
where $e_j$ denotes the $j$-th basis vector.
\end{assumption}

Under Assumption \ref{as:lips}, the outcome function is the sum of an individual-specific random function that is Lipschitz in prices and an additional term that  is Lipschitz in net demand. This allows for outcomes that are discontinuous in prices through net demand. Supplier profit, for example, is a special case of Assumption \ref{as:lips}, where in the single-good case, profit  is $Y_i(w, p) =  -(p - \Gamma_i(w) ) Z_i(w, \, p) \mathbbm{1}(Z_i(w, \, p) <  0)$ and $\Gamma_i(w)$ represents a firm's production costs under treatment $w$. In Section \ref{sec:sim}, outcomes are a measurement of height for young children, and the treatment is a household-level cash transfer. Assumption \ref{as:lips} allows for a variety of flexible data-generating processes for height; one example is a random coefficients model of height that is polynomial in consumption, where $\Gamma_i(w)$ are the random coefficients and noise term in the model.

\begin{assumption}
\label{as:lips}
The outcome function is the sum of a random Lipschitz function of $p$, $H_i(w, \, p)$, and a fixed transformation $\psi(\cdot)$ of $Z_i(w, \, p)$, prices $p$, and a (possibly unobserved) bounded random variable $\Gamma_i(w) \in \mathcal G \subset \mathbb R^m$, that is Lipschitz in each of its arguments: 
\begin{equation}
Y_i(w,\, p) = H_i(w, \, p) + \psi(\Gamma_i(w), \, Z_i(w, \, p), \, p). 
\end{equation}
\end{assumption}

Finally, in Assumption \ref{as:regularity} we list some additional regularity assumptions, which are standard in the literature on asymptotic statistics. Although individual net demand and outcome functions may be discontinuous, we require their
expectation to vary smoothly in prices. The weak continuity assumption in Part 2 also limits individual-level discontinuity by requiring 
that discontinuity points cannot concentrate at specific values of $p$. In our profit example, these two smoothness assumptions require  the distribution of costs at a firm-level to be sufficiently smooth, even if production is discontinuous in prices. 


\begin{assumption}
\label{as:regularity}
The following regularity conditions hold:
\begin{enumerate}
\item Net demand and outcome functions are uniformly bounded, i.e.,
there is a constant $M < \infty$ such that, almost surely,
$\abs{Y_i(w, \, p)} \leq M$ and $\abs{Z_{ij}(w, \ p)} \leq M$
for all $w \in \{0, 1\}$, $p \in \mathcal S$ and $j \in \{1, \ldots, J \}$.
\item Net demand is weakly continuous in $p$. There is a constant $L > 0$ such that for all pairs of prices
$p ,\, p'$, all $w$, and all $j$, we have $\mathbb E[ (Z_{ij}(w, p) - Z_{ij}(w, p'))^2] \leq L \Norm{p - p'}_2$.
\item For all $p \in \mathcal S$ and $w \in \{0, 1\}$, $y(w, p)$,
$z(w, p)$, $ y(p)$ and $z(p)$ are twice continuously
differentiable in $p$ with bounded first and second derivatives.
\item We have non-trivial variation in responses:
For all $w \in \cb{0, \, 1}$ and $p \in \set$, $\Var{Y_i(w, \ p)} > 0$, and $\Var{Z_i(w, \ p)}$ is
positive definite, i.e., $\Var{Z_i(w, \ p)} \succ 0$.
\end{enumerate}
\end{assumption}

Our first result is that, given these assumptions, the sample equilibrium price
concentrates on $p^*_\pi$ as the sample size grows, and has an asymptotically linear representation
(i.e., to first order, random price fluctuations can be written as a sum of additive contributions
from each unit). We can write $P_n(\bm W)$ as a method-of-moments estimator, so its asymptotic representation has the usual form of Z-estimators \citep{van1996weak}. All proofs are given in the appendix. 


\begin{theorem}
\label{theo:prate}
Under Assumptions \ref{as:sampling},  \ref{as:interference}, \ref{as:market},  \ref{as:monotone} and \ref{as:regularity} and either Design \ref{def:rct} or \ref{def:aug}, 
the equilibrium price satisfies
\begin{equation}
\label{eq:prate_expansion}
\begin{split}
&P_n(\bm W)  - p^*_{\pi} =  -   \xi_z^{-1} \, \frac{1}{n}\sum \limits_{i=1}^n Z_i(W_i, p^*_{\pi}) + o_p(n^{-1/2}), \\
&\sqrt{n}\p{P_n(\bm W)  - p^*_{\pi}} \Rightarrow \nn\p{0, \,  \xi_z^{-1} \Var{Z_i(W_i, p^*_{\pi})} \p{ \xi_z^{-1}}^\top},
\end{split}
\end{equation}
where $\xi_z$ is as defined in Assumption \ref{as:market}.
\end{theorem}

This result is a crucial building block towards the rest of the theory in the paper, since it provides a representation of $P_n(\bm W)$  in terms of the fixed price $p^*_{\pi}$ and an average of IID terms. This asymptotic representation is valid whether or not the Bernoulli-randomized trial is augmented with price perturbations. It will be helpful in characterizing the asymptotic variance of various random variables in our model. Before introducing our estimators in Section \ref{sec:estimation}, we next apply this result to characterizing the relationship between the sample and population-level average direct and indirect effect.
To this end, we first show that $\tau^*_{\text{ADE}}$ and $\tau^*_{\text{AIE}}$ exist and have a simple representation in terms of moments of the population distribution. Our technical results rely heavily on
concentration results from empirical process theory as described in \citet{van1996weak}.


 \begin{theorem} \label{theo:pop}
Under Assumptions \ref{as:sampling},  \ref{as:interference}, \ref{as:market}, \ref{as:monotone}, \ref{as:lips} and \ref{as:regularity} and either Design \ref{def:rct} or \ref{def:aug},
the population estimands are
\begin{equation}
\begin{split}
 & \tau^*_{\text{ADE}} = y(1, p^*_{\pi}) - y(0, p^*_{\pi}), \\
& \tau^*_{\text{AIE}} =  - \xi_y^{\top}  \xi_z^{-1} [z(1, p^*_{\pi}) - z(0, p^*_{\pi})],
\end{split}
\end{equation}
where $\xi_y = \nabla_p \mathbb E[Y_i(W_i, p^*_{\pi})]$ is a $J \times 1$ vector
and $\xi_z$ is as defined in Assumption \ref{as:market}.
\end{theorem}

The population direct effect is the difference in expected outcomes at $p^*_{\pi}$ evaluated at $W_i = 1$ and $W_i = 0$. In the next section, we show that this can be estimated with a differences in means estimate. This finding is in line with \citet{savje2021average}, who show that estimators that target the average treatment effect under no-interference settings generally recover the average direct effect under interference. However, our results will go beyond those of \citet{savje2021average}, since we will also provide a central limit theorem for estimators of $\tau^*_{\text{ADE}}$.

The indirect effect, which is a key component of the policy counterfactual $\tau_{\text{MPE}}$, is not estimable using variation in treatment only. The population indirect effect depends on the product of three terms: how the treatment affects net demand, how net demand is affected by prices ($\xi_z$), and how changes in prices affect outcomes ($\xi_y$). In settings where the treatment impacts market-clearing prices through net demand, and outcomes are sensitive to market prices, then spillover effects and the indirect effect are stronger.  In Section \ref{sec:estimation} we will show how each component of $\tau^*_{\text{ADE}}$ and $\tau^*_{\text{AIE}}$ can be estimated using unit-level experiments.

Finally, we connect the limiting estimands $\tau^*_{\text{ADE}}$ and $\tau^*_{\text{AIE}}$ to their finite-sample
counterparts $\tau_{\text{ADE}}$ and $\tau_{\text{AIE}}$. In the case of the direct effect, Theorem \ref{theo:sade}
provides an asymptotically linear expansion for $\tau_{\text{ADE}}$ around $\tau^*_{\text{ADE}}$, thus allowing
us to disambiguate between the sample and population estimands when conducting inference. This
mirrors the well-known relationship between the sample- and population-average treatment effects
in the no-interference setting \citep{imbens2004nonparametric}.


\begin{theorem}
\label{theo:sade}

Under the conditions of Theorem \ref{theo:pop},
\begin{equation}
\label{eq:ade_asymp}
\tau_{\text{ADE}}  =  \frac{1}{n} \sum \limits_{i=1}^n  \Big (Y_i(1, p^*_{\pi}) - Y_i(0, p^*_{\pi})  - \p{\pi_i \Delta_i(1, \, p^*_\pi) + (1 - \pi_i) \Delta_i(0, \, p^*_\pi) } \Big ) + o_p(1/\sqrt{n}),
\end{equation}
where
\begin{equation}
\label{eq:Delta}
\Delta_i(w, \, p) = \nabla_p^{\top}[  y(1, p) - y(0, p) ] \xi_z^{-1}  Z_i(w, \, p).
\end{equation}
Furthermore, writing $\varepsilon_i(w) = Y_i(w, p^*_{\pi}) - y(w, p^*_{\pi})$, we have
\begin{equation}
\label{eq:ade_clt}
\sqrt{n}\p{\tau_{\text{ADE}} -  \tau^*_{\text{ADE}}} \Rightarrow \nn\p{0, \, \Var{ \varepsilon_i(1)   - \varepsilon_i(0)  - \p{\pi_i \Delta_i(1, \, p^*_\pi) + (1 - \pi_i) \Delta_i(1, \, p^*_\pi) }}}.
\end{equation}
\end{theorem}

Relative to the no-interference setting, the expansion in Theorem \ref{theo:sade} includes additional $\Delta_i$ terms  that depend on individual-level net demand. In a finite market where treatment affects demand and supply, there is variation in realized market prices that then impacts $\tau_{\text{ADE}}$. The $\Delta_i$ terms capture this additional source of variation.
We also note that, as in the no-interference setting, the variance of $ \tau_{\text{ADE}}$ is generally not identified, as it depends on the covariance of individual treated and control potential outcomes; however, as shown in Proposition \ref{prop:neyman}, it will still be possible to provide asymptotically conservative inference (i.e., construct confidence intervals with potentially greater-than-nominal coverage).



Meanwhile, for the indirect effect, $\tau_{\text{AIE}}$ converges to $\tau_{\text{AIE}}^*$, but Theorem \ref{theo:saie}
does not provide a rate of convergence. We conjecture that establishing useful rates of convergence here
would require replacing Assumptions \ref{as:market} and \ref{as:monotone} with stronger assumptions that
are more explicit about finite-sample price formation. Under our current assumptions, the convergence rate
of $\tau_{\text{AIE}}$ can be slower than $\sqrt n$; see Appendix \ref{ap:aierate} for additional explanation
in the context of a simple example.  Since our estimator in Section \ref{sec:estimation} targeting $\tau_{\text{AIE}}^*$
also converges at a slower than $\sqrt n$ rate, we conjecture that confidence intervals built for
$\tau^*_{\text{AIE}}$ will also generally cover $\tau_{\text{AIE}}$; we also note that this holds
in our numerical experiments.

\begin{theorem} \label{theo:saie}
Under the assumptions of Theorem \ref{theo:pop},
$\tau_{\text{AIE}} \overset{p}{\to}    \tau^*_{\text{AIE}}$,
\end{theorem}

\section{Estimation and Inference}
\label{sec:estimation}

We now move to estimation of the quantities defined in the previous section using data from unit-level randomized experiments. We first derive the limiting distribution of a differences-in-means estimator that is valid when data is generated from a standard RCT that is run on an entire market. This estimator is consistent for the average direct effect, but its asymptotic variance depends on price-elasticity terms that cannot be estimated using data from an RCT that randomizes treatment only. We show that if confidence intervals are constructed using a variance estimator based on the asymptotic variance without price interference, then coverage for the population direct effect will not be asymptotically exact.


In order to perform inference for the direct effect and to perform estimation and inference for the indirect effect, we need non-zero price shifters.
Algorithmically, the estimator for the indirect effect looks like a combination of differences-in-means estimator and an instrumental variables estimator. Under an augmented randomized experiment, we show how to construct asymptotically valid confidence intervals for population versions of both effects. In this section, we assume that experiments are run on the entire market of $n$ participants. However, as described in more detail in Appendix \ref{sec:subsample}, the extension to settings where only a sub-sample of the market is observed is straightforward.

\subsection{Estimation for the Direct Effect in a Standard RCT}

Following a number of recent papers, including \citet{savje2021average} and \citet{li2022random},
we consider estimating the direct effect using a differences-in-means estimator
\begin{equation}
\hat \tau_{\text{ADE}} = \frac{1}{n}  \sum \limits_{i=1}^n \left [  \frac{W_i Y_i }{\hat \pi}- \frac{(1-W_i)Y_i}{ 1 - \hat \pi} \right], \ \ \ \ \hat \pi = \frac{1}{n} \sum \limits_{i=1}^n W_i.
\end{equation}
Our first result is that---even under interference that occurs through market prices---the standard differences-in-means estimator converges to the population direct effect at a $\sqrt n$ rate and is asymptotically normal. The variance of the difference-in-means estimator depends both on the variance of the potential
outcomes---as in a standard no-interference setting---as well as an additional term ($\Delta_i(w, p^*_{\pi})$)
that is a function of net demand and price sensitivity of outcomes and net
demand.  \footnote{Our finding that the difference in means is consistent for the ADE mirrors general
findings in \citet{savje2021average}; however, the fact that we get a $\sqrt n$ rate of convergence
and a central limit theorem depends on our marketplace model.}

\begin{theorem}
 \label{theo:adeinf}
 Under the conditions of Theorem \ref{theo:pop} and with uniform treatment randomization probabilities $\pi_i=\pi$ for all $i$, under Design \ref{def:rct} or Design \ref{def:aug},
 \begin{equation}
 \hat \tau_{\text{ADE}} = \tau^*_{\text{ADE}} + \frac{1}{n} \sum \limits_{i=1}^n \p{\frac{W_i \varepsilon_i(1)}{\pi} - \frac{(1 - W_i) \varepsilon_i(0)}{1 - \pi} - \Delta_i(W_i, \, p^*_\pi)} + o_p(1),
 \end{equation}
 where $\varepsilon_i(w) = Y_i(w, p^*_\pi) - y(w, p^*_\pi)$ and $\Delta_i(w, \, p)$ is as in Theorem \ref{theo:sade}.
Furthermore,
\begin{equation}
 \sqrt{n} \left ( \hat \tau_{\text{ADE}}  - \tau^*_{\text{ADE}} \right)  \Rightarrow \mathcal{N} (0, \sigma^2_D), \ \ \ \
 \sigma^2_D = \EE{\p{ \frac{W_i \varepsilon_i(1)}{ \pi}  - \frac{( 1- W_i)  \varepsilon_i(0)}{  1- \pi }  - \Delta_i(W_i, p^*_{\pi})}^2}.
\end{equation}
\end{theorem}

Although we have shown that the familiar difference-in-means is asymptotically normal around the ADE, the
above result also implies that the usual asymptotic variance derived under the no-interference setting
does not match $\sigma^2_D$ in our market equilibrium model, unless $\Delta_i(W_i, p^*_{\pi}) = 0$. This term is zero if there is homogeneity in price derivatives under treatment and control, so that $\nabla_p[y(1, p) - y(0, p)] = 0$. In general, to construct confidence intervals that
are asymptotically exact for the population direct effect, we need estimates of the price elasticities that appear in the
$\Delta_i(w, p^*_{\pi})$ terms. In the next section, we will show how non-zero price perturbations under Design \ref{def:aug} 
allows us to estimate these price sensitivities as well as the AIE, which is not
estimable with treatment randomization only.

Finally, we note that Theorem \ref{theo:adeinf} quantifies the errors of $\hat \tau_{\text{ADE}}$
as an estimator of the population estimand $\tau^*_{\text{ADE}}$. In many settings, however, it is of primary interest to provide inference about the sample direct effect $\tau_{\text{ADE}}$, which is defined in a finite-sized market. Combining
Theorems \ref{theo:sade} and \ref{theo:adeinf}, we can verify that
\begin{equation} \label{eq:saded} 
\hat \tau_{\text{ADE}} - \tau_{\text{ADE}}  = \frac{1}{n} \sum \limits_{i=1}^n  (W_i - \pi) \left (\frac{ \varepsilon_i(1) }{\pi} + \frac{ \varepsilon_i(0)}{1 - \pi}  - \p{\Delta_i(1, p^*_{\pi}) - \Delta_i(0, p^*_{\pi})} \right ) + o_p\p{\frac{1}{\sqrt{n}}},
\end{equation}
and furthermore that
\begin{equation}
\begin{split}
&\sqrt n \p{\hat \tau_{\text{ADE}} - \tau_{\text{ADE}}} \Rightarrow \nn\p{0, \, \bsigma_D^2}, \\
&\bsigma_D^2 = \pi \p{1 - \pi} \EE{ \left (\frac{ \varepsilon_i(1) }{\pi} + \frac{ \varepsilon_i(0)}{1 - \pi}  - \p{\Delta_i(1, p^*_{\pi}) - \Delta_i(0, p^*_{\pi})} \right )^2}.
\end{split}
\end{equation}
The asymptotic variance here still depends on the market equilibrium effects; furthermore,
it depends on the correlation of $\varepsilon_i(1)$ and $\varepsilon_i(0)$ and so is generally
not identified. However, the following result shows that, in an extension of the classic result
of \citet{neyman1923applications}, $\sigma_D$ is a conservative upper bound for $\bsigma_D$ and
so confidence intervals built using Theorem \ref{theo:adeinf} that are exact for $\tau^*_{\text{ADE}}$
will also be conservative for $\tau_{\text{ADE}}$.\footnote{If
$\tau_{\text{ADE}}$ is the only estimand of interest, we can adapt results from
\citet{aronow2014sharp} to derive a tighter bound for $\bsigma^2_D$ that is estimable using price
perturbations as discussed in the following section; see  Appendix \ref{ap:bound} for details.}

\begin{proposition}
\label{prop:neyman}
Under the conditions of Theorem \ref{theo:adeinf}, $\bsigma^2_D \leq \sigma^2_D$.
\end{proposition}

\subsection{Augmented Randomized Experiment}
\label{sec:aug} 

%
%

We now consider use of randomized individual-level price-shifters as defined in Design \ref{def:aug} to accomplish further estimation and inferential tasks.
These price perturbations can be interpreted as experimenter-created instruments for estimating price derivatives locally.
They shift demand and supply in the neighborhood of the market-clearing price, and since they are randomized, they are independent of other variables that determine demand or outcomes.\footnote{The fact that we only consider
small, unobstrusive perturbations to the market also means we do not get to observe behaviors
at prices far from the large-sample equilibrium $p^*_\pi$, and so cannot identify or
consistently estimate full demand curves.}
 In a typical two-sided market, implementing the experiment requires introducing a small random discount, fee, or subsidy to both producers and consumers for any products where the treatment is expected to significantly affect supply or demand.\footnote{Our theory requires that the size of the price perturbation decreases with sample size. For a fixed market size, as the perturbation size increases, the variance of $\hat \tau_{\text{AIE}}$ decreases, but large price perturbations distort production and would be likely be viewed unfavorably by various market stakeholders. The simplest guidance for choosing the size of the price perturbation is to choose a value of $c$ so that for the sample size of the experiment, the size of the price perturbation is noticeable to market participants but limited to a small percentage of the current market price (e.g. 1-2\%). For markets where it is not possible for reputational or legal reasons to introduce any individual-level price variation, the components of $\tau^*_{\text{AIE}}$ can be estimated using price variation across products or time instead, at the cost of introducing additional structural assumptions on the environment.}

In Theorem \ref{theo:pop}, we found that $\tau^*_{\text{AIE}}$ can be expressed in terms of certain price elasticities and the direct effect of
treatment on net demand. Our first result using randomized price-shifters is that, using data collected under Design \ref{def:aug}, we can produce
a simple and consistent plug-in estimator for the indirect effect based on the functional form for $\tau^*_{\text{AIE}}$ derived in Theorem \ref{theo:pop}.
Let $\bm Y$ be the $n$-length vector of observed outcomes, where $Y_i = Y_i(W_i, P_n(\bm W), + U_i),$ $\bm U$ is the $n \times J$ matrix of price perturbations, and $\bm Z$ is the $n \times J$ matrix of observed net demand, where $Z_i = Z_i(W_i, \bm P_n(\bm W) + U_i)$. The estimator is
\begin{equation}
\hat \tau_{\text{AIE}} =  -  \hat \gamma^{\top} \cdot \hat {\tau}^{z}_{\text{ADE}}, \ \ \ \
\hat {\gamma} = (\bm U^{\top} \bm Z)^{-1} (\bm U^{\top} \bm Y),
\end{equation}
where $\hat {\gamma}$ is a $J \times 1$ vector that estimates $[\xi^{\top}_z]^{-1} \xi_y $.
The direct effect of the treatment on net demand is estimated via a difference-in-means estimator,
\begin{equation}
\label{eq:HTZ}
\hat {\tau}^z_{\text{ADE}} =  \frac{1}{n} \sum \limits_{i=1}^n \left [ \frac{W_i { Z}_i}{\hat \pi }-  \frac{(1- W_i) {Z}_i}{ 1 - \hat \pi } \right ] .
\end{equation}
We provide a central limit theorem for our indirect effect estimator below.
Its rate of convergence depends on the magnitude of the price perturbations $h_n$,
and is always slower than the $\sqrt{n}$-rate obtained for the direct effect.
In deriving this result, it is helpful to note that algorithmically $\hat \tau_{\text{AIE}}$ is the product of a rescaled instrumental variables estimator and a differences in mean estimator.
We can then use standard results on the asymptotic behavior  of IV estimators to guide our analysis.

  \begin{theorem}
  \label{theo:aieinf}
Suppose the Assumptions of Theorem \ref{theo:pop} hold and that treatment and price-shifters are assigned according to Design \ref{def:aug}. Then, the estimated average indirect effect can be written as:
   \begin{equation}
   \hat \tau_{\text{AIE}} = \tau^*_{\text{AIE}} -    \frac{1 }{n h_n^2} \, \sum \limits_{i=1}^n  Q_z^{\top} U_{i}  \nu_i(W_i)  + o_p\p{\frac{1}{\sqrt{n} h_n}},
   \end{equation}
  where $\nu_i(W_i) = Y_i(W_i, p^*_{\pi}) - Z_i(W_i, p^*_{\pi}) ^{\top} [\xi_z^{-1}]^{\top} \xi_y $ and $Q_z = \xi_z^{-1} [\tau^{*,z}_\text{ADE}]$.
Furthermore, this estimator satisfies a central limit theorem
\begin{equation}
\sqrt{n}{h_n} \left ( \hat \tau_{\text{AIE}}  - \tau^*_{\text{AIE}} \right)  \Rightarrow \mathcal{N}(0,  \sigma^2_I), \ \ \ \ \sigma^2_I = Q_z^{\top} \mathbb E[\nu^2_i(W_i) \, I_{J \times J}] Q_z
\end{equation}
where  $I_{J \times J}$ is the $J \times J$ identity matrix.
  \end{theorem}

Under general patterns of interference, inference on the indirect effect is challenging \citep{savje2021average, li2022random}, and consistent estimators for the variance of the indirect effect are generally not available. There are two reasons why our paper overcomes this difficulty. Although our interference pattern is dense, it is structured in that all interference happens through the market price, and the market price forms by satisfying a score condition. This structure leads to an analytical functional form for the variance of the indirect effect, as given above. Second, with data from a richer randomized experiment that includes small price perturbations, we are able to estimate each component of this variance, and the variance for the indirect effect.

We estimate $\sigma_D^2$ and $\sigma_I^2$ via natural plug-in estimators.
Let $n_w$ denote the number of units with $W_i = w$. The variance of the direct effect can be estimated as
\begin{equation}
\label{eq:sigmaD}
\begin{split}
& \hat \sigma^2_D = \frac{1}{n} \sum \limits_{i=1}^n \left [ \frac{W_i \hat \varepsilon_i(1) }{ \hat \pi} - \frac{(1 - W_i )\hat \varepsilon_i(0)} {1 - \hat \pi} -   (\hat {\xi}_{y1} - \hat {\xi}_{y0})^{\top} [\hat {\xi}_z]^{-1} Z_i \right ]^2, \\
& \hat \varepsilon_i(w) = Y_i  - \frac{1}{n_w} \sum \limits_{i : W_i = w}  Y_i, \\
\end{split}
\end{equation}
where,
for $w \in \{0, 1\}$, $\hat {\xi}_{yw}$ is a  $J \times 1$ vector estimated from regressions of $Y_i$ on $U_i$ using only observations with indices in the set $\{ i: W_i = w \}$, and $\hat {\xi}_z $ is a $J \times J$ matrix which is computed via regressions of net demand $Z_{ij}$ on price perturbations $U_i$ for $j \in \{1, \, \ldots, J\}$ and $\hat {\xi}_y$ is a $J \times 1$ vector computed via a regression of $Y_i$ on $U_i$. Meanwhile, for the variance of the indirect effect, the natural plug-in estimator is\footnote{In Appendix \ref{ap:2ndorder}, we also describe a second order adjustment to $\hat \sigma^2_I$ that is asymptotically negligible but yields better coverage under in our experiments when the sample size is moderate.}
\begin{equation}
\label{eq:sigmaI}
\hat \sigma^2_I =   \frac{1}{n \sqrt h_n} \sum \limits_{i=1}^n  \left ( (Y_i -{ Z}_i^{\top} [\hat {\xi}^{\top}_z]^{-1}\hat { \xi}_y) U^{\top}_{i} \hat {\xi}_z ^{-1} \hat {\tau}^z_{\text{ADE}} \right)^2
\end{equation}
These variance estimates can then be paired with our asymptotic normal approximation to build confidence intervals for the direct and indirect estimates; for example, for a confidence level of 95\%, the confidence intervals are constructed as
\begin{equation}
\hat{\tau_D} \pm 1.96\cdot \frac{ \hat \sigma_D}{\sqrt n}, \ \ \ \ \ \hat{\tau_I} \pm 1.96 \cdot \frac{\hat \sigma_I}{\sqrt n h_n}.
\end{equation}
The following result verifies validity of these variance estimators.

\begin{theorem}
  \label{theo:varest}
 Under the Assumptions of Theorem \ref{theo:aieinf},  $ \hat {\sigma}^2_D \overset{p}{\to} \sigma^2_D$ and $ \hat {\sigma}^2_I \overset{p}{\to} \sigma^2_I$.
  \end{theorem}

\section{Heterogeneous Treatment Effects}
\label{sec:het}

So far, we defined a potential outcomes model that captured equilibrium interference but reduced to the Neyman-Rubin model without interference. We then used this model to define average direct and indirect treatment effects under general treatment allocation rules. In the previous section, we proposed estimators for these effects that relied on data generated from randomized trials where the treatment was assigned with constant probability. In this section, we return to general treatment rules and discuss heterogeneous effects and optimal targeting when there is interference through an equilibrium statistic.

The planner controls the (potentially randomized) treatment allocation function $\pi(\cdot): \mathcal X \rightarrow [0, 1]$ where $\pi(x)= \PP{W_i = 1 \cond X_i = x}$. The conditional expectation functions are defined as $y(w, p, x) = \mathbb E[ Y_i(w, p) \cond X_i = x]$ and $z(w, p, x) = \mathbb E[Z_i(w, p) \cond X_i = x]$. We are interested both in quantifying how relevant treatment effects vary with $x$, and how this information can be used to guide choices of $\pi(\cdot)$ that achieve better outcomes. 

\subsection{Definitions of Conditional Treatment Effects Under Interference}

Under SUTVA, the conditional average treatment effect (CATE) is defined as $\tau(x) = \mathbb E[Y_i(1) - Y_i(0) | X_i = x]$ \citep{imbens2004nonparametric}.  Without interference, the optimal unconstrained targeting rule allocates treatments only to those with a positive CATE \citep{manski2004statistical}, or potentially to those whose CATE exceeds a budget-informed threshold \citep{bhattacharya2012inferring}.
Such treatment assignment rules, however, are no longer optimal under interference---and the CATE is no longer even well defined.

We begin this section by proposing two definitions of conditional estimands that play a similar role to the average direct and indirect effect, but for targeted treatments:
The Conditional Average Direct Effect (CADE) and the Conditional Average Indirect Effect (CAIE). The CADE
 \begin{equation}
 \label{eq:CADE}
  \bar \tau_{\text{CADE}}(x)  = \mathbb E[Y_i(W_i = 1; W_{-i}) - Y_i(W_i = 0; W_{-i}) | X_i = x]
 \end{equation}
 is the expected effect of treating an individual with covariate value $x$ on their own outcomes in a system of $n$ individuals. The CAIE
 \begin{equation}
  \bar  \tau_{\text{CAIE}}(x)  = (n-1) \mathbb E[Y_{j}(W_i = 1; W_{-i}) - Y_{j} (W_i =0 ; W_{-i}) | X_i = x]
 \end{equation}
 is the expected effect of treating an individual with covariate value $x$ on everyone else's outcomes in a system of $n$ individuals. These estimands are non-random quantities that depend both on the population distribution and on the market size. They are policy-relevant for settings where we expect to deploy a policy in a market of similar size and composition to the observed market. When interference is removed, the CADE is equal to the CATE, and the CAIE is zero.


We next connect these definitions of heterogeneous effects to conditional marginal effects; this result is valid under general patterns of interference. In Section \ref{sec:method}, we reported the results of  \citet{hu2021average}, showing the sum of $\tau_{\text{ADE}}$  and $\tau_{\text{AIE}}$ is equal to the effect on average outcomes of a marginal increase in each individual's treatment probability.  Proposition \ref{prop:hte} extends this result to our proposed heterogeneity measures.

\begin{proposition} \label{prop:hte}
Let $Y_i(\wvec)$ be potential outcome functions with an arbitrary sampling distribution,
and let treatment be generated as $W_i \sim \text{Bernoulli}(\pi_i)$ with treatment assignment
probabilities $0 < \pi_i < 1$ that may be dependent on the $Y_i(\wvec)$. Then,
\begin{equation}
\bar \tau_{\text{MPE}}(x) = \bar \tau_{\text{CADE}}(x) + \bar \tau_{\text{CAIE}}(x), \ \ \ \
\bar \tau_{\text{MPE}}(x) := \mathbb E \left [ \frac{\partial}{\partial \pi_k} \sum \limits_{i=1}^n \mathbb E_{\pi}[Y_i(\bm W) ]\Big | X_k = x\right].
\end{equation}
\end{proposition}

A policymaker may be interested in taking advantage of heterogeneous responses to treatment in the population by implementing a targeting rule, rather than a treatment rule with uniform probability. Proposition \ref{prop:hte} shows that the sum of the CADE and the CAIE is relevant for making the decision on which group's treatment probability to increase, and which to decrease, when the objective is maximizing expected outcomes.

The next step, in Theorem \ref{theo:hte}, is to derive the population versions of the CADE and CAIE. These are purely population estimands, defined as the limit of $\bar \tau_{\text{CADE}}$ and $\bar \tau_{\text{CAIE}}$ as the market size grows to infinity. We will show below that these quantities are relevant to treatment targeting in a market that is larger but otherwise similar to the observed market. The population estimands are also useful for deriving estimators that are consistent for both the large-market and finite-market estimands. Theorem \ref{theo:hte} shows that the population estimands have a simple and interpretable form and implies that estimators for one class of estimands are consistent for the other.

\begin{theorem} \label{theo:hte}
Under the Assumptions of Theorem \ref{theo:pop}, and the additional assumption that for all $x \in \mathcal S$, $y(w, p, x)$ and $z(w, p, x)$ is continuously differentiable in $p$ for all $p \in \mathcal S$. Then, the population conditional average direct effect is:
\begin{equation}
 \lim \limits_{n \rightarrow \infty} \bar \tau_{\text{CADE}}(x) = \tau^{*}_{\text{CADE}}(x) =  y(1, p^*_{\pi}, x) - y(0, p^*_{\pi}, x).
\end{equation}
The population conditional average indirect effect is:
\begin{equation}
\tau^*_{\text{CAIE}}(x) =   -\xi_y^{\top} \xi_z^{-1}\tau^{*,z}_{\text{CADE}}(x) , \ \ \ \
\tau^{*,z}_{\text{CADE}}(x) =  z(1, p^*_{\pi}, x) - z(0, p^*_{\pi}, x),
\end{equation}
and is the limit of $ \bar \tau_{\text{CAIE}}(x)$, where the convergence is over all sets with positive measure:
\begin{equation}
\lim \limits_{n \rightarrow \infty}  \EE{\bar \tau_{\text{CAIE}}(X) - \tau^*_{\text{CAIE}}(X) \cond X \in S}  = 0
\end{equation}
for all sets $S \subseteq \xx$ such that $\PP{X \in S} > 0$.
\end{theorem}

The population estimands have a simple form that suggests estimation strategies for the estimands $\bar \tau_{\text{CADE}}(x)$ and $\bar \tau_{\text{CAIE}}(x)$. The limit of the CADE is the direct treatment effect on outcomes conditional on $x$, holding the equilibrium price fixed. The limit of the CAIE is the direct treatment effect on net demand conditional on $x$, multiplied by an elasticity correction that does not depend on $x$. The augmented randomized experiment from Design \ref{def:aug} can be used to estimate the elasticity corrections and the conditional average treatment effects required to estimate $ \bar \tau_{\text{CADE}}(x)$ and $\bar \tau_{\text{CAIE}}(x)$.

For the elasticity corrections, the market price is an aggregate statistic, so individuals with different covariates all respond to the same market prices. A change in net demand of a given size always has the same impact on the market price. Although individuals' responses to the treatment through outcomes or net demand are heterogeneous, the elasticity correction that transforms the $\tau_{\text{CADE}}^{*,z}(x)$ to $\tau_{\text{CAIE}}^*(x)$ is unconditional. This implies that a group of individuals' effect on the system depends on their covariates only through conditional direct effects. Estimators for $\xi_y$ and $\xi_z$ from the previous section of the paper apply directly.

Below, we show that the $k$-nearest neighbor estimator is consistent for
$\tau_{\text{CADE}}^{*}(x)$; the same result also immediately holds for $\tau_{\text{CADE}}^{*,z}(x)$.
The salient fact in establishing this result is that the $k$-nearest neighbor estimator is
a difference-in-means estimator that has been localized in $X$-space; and the
proof suggests that other standard CATE estimators that are effectively localized
difference-in-means estimators, such as the causal trees of \citet{athey2016recursive},
are also consistent for the CADE. In our experiments, we use causal forests as implemented
in the \texttt{grf} package of \citet{athey2019generalized} to estimate the CADE.

\begin{theorem}
\label{theo:cade_est}
Let $N_k(x)$ be the $k$ closest observations to $x$ in terms of
covariate distance $\Norm{X_i - x}_2$, breaking ties randomly if needed.
Suppose that we collect data under Design \ref{def:rct}, and
construct the $k$-nearest neighbor estimator for $\tau_{\text{CADE}}(x)$ as
\begin{equation}
\htau_{\text{CADE}}(x) = \frac{\sum_{\cb{i \in N_k(x) : W_i = 1}} Y_i}{\abs{\cb{i \in N_k(x) : W_i = 1}}} -
\frac{\sum_{\cb{i \in N_k(x) : W_i = 0}} Y_i}{\abs{\cb{i \in N_k(x) : W_i = 0}}}.
\end{equation}
Under the assumptions of Theorem \ref{theo:hte}, suppose furthermore
that the conditional distribution of $\cb{Y_i(w, \, p), \, Z_i(w, \, p)}$ given
$X_i = x$ varies continuously in $x$. Then, given
any sequence $k \rightarrow \infty, \, k/n \rightarrow 0$, the $k$-nearest
neighbor estimator is consistent,
\begin{equation}
\htau_{\text{CADE}}(x) \rightarrow_p \tau_{\text{CADE}}^*(x),
\end{equation}
at any point $x \in \xx$ with positive local mass, i.e.,
with $\PP{\Norm{X_i - x}_2 \leq \varepsilon} > 0$ for any $\varepsilon > 0$.
\end{theorem}




\subsection{Equilibrium-Stable Targeting}
\label{sec:eqmstable}

The results of the previous section imply that estimates of $\tau_{\text{CADE}}^*(x)$ and $\tau_{\text{CAIE}}^*(x)$
could be used to (locally) optimize an unconstrained targeting policy. There are a variety of constrained targeting rules that may be of interest. Here, we focus on one specific question of this type, namely what is the optimal
treatment assignment policy that does not move equilibrium prices relative to those seen in the experiment?
There are two reasons to consider this class of targeting rules. First, a conservative policymaker may be
reluctant to significantly modify the equilibrium---even if it is beneficial on average to individuals---and
so they may want to know how much they can improve outcomes without changing the equilibrium. Second, from a
practical point of view, we find that the answer to this question admits a simple econometric strategy, and
can be answered without needing to estimate price elasticities and without recourse to an augmented experimental
design. Instead, the optimal equilibrium-stable targeting rule (and its performance) can be estimated using
data from a baseline RCT following Design \ref{def:rct}, as long as the RCT collects outcome and relevant
supply and demand data at an individual level.

The optimization problem for the equilibrium-stable policy is to find a new policy $\nu(\cdot)$ that maximizes expected outcomes in the population, while maintaining the population equilibrium price obtained under the policy $\pi(\cdot)$, i.e., $p^*_{\pi}$. The value of a given policy $\nu(\cdot)$ is: 
\begin{equation}
\label{eq:target_mean_field}
V(\nu) =  \limn \EE{\EE[\nu]{Y_i}} = \EE{(1 - \nu(X_i)) Y_i(0, \, p^*_\nu) + \nu(X_i) Y_i(1, \, p^*_\nu) }, 
\end{equation}where $\EE[\nu]{Y_i}$ denotes expected rewards under the considered new policy $\nu(\cdot)$. The optimal policy is defined as 
\begin{equation}\label{eqn:target}
\nu^*(\cdot) = \stackrel[{\nu \, : \, \xx \rightarrow [0, \, 1]}]{}{\argmax} \cb{  V(\nu):  p^*_{\nu} = p^*_{\pi} },
\end{equation}
where $p^*_{\nu}$ is the population equilibrium price under $\nu(\cdot)$.

When the targeting policy is restricted to have
the same equilibrium effect as a baseline policy, then solving for that optimal policy takes
the form of a linear optimization problem.

\begin{proposition}\label{prop:opt}
Under the assumptions of Theorem \ref{theo:pop},
optimizing the target \eqref{eq:target_mean_field} across all asymptotically
equilibrium-stable policies is equivalent to solving the following linear optimization problem:
\begin{equation}
\label{eq:LP_raw}
\nu^*(\cdot) = \stackrel[{\nu \, : \, \xx \rightarrow [0, \, 1]}]{}{\argmax} \cb{\EE{\nu(X_i) \, \tau^*_{\text{CADE}}(X_i)} : \EE{\p{\nu(X_i) - \pi(X_i)} \tau^{*,z}_{\text{CADE}}(X_i)} = 0}.
\end{equation}
\end{proposition}

The above reveals that we can solve for the optimal price-stable policy without
access to price elasticities. Thus, we can obtain a plug-in estimate for the
optimal rule using CADE estimates derived from an RCT without price perturbations
as in Design \ref{def:rct}; recall that the CADE itself can be identified without
price perturbations (Theorem \ref{theo:cade_est}).
Furthermore, following \citet{dantzig1951fundamental}, we can verify that
its solution takes on a simple parametric
form: The optimal price-stable policy is a thresholding rule that compares
the CADE for the outcomes against a shadow cost of net demand effects
$ c \cdot \tau_{\text{CADE}}^{*,z}(x)$, where $c \in \RR^J$ can be interpreted
as a shadow-price vector.

\begin{theorem} \label{theo:ratio}
Under the conditions of Proposition \ref{prop:opt}, there exists a vector $c \in \RR^J$
for which the asymptotic equilibrium-stable targeting problem \eqref{eq:LP_raw} admits a solution $\nu(\cdot)$ with the following property:
$\nu(x) = 1$ whenever $\tau^{*}_{\text{CADE}}(x)  > c^{\top} \, \tau_{\text{CADE}}^{*,z}(x)$ and
$\nu(x) = 0$ whenever $\tau^{*}_{\text{CADE}}(x)  < c^{\top} \, \tau_{\text{CADE}}^{*,z}(x)$. Furthermore,
if there exists $b \in [0, 1]$ for which the policy
\begin{equation}
\label{eq:nuca}
\nu_{c,b}(x) = \begin{cases}
1& \text{if} \ \ \tau^{*}_{\text{CADE}}(x)  > c^{\top} \, \tau_{\text{CADE}}^{*,z}(x), \\
b  & \text{if} \ \ \tau^{*}_{\text{CADE}}(x)  = c^{\top} \, \tau_{\text{CADE}}^{*,z}(x), \\
0 & \text{else,}
\end{cases}
\end{equation}
satisfies the constraints in \eqref{eq:LP_raw}, then this policy is optimal.
\end{theorem}

We can estimate the optimal equilibrium-neutral rule by finding a rule that meets an empirical version of the conditions in Theorem \ref{theo:ratio}, where $\tau^{*}_{\text{CADE}}(x)$ and $ \tau_{\text{CADE}}^{*,z}(x)$ are replaced with appropriate estimators. With consistent estimators of the conditional average direct effects, such as the estimator proposed in Theorem \ref{theo:cade_est}, then the estimated equilibrium-neutral rule is consistent for the population equilibrium-neutral rule under some smoothness assumptions on the distribution of conditional average treatment effects. A description of the estimation procedure and a formal consistency result is provided in Appendix \ref{as:target_con}. 

The targeting rule in Theorem \ref{theo:ratio} takes on a particularly simple form
when we only have $J = 1$ good, and the intervention has a crowding-out type structure
where direct effects are positive but spillovers are negative,
$\tau^{*}_{\text{CADE}}(x) > 0$ and $\tau_{\text{CADE}}^{*,z}(x) < 0$. In this case,
the optimal price-stable targeting rule will be of the form
\begin{equation}
\nu^*(x) = 1\p{\cb{\frac{\tau^{*}_{\text{CADE}}(x)}{-\tau^{*,z}_{\text{CADE}}(x)} > c}}
\end{equation}
with potential random tie-breaking at the cutoff. In other words, we want to give treatment
to units whose direct effects are large relative to corresponding spillovers; we then
pick $c$ to satisfy the price-stability constraint. This
algorithmic structure is familiar from the literature on cost-sensitive treatment
targeting \citep[e.g.,][]{sun2021treatment}.

\section{Example: Cash-Transfer Experiments}
\label{sec:sim}

In this section, we estimate a simple model of the direct and indirect impacts of a conditional cash transfer on children's health outcomes in the Philippines using data from \citet{filmer2023}. We use data simulated from the model to illustrate the performance of the average direct and indirect treatment effect estimators and the estimated targeting rule proposed in this paper.
\citet{filmer2023} use a village-level, cluster-randomized experiment to analyze the effect of the Pantawid conditional cash transfer program on young children's health outcomes. The program provides a monthly sum for eligible families, where the amount depends on the number of children in the household. Families are eligible if they meet a proxy means test and have either children under 14 years of age or a pregnant woman in the household. The authors show that the health of eligible children in treated villages is better than in control villages. However, children whose families are ineligible for the transfer are worse off in treated villages compared to control villages.

The authors argue that this phenomenon is due to spillover effects, which largely arise from equilibrium effects in the market for perishable protein sources. Specifically, the authors provide evidence that eggs, which are a commonly consumed perishable protein source in the Philippines, have higher prices in treated compared to control villages where saturation of eligible households is high. This price effect is explained by the increase in demand for protein from eligible families and inelastic supply in remote villages. The higher prices in treated villages lead ineligible children to consume less protein, which explains a significant proportion of the negative effect of the cash transfer program on ineligible children. It thus appears that Assumption \ref{as:interference} is reasonable, in that the spillovers described in detail by the authors are restricted to those that occur through market prices.

\subsection{Treatment Effects in a Village Economy}
\label{sec:sim_exp}

Under the assumption that markets in remote villages are isolated from one another, the village-level randomized experiment allowed \citet{filmer2023} to recover treatment effects that include equilibrium effects.
The total effect $\tau_{\text{TOT}}$ could be estimated using data from the cluster-randomized experiment by comparing the average outcomes of all children in treated compared to control villages.
In this paper, in contrast, our focus is on randomized experiments where only a single market is observed, but randomization is possible at the household level. While our econometric setting is different from that of \citet{filmer2023}, we can use their dataset to conduct a calibrated evaluation of our approach. We first use moments from the cluster-randomized experiment in \citet{filmer2023} to estimate a model of supply, demand, and children's outcomes in a single remote village market. Then, we use this model to simulate a single-village, unit-randomized experiment, and assess the ability of our estimators of the average direct and indirect treatment effects to recover equilibrium effects from this simulated experiment.

Let $i \in \{1, \, \ldots, n \}$ index the people in the village and $h \in \{1, \, \ldots n_h \}$ index the households in the village. For each $h \in \{1, \, \ldots, n_h \}$, $A^h$ is a set of integers containing the indexes of the people that are part of household $h$ and $C^h$ is a set of integers containing the indexes of the children aged 0-5 that are part of household $h$. $h(\cdot)$ is a map that provides the household index given a person's index. The binary treatment $W_h \in \{0, 1\}$ is assigned at the household level. $E_h \in \{0, 1\}$ indexes the eligibility for a household.
We assume that individual demand in eggs per week is determined by the following linear equation, where $p$ is the price of eggs:
\[ D_i(W_h, p) = \theta_{d01} \cdot E_{h(i)} + \theta_{d00} \cdot ( 1- E_{h(i)})  + \theta_{dw} \cdot W_{h(i)} \cdot E_{h(i)} + \theta_{dp} \cdot p + \varepsilon_{d, h(i)} + \nu_{d, i}(W_i), \]
where the household and individual-level error terms are normally distributed. Demand for eggs is increased for eligible adults and children in households that receive the Pantawid conditional cash transfer. Demand also responds to the market price of eggs.

We assume that the aggregate supply  per person in the village is responds linearly
to prices, $S(p) = \theta_{s0} + \theta_{sp} \cdot p$.
This formulation assumes that the village supply is not directly affected by the treatment; in \citet{filmer2023}, the authors argue that egg supply in the Philippines is dominated by larger commercial producers, which makes this a reasonable assumption.

Our target outcome is the height-for-age Z-score for children between 0 and 5 years of age. For our
evaluation, we assume that these outcomes are generated as
\[ Y_i (W_{h(i)}, p)  = \theta_{y01}E_{h(i)} + \theta_{y00} (1 - E_{h(i)}) + \theta_{yd} D_i(W_{h(i)}, p) + \theta_{yw} W_{h(i)} + \varepsilon_{y, h(i)} + \nu_{yi}(W_i),  \]
where the household and individual-level error terms are normally distributed. Consumption of a perishable protein source increases child-level health outcomes. The conditional cash transfer also has impact on health metrics through other factors, which is captured by $\theta_{yw}$.

For the purpose of our analysis, we aggregate individual level outcomes at the household level:
\begin{equation}
\label{eqn:aggh}
 D_h(W_h, p) = \frac{n_h}{n} \sum \limits_{i \in A^h}  D_i(W_h, p), \qquad Y_h(W_h, p) = \frac{n_h}{n_c} \sum \limits_{i \in C^h} Y_i(W_{h(i)}, p).
 \end{equation}
The factor $\frac{n_h}{n}$ ensures that average treatment effects on household-level demand can be interpreted as the treatment effect on eggs consumed per-person in the village. The factor $\frac{n_h}{n_c}$ ensures that average treatment effects on household-level outcomes can be interpreted as the treatment effect per child aged 0 to 5 in the village. The realized egg price in the market matches aggregate supply and demand:
\[ \frac{1}{n_h} \sum \limits_{h=1}^{n_h} D_h(W_{h(i)}, P(\bm W)) = S(P(\bm W)) \] Given a treatment vector $\bm W$, the realized household-level outcomes are $Y_h(\bm W) = Y_h(W_h, P(\bm W))$ and household level net demand is $Z_h(\bm W) = D_h(\bm W, P(\bm W)) - S(P(\bm W))$.

The 10 parameters in this model are estimated by matching 10 moments from a subset of the data in \citet{filmer2023}, restricted to households in villages above the 70th percentile of the remoteness index defined in the paper, where equilibrium effects were most meaningful. The 10 moments are: 
\begin{itemize}
\item  Average outcomes and demand for eggs for eligible and ineligible children in control villages;
\item The average price of eggs in treated villages and control villages;
\item The elasticity of demand for eggs in the Philippines reported in the paper;
\item The total treatment effect on outcomes and demand for eligible children;
\item The ratio of total treatment effects on outcomes and demand for ineligible children.
\end{itemize}
The total treatment effect here is the difference in average outcomes (or demand) in treated versus control villages. The estimation details and the estimated parameters of the model are reported in Appendix \ref{app:sim}.

Figure \ref{fig:curve} shows that aggregate demand in the village is decreasing in price, but is shifted upward by the cash transfer when eligible families are treated. Supply is increasing in price but is inelastic enough to explain the shift in prices observed in the data in treated villages. Figure \ref{fig:gte}  shows the distribution of $\tau_{\text{ADE}}$ and $\tau_{\text{TOT}}$, simulated using repeated samples of a village of 1000 households from the model. In this linear model, $\tau_{\text{TOT}} = \tau_{\text{MPE}}$. An RCT estimates the direct effect, which is more than 50\% higher than the total effect. Figure \ref{fig:curve} explains why the direct effect and the total effect are so different in this setting. The treatment raises demand for eggs among treated families, which impacts the market-clearing price for the treated market compared to the control market. Holding  food prices fixed, the effect of the treatment is a large positive increase in health outcomes. However, when the effect on food prices is taken into account, which impacts both eligible and ineligible children, the total treatment effect is much lower.


\begin{figure}

\centering
  \begin{subfigure}[b]{0.8\textwidth}
         \centering
         \includegraphics[width=\textwidth]{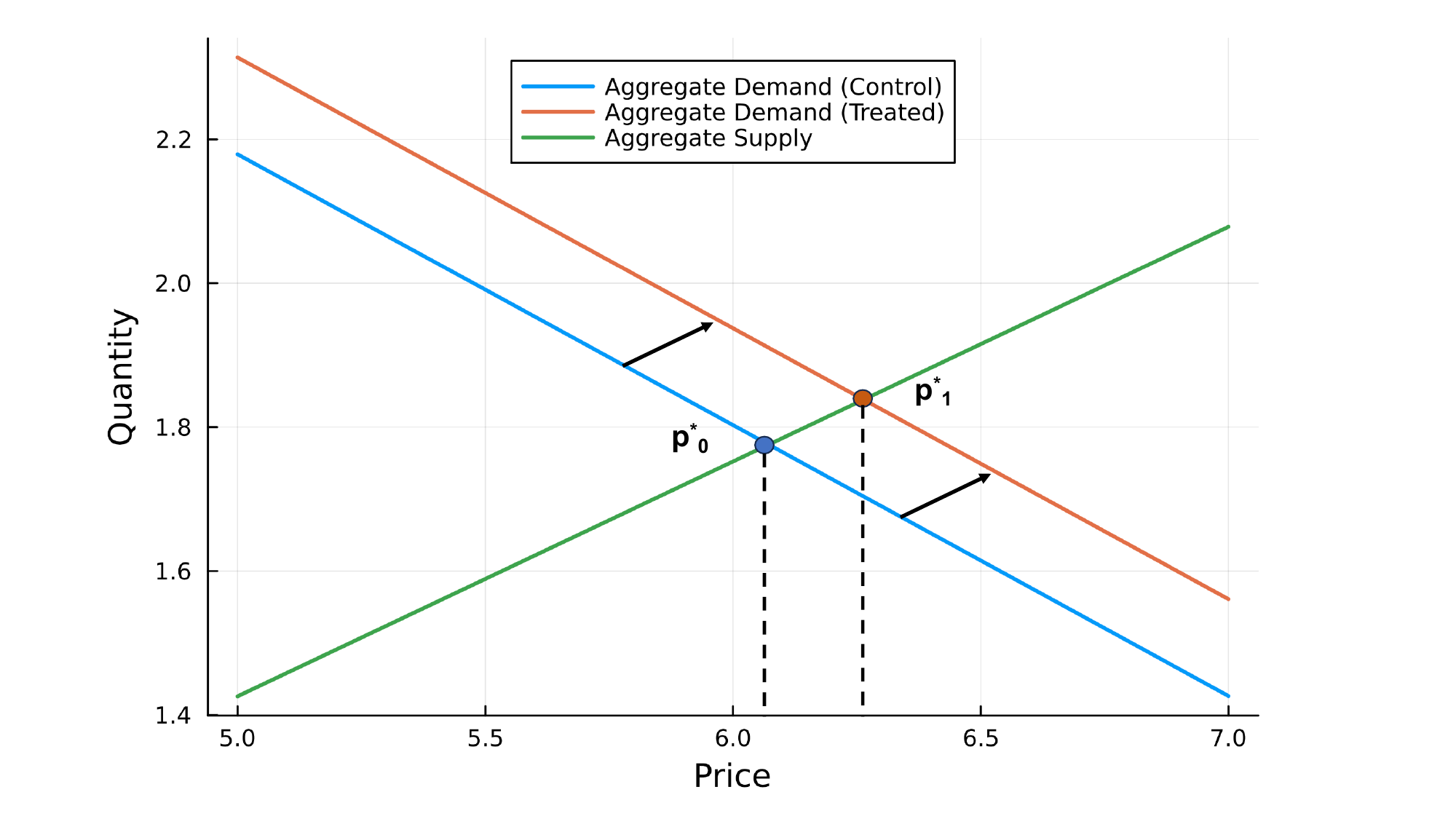}
         \caption{Estimated expected demand and supply under treatment and control. Prices are per egg in local currency. }
         \label{fig:curve}
     \end{subfigure}
     \begin{subfigure}[b]{0.7\textwidth}
         \centering
         \includegraphics[width=\textwidth]{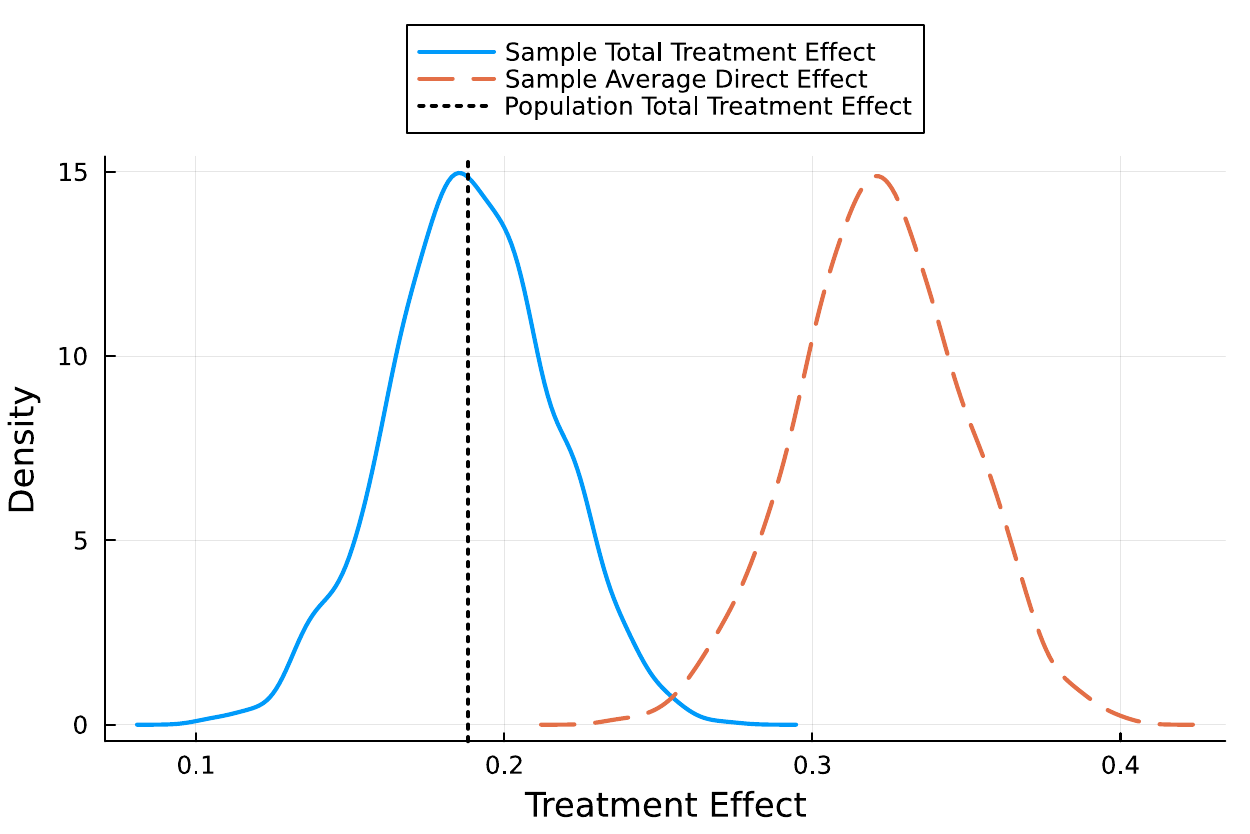}
         \caption{Smoothed empirical distribution of the average direct effect $\tau_{\text{ADE}}$ and the total effect $\tau_{\text{TOT}}$ for samples of a village of $1,000$ households.}
         \label{fig:gte}
     \end{subfigure}
     \hfill
     \caption{Illustration of the gap between $\tau_{\text{ADE}}$ and $\tau_{\text{TOT}}$ in the village market model. The intervention has both a direct effect on children's health and an offsetting indirect effect through increased food prices.}
\end{figure}

In Table \ref{tab:coverage}, we show that both the direct and the indirect effect are estimable in finite samples using our proposed unit level augmented experiment. The size of the price-shifters in the simulated experiment is 0.15 Philippine pesos per egg, which is less than 2.5\% of the market price. We report the results of a Monte Carlo simulation with 1,000 repetitions and a sample size of 2,000 households to evaluate the bias, variance, and coverage properties of the estimators and confidence intervals for $\hat \tau_{\text{ADE}}$ and $\hat \tau_{\text{AIE}}$. The table illustrates that the bias of the average direct effect and indirect effect estimator based on the augmented randomize experiment are low in finite samples. Furthermore, the coverage when $\hat \sigma_D$ and $\hat \sigma_I$ are used to construct confidence intervals in finite samples is slightly above the asymptotic confidence level of 95\% for both population estimands. As expected from Theorem \ref{theo:sade}, intervals that target $\tau^*_{\text{ADE}}$ are conservative for $\tau_{\text{ADE}}$.  For the AIE, in simulations the coverage for the sample estimand is also more conservative than for the population estimand. In this simple linear model, $\nabla_p y(1, p) = \nabla_p y(0, p)$, so $\Delta_i(W_i, p) = 0$ in \eqref{eq:saded}, and confidence intervals for differences in means estimation that ignore price fluctuations are asymptotically valid. The augmented randomized experiment, however, is required to estimated the AIE in this model and to build asymptotically valid confidence intervals for the ADE in more general settings with additional heterogeneity and non-linearity. 

\begin{table} [t]
\centering
\begin{tabular}{lrrrrr}
\toprule
 & Estimate & Bias & S.D. & Coverage for $\tau$ & Coverage for $\tau^*$ \\
\midrule
$\hat\tau_{\text{ADE}}$ & 0.317 & -0.005 & 0.123 & 0.965 & 0.959 \\
$\hat\tau_{\text{AIE}}$ & -0.132 & 0.002 & 0.082 & 0.964 & 0.961 \\
\bottomrule
\end{tabular}
\caption{Monte Carlo Simulation Results for $\hat \tau_{\text{ADE}}$ and $\hat \tau_{\text{AIE}}$ for a sample of $n=2,000$ and $1,000$ repetitions \label{tab:coverage}}
\end{table}

\subsection{Heterogeneous Treatment Effects in the Model}
\label{sec:sim_het}

In the setting of \citet{filmer2023}, there may be heterogeneity in both outcome and demand effects that is correlated with observed pre-treatment covariates. If so, a targeting rule can improve outcomes compared to a randomized rule that induces the same equilibrium. Since the replication package did not contain data on pre-treatment household characteristics that can be matched to post-treatment data, we cannot estimate heterogeneous treatment effects directly. Instead, we augment the model in Section \ref{sec:sim_exp} with an additional 10 generated household-level covariates that are observed by the planner, and a set of four heterogeneous household types that are correlated with the generated  covariates but unobserved. These household types are generated purely for this simulation exercise and are not calibrated to real data. Household Type A has the same data-generating process in the previous section. Household Type B is especially focused on their children, and purchases enough food that outcomes improve even more than for Type A. For Household Type C, which makes up a small percentage of the households, increased cash is not beneficial for child health outcomes, as it makes consuming some welfare-decreasing product possible for adults, and the budget share of food of the household decreases. For Household Type D, there is a small positive effect on children's health outcomes, but a large effect on demand for the perishable protein source; in these houses, adults consume most of the additional food that is purchased with the cash transfer. The details of the augmented model are in Appendix \ref{app:sim}. Then, we estimate the optimal equilibrium-stable targeting rule on data simulated from this model.

\begin{figure}
\centering
\includegraphics[width = 0.7\textwidth]{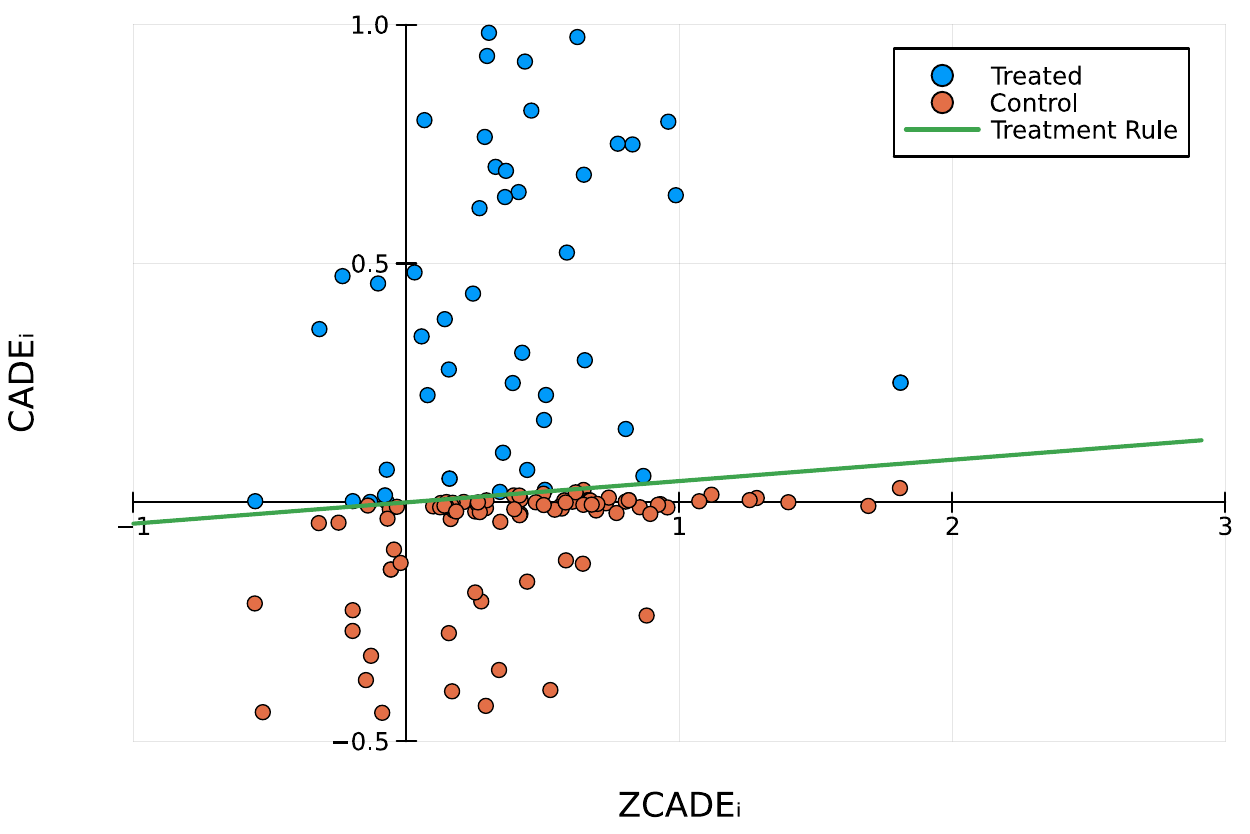}
\caption{A scatter-plot of an estimate of  $\tau_{\text{CADE}}(X_i)$ and $\tau^z_{\text{CADE}}(X_i)$ for a sample of 125 households, estimated using the causal forest of \citet{athey2019generalized}. Overlaid is the optimal equilibrium-stable targeting rule, computed by solving a plug-in version of the linear program in Proposition \ref{prop:opt} on a sample of 5,000 households, which takes the form of the rule in Theorem \ref{theo:ratio}.}
\label{fig:empirical}
\end{figure}

In Figure \ref{fig:empirical}, we illustrate the optimal rule estimated using a sample of 5,000 households. Ignoring equilibrium effects, a good targeting rule would allocate treatment to those individuals who are estimated to have a positive CADE. Here, however, this rule would result in a larger impact on perishable food demand compared to the RCT since, on average, those who increase their children's health using the cash transfer also increase their consumption of perishable food. In contrast, the optimal rule that respects the equilibrium constraint allocates treatment to those with $\hat \tau_{\text{CADE}}(X_i) > c\hat  \tau^z_{\text{CADE}} (X_i)$. We can see that for the most part, it is those with a positive CADE that are treated. However, households of Type D consume a lot of additional perishable food without raising a child's health outcomes  very much.  Dropping some of these households  from the equilibrium-stable targeting rule allows the planner to target more households from Type B while keeping the equilibrium food price stable.

\begin{table}
\centering
\begin{tabular}{lrrr}
\toprule
 & \textbf{Target-Optimal} & \textbf{Target-Direct} & \textbf{Target-Random} \\
\midrule
Estimated Height-For-Age Z-Score & -1.749 & -1.925 & -2.112 \\
Standard Deviation & 0.09 & 0.092 & 0.026 \\
\bottomrule
\end{tabular}
\caption{The first row is the average height-for-age Z-Score of children aged 0-5 in a test sample of 5,000 households, where outcomes are simulated from the model in Section \ref{app:het} under three different treatment rules, averaged across 50 simulations. For each of the 50 simulations, the treatment rules are estimated on a training sample of 5,000 samples. The second row is the standard deviation of the average outcomes across the 50 simulations. \textbf{Target-Random} treats a random 55\% of eligible households. \textbf{Target-Optimal} treats households according to treatment rule that solves an empirical version of the linear program in Section \ref{sec:eqmstable}, with $\hat \tau_{\text{CADE}}(x)$ and $\hat \tau^z_{\text{CADE}}(x)$ estimated using a causal forest. The equilibrium constraint ensures that the equilibrium price is the same as the Target-Random rule in the training sample. \textbf{Target-Direct} treats eligible households according to  $\tilde \pi(x) = \hat \alpha \cdot  \mathbbm{1} (\hat \tau_{\text{CADE}}(x) > 0)$. $\hat \alpha$ is chosen so that, in the training data, the total net demand under $\tilde \pi$ is equal to the total net demand under Target-Random.
\label{tab:empirical}}
\end{table}

We then evaluate the optimal equilibrium stable rule by estimating the rule using the procedure in Section \ref{sec:eqmstable} on a training sample of 5,000 households, and then evaluating the expected value of that rule numerically on a separated simulated dataset of 5,000 households. The results are reported in Table \ref{tab:empirical}. Note that our health outcome, which is average Z-Scores of children under  5 years in the simulated village, is negative since children living in remote and isolated villages in the Philippines are smaller than average. The closer Z-scores are to zero, the more a targeting rule has improved health outcomes for young children in the simulated village. 

Compared to Target-Random, Target-Direct avoids treating some of the few households for  which the cash transfer is not beneficial, which improves outcomes by 8.9\%. However, for equilibrium stability, $\hat \alpha < 1$, and the rule does not distinguish between households based on their demand for the perishable protein source. Target-Optimal has improved health outcomes since it treats more households with a large positive direct effect on outcomes and avoids treating households for which the treatment has a small impact on outcomes but a large impact on demand. It improves health outcomes by 17\% compared to Target-Random.


\section{Discussion}
\label{sec:conc}

Analyzing the performance of randomized control trials in settings with equilibrium effects is needed given the rapid growth of experimentation both in practice and in research studies. The Neyman-Rubin framework that relies on SUTVA rules out interaction effects that can have an important impact on decision-relevant treatment effects. A parametric structural model may capture a variety of complex equilibrium effects, but is not robust to misspecification, which can be problematic when individuals behave in complex and heterogeneous ways. A model of treatment effects under general patterns of interference is intractable without clustering or other assumptions.

This paper shows that it is fruitful to marry ex-ante knowledge about the structure of an economic environment with a non-parametric stochastic model of treatment effects under interference.  This leads to a characterization of asymptotic properties of treatment effects and estimators of those treatment effects based on new and existing experimental designs that are robust to a wide range of modeling choices. Results on estimation, inference, and optimal targeting in complex environments with some economic structure imposed can then be easily compared with the large body of work that studies causal inference under SUTVA.

Our results on the direct effect and on targeting rely on the assumption that all
spillovers are mediated by the prices of a finite number of traded goods; our
results for the indirect effect and for inference further rely on us being able to
exogenously apply small shifts to the prices that different market participants
are exposed to. As discussed in the previous section, such assumptions may be
relevant to experiments run in communities with self-contained markets.
\citet{aouad2024digitized} describe an experimental setting where researchers
open a subsidy store that they then use to randomize subsidy eligibility at
the unit level; and experiments of this type could plausibly also integrate
small, random price perturbations. Another class of applications our methods
are a natural fit for are online platforms for ride sharing, freelance labor,
short-term rentals, etc. Price-mediated spillovers are of central importance
when experimenting in such systems; for example, in ride sharing, both drivers and riders
respond to the average price of a mile and/or minute of transportation, and
both supply- and demand-side interventions may alter the market-clearing price
for transportation. Furthermore, technology companies running such platforms
have already documented willingness to run experiments comparable to the price
perturbations we describe, e.g., via random discounts or bonuses \citep{castillo2023,holtz2024reducing}.
There remains, however, a large class of settings where our methods cannot
be applied, e.g., when spillovers are not mediated via observed equilibrium
quantities \citep[e.g.,][]{cai2015social,manski1993identification}, or in
matching problems where equilibrium dynamics play a key role \citep{li2023experimenting}.

There are a variety of avenues for future work  possible. One limitation of our approach is that we analyze the large sample limit of the market place where the number of suppliers and the number of buyers grow large; an analysis of experiments in settings where firms have significant market power would likely require different techniques. A related potential direction for extending our results would be to allow for the market to depend on prices of a continuum of goods, e.g., localized prices that vary continuously in space. In general, extending our results to a broader class of equilibrium mechanisms would be of considerable interest.


\setlength{\bibsep}{0.2pt plus 0.3ex}

\singlespacing
\bibliographystyle{plainnat-abbrev}
{\footnotesize
\bibliography{../../glsBib.bib}
}

\newpage


\appendix

\section{Details from the Simulations in Section \label{app:sim}  }

\subsection{Estimation}

The model described in the main text captures the impact of a conditional cash transfer on children's health outcomes and the market for a perishable protein source in a remote village in the Philippines. It is designed based on the results in \citet{filmer2023}. 

There are 10 coefficients of the model $\bm \theta = \begin{bmatrix} \theta_{d00} & \theta_{d01} & \theta_{dw} & \theta_{dp} & \theta_{s0} & \theta_{sp} & \theta_{y00} & \theta_{y01} & \theta_{yd} & \theta_{yw} \end{bmatrix}^{\top}$. The model is estimated to match the following 10 moments from the subsample of remote villages in \citet{filmer2023}. These moments are computed using data files found in the replication package for the paper, see \citet{filmerdata}.
\begin{itemize}
\item The average demand for eggs per week among eligible and ineligible children in control villages $(\hat \mu^d_{elig}, \hat \mu^d_{inelig})$.
\item The average height-for-age Z-score among eligible and ineligible children in control villages ($\hat \mu^y_{elig}, \hat \mu^y_{inelig})$.
\item The average price of eggs in treated and control villages ($\hat p^0$ and $\hat p^1$). The elasticity of demand for individuals in control villages in the model matches the elasticity of egg demand cited, although not directly estimated in \citet{filmer2023} $(\eta_{egg})$.
\item The effect of the treatment on eligible children on height-for-age Z-score and eggs per week ($\hat \tau^y_{elig}, \hat \tau^d_{elig})$,  where treatment effects are defined as the differences-in-means for ineligible children in treated versus control villages.
\item The ratio of treatment effects on height-for-age Z-score and eggs-per-week demand for ineligible children ($\hat \pi = \hat \tau^y_{inelig}/\hat \tau^d_{inelig}$)\footnote{It is possible to estimate the model by matching the treatment effects on outcomes and demand directly for ineligible children, rather than their ratio. This would avoid using the elasticity of egg demand, which is taken from the literature rather than estimated from the data. However, the sample size of ineligible children in remote villages is small, so estimating the model to match these individual moments, which are imprecisely estimated, rather than their ratio, leads to an implied elasticity that does not seem reasonable.}. Again, treatment effects are defined as the differences-in-means for ineligible children in treated versus control villages.
\end{itemize}

We use these moments to estimate the model. The randomization of treatment makes estimating coefficients on treatments straightforward, because the errors in the demand equation are independent of $W_{h(i)}$. Under the assumptions in \citet{filmer2023}, the randomization of treatment also provides enough variation to estimate coefficients on prices in a linear demand and supply model. The coefficient on price in the demand model is provided directly by the elasticity of demand taken from the literature in \citet{filmer2023}. Given the demand elasticity, and under the assumption that $W_{h(i)}$ does not enter into the supply equation, then the change in prices in treated in and control villages provides an estimate for the supply elasticity under the market-clearing condition. The authors argue that treatments do not affect supply meaningfully, because most eggs sold in small villages are produced outside each village. Formally, each parameter has an estimator, which is a simple function of the moments. 

 \begin{equation*}
 \begin{split}
&  \hat \theta_{d00} = \hat \mu^d_{inelig} -   \hat \theta_{dp} \cdot \hat p^0, \qquad \hat \theta_{d01} =  \hat \mu^d_{elig} - \hat \theta_{dp} \cdot \hat p^0, \\
 & \hat \theta_{dp} = \frac{\eta_{egg}}{  1.01 \cdot \hat p_0 } \cdot \frac{1}{n} \sum \limits_{i=1}^n (E_i \hat \mu^d_{elig} + ( 1- E_i) \hat \mu^d_{inelig}), \qquad \hat \theta_{dw} = \hat \tau^d_{elig}  - \hat \theta_{dp} ( \hat p_1 - \hat p_0),  \\
 & \hat \theta_{sp} = \left (\frac{1}{n} \sum \limits_{i=1}^n E_i \hat \tau^d_{elig} + \frac{1}{n} \sum \limits_{i=1}^n ( 1- E_i) (\hat p^1 - \hat p^0) \hat \theta_{dp} \right ) \Big / (\hat p^1 - \hat p^0) \\
 & \hat \theta_{s0} = \frac{1}{n} \sum \limits_{i=1}^n (E_i \hat \mu^d_{elig} + ( 1- E_i) \hat \mu^d_{inelig}) - \hat \theta_{sp} \hat p_0, \\
 & \hat \theta_{yd} = \frac{ \hat \tau^y_{inelig}} { \hat \tau^d_{inelig} }, \qquad \hat \theta_{yw} = \hat \tau^y_{elig} - \hat \theta_{yd} \hat \tau^d_{inelig}, \\
 & \hat \theta_{y00} = \hat \mu^y_{inelig} - \hat \theta_{yd} \hat \mu^d_{inelig}, \qquad   \hat \theta_{y01} = \hat \mu^y_{elig} - \hat \theta_{yd} \hat \mu^d_{inelig}.
 \end{split}
 \end{equation*}

 The moments used for estimating the model and the corresponding parameter estimates are described in Table \ref{tab:sim_estimates}.

\begin{table} [ht]
\centering
\begin{tabular}{llllllllll}
\toprule
\multicolumn{10}{c}{Estimated Moments from \citet{filmer2023}} \\
\midrule
$\hat \mu^d_{elig}$ & $\hat \mu^d_{inelig}$ & $\hat \mu^y_{elig}$ & $\hat \mu^y_{inelig}$ & $\hat p^0$ & $\hat p^1$ & $\eta_{egg}$ & $\hat \tau^y_{elig}$ & $\hat \tau^d_{elig}$ & $\frac{ \hat \tau^y_{inelig}} {\hat \tau^d_{inelig}}$ \\
\midrule
1.6 & 2.2 & -2.47 & -1.44 & 6.07 & 6.26 & -1.3 & 0.32 & 0.12 & 1.85 \\
\midrule
\midrule
\multicolumn{10}{c}{Estimated Parameters} \\
\midrule
$\hat \theta_{d00}$ & $\hat \theta_{d01}$ & $\hat \theta_{dw}$ & $\hat \theta_{dp}$ & $\hat \theta_{s0}$ & $\hat \theta_{sp}$ & $\hat \theta_{y00}$ & $\hat \theta_{y01}$ & $\hat \theta_{yd}$ & $\hat \theta_{yw}$ \\
\midrule
4.49 & 3.89 & 0.19 & -0.38 & -0.21 & 0.33 & -5.51 & -5.43 & 1.85 & 0.1 \\
\bottomrule
\end{tabular}
\caption{Estimation of the Model in Section \ref{sec:estimation}}
\label{tab:sim_estimates}
\end{table}

Last, for simulating data from the model, $C_i$  and $h(i)$ are determined by random draws from the empirical distribution of households with children between 0 and 14 (who are in eligible for the transfer if they meet a proxy means test) from the baseline survey in \citet{filmer2023}. $E_i$ is drawn randomly to match the average percentage of eligible households in remote villages. The household and individual-level error terms in the demand equation are independent and mean-zero normally distributed variables with a standard deviation of 1/3rd. Similarly, the household level error terms in the outcome equation are independent and mean-zero normally distributed variables with a standard deviation of 1/3rd and the individual-level error terms have standard deviation of 1.


\subsection{Adding Heterogeneity}
\label{app:het}
The model in Section \ref{sec:sim_exp} is augmented to add some additional heterogeneity. For each household, a set of covariates $X_h \sim \mathcal{N}(0, 1)^{10}$ is drawn. Households now have type $\zeta_h \in \{A, B, C, D\}$. Households of type $\zeta_h = A$ have data generated by the model in the previous section. In response to the treatment, they increase their purchase of a perishable protein source, and children's health outcomes are improved on average.
\begin{equation*}
\begin{split}
& D^A_i(W_h, p) = \theta_{d01} \cdot E_{h(i)} + \theta_{d00}( 1- E_{h(i)})  + \theta_{dw} W_{h(i)}E_{h(i)}   +  \theta_{dp} \cdot p + \varepsilon_{d, h(i)} + \nu_{d, i}(W_i)\\
& Y^A_i (W_{h(i)}, p)  = \theta_{y01}E_{h(i)} + \theta_{y00} (1 - E_{h(i)}) + \theta_{yd} D^A_i(W_{h(i)}, p) + \theta_{yw} W_{h(i)}E_{h(i)}   +  \varepsilon_{y, h(i)} + \nu_{yi}(W_i)
\end{split}
\end{equation*}
For the other three types, treatment effects are distinct from Type A.  Households of type $\zeta_h = B$ increase their consumption of a perishable protein source and purchase a more expensive non-perishable protein source (such as canned fish). We assume for this second type of good, because it is non-perishable, supply is much more elastic and equilibrium effects are negligible. So, for these households, the data generating process is as follows:

\begin{equation*}
\begin{split}
& D^B_i(W_h, p) = D^A_i(W_h, p) + 0.75 \theta_{d01} \cdot E_{h(i)} \cdot W_{h(i)} \\
& Y^B_i (W_{h(i)}, p)  = \theta_{y01}E_{h(i)} + \theta_{y00} (1 - E_{h(i)}) + \theta_{yd} D^A_i(W_{h(i)}, p) + (\theta_{yw} + 0.2) W_{h(i)}E_{h(i)}   +  \varepsilon_{y, h(i)} + \nu_{yi}(W_i)
\end{split}
\end{equation*}

For households of type $\zeta_h = C$, the cash transfer does not have its intended positive effect. In response to the cash transfer, an increased ability of parents to afford addictive goods (such as cigarettes, or alcohol) leads to a decrease in the purchase of perishable protein, which has a large negative effect on children's health. Let $\mbox{Child}_i$ be an indicator if individual $i$ is under 5 years of age.

\begin{equation*}
\begin{split}
& D^C_i(W_h, p) = \theta_{d01} \cdot E_{h(i)} (1 - 0.75 W_{h(i)} \mbox{Child}_i) + \theta_{d00}( 1- E_{h(i)})   +  \theta_{dp} \cdot p + \varepsilon_{d, h(i)} + \nu_{d, i}(W_i)\\
& Y^C_i (W_{h(i)}, p)  = \theta_{y01}E_{h(i)} + \theta_{y00} (1 - E_{h(i)}) + \theta_{yd} D_i(W_{h(i)}, p) +  \theta_{yw} W_{h(i)}E_{h(i)}   +  \varepsilon_{y, h(i)} + \nu_{yi}(W_i)
\end{split}
\end{equation*}

For households of type $\zeta_h = D$, there is an increase in food consumption of adults, but not of children.

\begin{equation*}
\begin{split}
& D^D_i(W_h, p) = \theta_{d01} \cdot E_{h(i)} + \theta_{d00}( 1- E_{h(i)})   +  \theta_{dp} \cdot p + \varepsilon_{d, h(i)} +  1.5\theta_{dw} W_{h(i)}E_{h(i)} ( 1- \mbox{Child}_i)+ \nu_{d, i}(W_i)\\
& Y^D_i (W_{h(i)}, p)  = \theta_{y01}E_{h(i)} + \theta_{y00} (1 - E_{h(i)}) + \theta_{yd} D_i(W_{h(i)}, p)  + 3 \cdot \theta_{yw} W_{h(i)}E_{h(i)}   +  \varepsilon_{y, h(i)} + \nu_{yi}(W_i)
\end{split}
\end{equation*}

The household type is correlated with observed covariates as follows. The score for each household type is:
\begin{align*}
U_{hA} & = 1+ X_{2h}  - 0.5 X_{3h} + \varepsilon_{hA},   \\
U_{hB} & =  X_{1h} + \varepsilon_{hB}, \\
U_{hC} & =  -1+ X_{3h} + \varepsilon_{hC}, \\
U_{hD} & =  X_{4h} + \varepsilon_{hD},
\end{align*}

where each of the noise terms are drawn from a Type I extreme value distribution. $\zeta_h = \arg \max \limits_{T \in \{A, B, C, D \} } U_{hT}$. The coefficients of the model are the same as in Table \ref{tab:sim_estimates}. In the simulation, 44\% of the households are Type A, 22\% are type B,  22\% are Type D, and 12\% are Type C. Demand and outcomes are aggregated by household as in Equation \eqref{eqn:aggh}.

\section{Proof of Main Results}

We use the following notation through the proof. For a deterministic sequence $X_n$
and a positive, deterministic sequence $a_n$,
we say that $X_n = \oo(a_n)$ if $\limsup_{n \rightarrow \infty} X_n / a_n < \infty$,
and $X_n = o(a_n)$ if $\limsup_{n \rightarrow \infty} X_n / a_n < 0$. For a sequence
of random variables $X_n$, and $a_n$ as before, we say that
$X_n = \oo_p(\alpha_n)$ if for every $\varepsilon$, there exists an $c < \infty$ such
that $\lim \limits_{n\rightarrow \infty }   \PP{ \| A_n \| > \alpha_n c} < \varepsilon$,
and we say that $X_n = o(\alpha_n)$ if, for any $ c>0 $,
$\lim \limits_{n \rightarrow \infty} \PP{ \|X_n\| > \alpha_n c}  = 0$.

\subsection{Proof of Theorem \ref{theo:prate}}
\label{app:prate}

We start by proving the following more general central limit theorem for market prices
$P_n(\bm W)$ in a system with potential price perturbations. Theorem \ref{theo:prate} follows from Lemma \ref{lem:ptrate}.

\begin{lemma}\label{lem:ptrate}
Under the setting of Theorem \ref{theo:prate}, suppose furthermore that the
market may have random price perturbations $U_{i}$ as in Design \ref{def:aug}
with $\EE{U_{i}} = 0$ and $\Norm{U_{i}} \leq  h_n$
almost surely with $h_n = o(n^{-1/4})$. Then,
\begin{equation}
\label{eq:ptrate_expansion}
P_n(\bm W)  - p^*_{\pi} =  -\xi_z^{-1}\, \frac{1}{n}  \sum \limits_{i=1}^n Z_i(W_i, p^*_{\pi}+U_i) + o_p(n^{-1/2}).
\end{equation}

and $\mathbb E\left [   \left | \left | P_n(\bm W) - p^*_{\pi}) + \xi_z^{-1}\, \frac{1}{n}  \sum \limits_{i=1}^n Z_i(W_i, p^*_{\pi}+U_i) \right | \right|_2^2  \right ] = o(1/n)$. Furthermore, $\sqrt{n}\p{P_n(\bm W)  - p^*_{\pi}} \Rightarrow \nn\p{0, \, \Var{Z_i(W_i, p^*_{\pi})}}$.
\end{lemma}

\begin{proof}
Because $\set$ is compact, Assumptions \ref{as:market} and  \ref{as:regularity} together
imply that not only is there is a unique population equilibrium price $p^*_\pi$ with $z(p^*_\pi)$,
but there is also a critical radius $\varepsilon^* > 0 $ such that
\begin{equation}
\label{eq:strong_opt}
\Norm{z(p)}_2 \geq \frac{\lambda_{\min}\p{\xi_z}}{2} \min\cb{\Norm{p - p^*_\pi}_2,\, \varepsilon^*}.
\end{equation}
To verify this result, we note that, by twice differentiability of $z$, 

\[ z(p) - z(p^*_{\pi}) = \xi_z (p - p^*_{\pi}) + h(p - p^*_{\pi}) || p - p^*_{\pi}) ||  \] 
for a continuous function $h(p - p^*_{\pi})$ with $h(0) = 0$. Let $\lambda_{\min}\p{\xi_z}$ be the minimum eigenvalue of $\xi_z$ and let $\delta^*$ be such that $h(p - p^*_{\pi}) \leq \lambda_{\min}\p{\xi_z}/2$ for all $|| p - p^*_{\pi} || \leq \delta^*$. Then, for all $|| p - p^*_{\pi} || \leq \delta^*$, 
\[ z(p) - z(p^*_{\pi} ) \geq  \lambda_{\min}\p{\xi_z} || p - p^*_{\pi} ||_2  -  \lambda_{\min}\p{\xi_z}/2 || p - p^*_{\pi} ||_2 = \frac{\lambda_{\min}\p{\xi_z}}{2} || p - p^*_{\pi} ||_2. \] 
Thus, \eqref{eq:strong_opt} must hold with
$$ \varepsilon^* = \min\cb{\delta^*, \, \frac{2}{\lambda_{\min}\p{\xi_z}} \inf\cb{\Norm{z(p)}_2 : \Norm{p - p^*_\pi}_2 > \delta^*, \, p \in \set}}, $$
where the latter term must exist because $\set$ is compact, $z(p)$ is continuous over $\set$
and $p^*_\pi$ is the unique equilibrium point over $\set$. 

We next use this fact to establish tail bounds on $\sqrt{n}\Norm{P_n(\bm W)  - p^*_{\pi}}_2$; we will
then sharpen these into a central limit theorem in a second step. Write
$$ \bZ_n( p ) = \frac{1}{n} \sum \limits_{i=1}^n Z_i(W_i, p + U_{i}), \ \ \ \ z_n(p) = \EE{Z_i(W_i, p + U_{i})}. $$
By Lemma \ref{lemm:subgaussian}, there is a constant $C$ such that, for all $t \geq 1$,
$$ \PP{\sup_{p \in \set} \Norm{\bZ_n( p ) - z_n( p )}_2 > \frac{t}{\sqrt{n}}} \leq C t^{2J} e^{-2t^2}. $$
Furthermore, because $z(p)$ is twice continuously differentiable in $p$ and
we have $\EE{U_{i}} = 0$ and $\Norm{U} \leq h_n$ almost surely
\begin{equation}
\label{eq:noise_pert_bound}
z_n(p) - z(p) = \int (z(p + u) - z(p)) dF(u)
= \int u \cdot z'(p) dF(u) + \oo(h_n^2) = \oo(h_n^2) = o\p{\frac{1}{\sqrt{n}}}.
\end{equation}
Thus, for $n$ large enough, 
$$ \PP{\sup_{p \in \set} \Norm{\bZ_n( p ) - z( p )}_2 > \frac{1 + t}{\sqrt{n}}} \leq C t^{2J} e^{-2t^2}. $$
Meanwhile, by Assumption \ref{as:interference}, $\Norm{\bZ_n(P_n(\bm W))}_2 \leq a_n = o(1/\sqrt{n})$
with probability $1 - e^{-c_1n}$, and so for $n$ large enough 
\begin{equation}
\label{eq:ztb} 
 \PP{ \Norm{z( P_n(\bm W) )} > \frac{2 + t}{\sqrt{n}}} \leq C t^{2J} e^{-2t^2} + e^{-c_1n}. 
 \end{equation} 
\eqref{eq:ztb} together with \eqref{eq:strong_opt} implies that the probability that $||z(P_n(\bm W)) ||_2 \geq  \epsilon^*$ decreases in $n$ exponentially. So \eqref{eq:strong_opt} then implies that for $n$ large enough, that $|| P_n(\bm W) - p^*_{\pi}||_2 \leq z || P_n(\bm W)) ||_2$ with high probability, and we have the following tail bound available for $P_n(\bm w)$ for large enough $n$
\begin{equation}
\label{eq:Pn_tail_bound}
\PP{\Norm{P_n(\bm W) - p^*_\pi} > \frac{1}{\sqrt{n}} \frac{4 + 2t}{\lambda_{\min}\p{\xi_z}}} \leq C t^{2J} e^{-2t^2} + e^{-c_1n}, 
\end{equation}
which in particular implies that $P_n(\bm W) = p^*_\pi + \oo_p(1/\sqrt{n})$.


Now, to sharpen our result into a central limit theorem, we invoke
Lemma \ref{lemm:asymp_equicontinuity} to verify that, because
$P_n(\bm W)$ is consistent for $p^*_\pi$, we have asymptotic equicontinuity
\begin{equation} \label{eq:expz1}
\bZ_n( P_n(\bm W) ) - z_n( P_n(\bm W) ) = \bZ_n( p^*_\pi ) - z_n( p^*_\pi ) + o_p(1/\sqrt{n}).
\end{equation}
Together with \eqref{eq:noise_pert_bound} this becomes
\begin{equation}
\bZ_n( P_n(\bm W) ) - z( P_n(\bm W) ) = \bZ_n( p^*_\pi ) - z( p^*_\pi ) + o_p(1/\sqrt{n}).
\end{equation}
Again invoking twice continuous differentiability of $z(p)$, we get
\begin{equation}  \label{eq:expz2}
\bZ_n( P_n(\bm W) ) - \bZ_n( p^*_\pi ) =    \xi_z (P_n(\bm W) - p^*_\pi) + \oo\p{\Norm{P_n(\bm W) - p^*_\pi}^2_2} + o_p(1/ \sqrt n ), 
\end{equation}
and, thanks to invertibility of $\xi_z$ and the fact that $\bZ_n( P_n(\bm W) ) = o_p(1/\sqrt{n})$,
this implies our first desired claim given that
we already showed that $P_n(\bm W) = p^*_\pi + \oo_p(1/\sqrt{n})$.

By the tail bound on $P_n(\bm W) - p^*_{\pi}$ in \eqref{eq:Pn_tail_bound}, the bound on the expectation of the remainder term in Lemma \ref{lemm:asymp_equicontinuity}, and boundedness of outcome and net demand, then we also have the remainder term  $R_n = P_n(\bm W) - p^*_{\pi} - \bar Z_n( p^*_{\pi})$ is such that $ \mathbb E[ || R_n ||^2_2 ]  = o(1/n)$. The details of this argument follow the same approach as \eqref{eq:formalize_residual_bound}. 

Finally recall that, by Assumption \ref{as:monotone}, $\Var{Z_i(W_i, \, p^*_\pi)} \succ 0$
and so weak continuity implies that $\lim \limits_{n \to \infty} \text{Corr}\sqb{Z_i(W_i, \, p^*_\pi), \, Z_i(W_i, \, p^*_\pi + U_{i})} = 1$.
The claimed central limit theorem then follows by the standard Lindeberg
argument thanks to boundedness of $Z_i(w, \, p)$.
\end{proof}

\subsection{Proof of Theorems \ref{theo:pop}, \ref{theo:sade} and \ref{theo:saie}}
\label{ap:pop}

We structure the bulk of the proof in terms of two lemmas given below, which also allow for
potential price perturbations in anticipation of later needs.
Given these two results, we first note that Theorem \ref{theo:sade} is in fact a special case
of Lemma \ref{lem:ade_ap} with $U_{i} = 0$ almost surely. Moreover, the claim
about the ADE in Theorem \ref{theo:pop} follows immediately from \eqref{eq:ade_asymp} by
noting that $\tau_{\text{ADE}}$ is almost surely bounded (because the outcomes are), and
that $\EE{Z_i(W_i, \, p^*_\pi)} = 0$.

For our results about the AIE, we rely on a result of \citet[Theorem 1]{hu2021average}, who show that
in Bernoulli trials with independent $W_i$ (but not necessarily constant treatment probabilities
$\pi_i$), we have
\begin{equation}
\label{eq:HLW}
\tau_{\text{MPE}} = \tau_{\text{ADE}} + \tau_{\text{AIE}}.
\end{equation}
Lemmas \ref{lem:inf_ap} shows that $\tau_{\text{MPE}}$ converges to $\tau^*_{\text{MPE}}$, both
in expectation and in probability, and we've already established the same for the ADE above.
Thus, by \eqref{eq:HLW}, $\tau_{\text{AIE}}$ must also converge to
$\tau^*_{\text{AIE}} = \tau^*_{\text{MPE}} - \tau^*_{\text{ADE}}$, both in expectation (Theorem \ref{theo:pop})
and in probability (Theorem \ref{theo:saie}).

\begin{lemma}\label{lem:ade_ap}
Under the setting of Theorem \ref{theo:pop}, suppose furthermore that the
market may have random price perturbations $U_{i}$ as in Design \ref{def:aug}
with $\EE{U_{i}} = 0$ and $\Norm{U_{i}} \leq h_n$
almost surely with $h_n = o(n^{-1/4})$. Then,
\begin{equation}
\label{eq:ade_ap_asymp}
\tau_{\text{ADE}}  =  \frac{1}{n} \sum \limits_{i=1}^n  \p{ Y_i(1, p^*_{\pi}) - Y_i(0, p^*_{\pi})  - \pi_i \Delta_i(1, \, p^*_\pi)-  (1 - \pi_i) \Delta_i(1, \, p^*_\pi) }  + o_p(1/\sqrt{n}),
\end{equation}
with $\Delta_i(w, \, p)$ as defined in \eqref{eq:Delta}.
Furthermore, writing $\varepsilon_i(w) = Y_i(w, p^*_{\pi}) - y(w, p^*_{\pi})$, we have
\begin{equation}
\label{eq:ade_ap_clt}
\sqrt{n}\p{\tau_{\text{ADE}} -  \tau^*_{\text{ADE}}} \Rightarrow \nn\p{0, \, \Var{ \varepsilon_i(1)   - \varepsilon_i(0)  -\pi_i \Delta_i(1, \, p^*_\pi) - (1 - \pi_i) \Delta_i(1, \, p^*_\pi)}}.
\end{equation}
\end{lemma}

\begin{lemma}\label{lem:inf_ap}
Under the setting of Lemma \ref{lem:ade_ap},
\begin{equation}
\label{eq:inf_ap_asymp}
\begin{split}
&\EE{\abs{\tau_{\text{MPE}} - \tau^*_{\text{MPE}}}} = o(1), \\
&\tau^*_{\text{MPE}} = y(1, \, p^*_\pi) - y(0, \, p^*_\pi) + \xi_y^\top \xi_z^{-1} \p{z(1, \, p^*_\pi) - z(0, \, p^*_\pi)}.
\end{split}
\end{equation}
\end{lemma}

\begin{proof}[Proof of Lemma \ref{lem:ade_ap}]
By definition of the ADE, we have
\begin{align*}
\tau_{\text{ADE}}
&= \EE[\pi]{\frac{1}{n} \sum_{i = 1}^n \p{Y_i(1, \, P_n(w_i = 1; \, \bm W_{-i}) + U_{i}) - Y_i(0, P_n(w_i = 0; \, \bm W_{-i}) + U_{i})}} \\
&= \EE[\pi]{\frac{1}{n} \sum_{i = 1}^n \p{Y_i(1, \, P_n(\bm W) + U_{i}) - Y_i(0, \, P_n(\bm W) + U_{i})} + R_n(1) - R_n(0)}, \\
R_n(w) &= \frac{1}{n} \sum_{i = 1}^n \p{Y_i(w, \, P_n(w_i = w; \, \bm W_{-i}) + U_{i}) - Y_i(w, \, P_n(\bm W) + U_{i})}.
\end{align*}
The residual terms $R_n(w)$ have $\mathbb E_{\pi}[R_n(w)] = o_p(1/\sqrt{n})$ -- we will defer showing this to the end of this proof.
By Lemma \ref{lem:ptrate}, we know that $P_n(\bm W) \rightarrow_p p^*_\pi$; thus
by Lemma \ref{lemm:asymp_equicontinuity}, for $w = 0, \, 1$,
\begin{align*}
&\frac{1}{n} \sum_{i = 1}^n \p{Y_i(w, \, P_n(\bm W) + U_{i}) - y_n(w, \, P_n(\bm W))} \\
&\quad\quad\quad\quad= \frac{1}{n} \sum_{i = 1}^n \p{Y_i(w, \, p^*_\pi + U_{i}) - y_n(w, \, p^*_\pi)} + o_p(1/\sqrt{n}),
\end{align*}
with $y_n(w, \, p) = \EE{Y_i(w, \, p + U_{i})}$. Furthermore, arguing via twice differentiability
of $y(w, \ p)$ like in the proof of Lemma \ref{lem:ptrate}, and noting the established rate of convergence
$\Norm{P_n(\bm W) - p^*_\pi}_2 = \oo_p(1/\sqrt{n})$ in that result, we get
$$ \frac{1}{n} \sum_{i = 1}^n \p{Y_i(w, \, P_n(\bm W) + U_{i}) - Y_i(w, \, p^*_\pi + U_{i})} = \p{P_n(\bm W) - p^*_\pi} \cdot \nabla_p y(w, \, p^*_\pi) + o_p(1/\sqrt{n}). $$
Combining our results so far, we get
\begin{align*}
&\tau_{\text{ADE}} = \mathbb{E}_{\pi}\Bigg[\frac{1}{n} \sum_{i = 1}^n \p{Y_i(1, \, p^*_\pi + U_{i}) - Y_i(0, \, p^*_\pi + U_{i})}  \\
&\quad\quad\quad\quad\quad\quad + \p{P_n(\bm W) - p^*_\pi} \cdot \nabla_p \p{y(1, \, p^*_\pi) - y(0, \, p^*_\pi)}\Bigg] + o_p(1/\sqrt{n}),
\end{align*}
Finally, plugging in the asymptotic expansion \eqref{eq:ptrate_expansion} for $P_n(\bm W) - p^*_\pi$ into
this expression gives
\begin{align*}
&\tau_{\text{ADE}} = \mathbb{E}_{\pi}\Bigg[\frac{1}{n} \sum_{i = 1}^n \p{Y_i(1, \, p^*_\pi + U_{i}) - Y_i(0, \, p^*_\pi + U_{i})}  \\
&\quad\quad\quad\quad\quad\quad + \nabla_p^\top \p{y(1, \, p^*_\pi) - y(0, \, p^*_\pi)} \xi_z^{-1} Z_i(W_i, \, p^*_\pi) \Bigg] + o_p(1/\sqrt{n}),
\end{align*}
which in turn yields \eqref{eq:ade_ap_asymp} by integrating over $W_i$. Finally, the central limit theorem follows by an analogous
argument to the one made in Lemma \ref{lem:ptrate}.

We now return to arguing that the residual terms $R_n(w)$ are small. Thanks to
Assumption \ref{as:lips}, there exists a constant $C$ such that
\begin{align*}
\abs{R_n(w)}
&\leq \frac{1}{n} \sum_{i = 1}^n \abs{Y_i(w, \, P_n(w_i = w; \, \bm W_{-i}) + U_{i}) - Y_i(w, \, P_n(\bm W_{-i}) + U_{i})} \\
&\leq \frac{1}{n} \sum_{i = 1}^n \abs{Y_i(w, \, P_n(w_i = 1; \, \bm W_{-i}) + U_{i}) - Y_i(w, \, P_n(w_i = 0; \, \bm W_{-i}) + U_{i})} \\
&\leq C \, \frac{1}{n} \sum_{i = 1}^n \sum_{j = 1}^J \abs{Z_{ij}(w, \, P_n(w_i = 1; \, \bm W_{-i}) + U_{i}) - Z_{ij}(w, \, P_n(w_i = 0; \, \bm W_{-i}) + U_{i})} \\
&\quad\quad\quad\quad  + C \, \frac{1}{n} \sum_{i = 1}^n \Norm{P_n(w_i = 1; \, \bm W_{-i}) - P_n(w_i = 0; \, \bm W_{-i})}_2. \\ 
\end{align*}
To finish controlling the remainder terms, we work with the two terms in this bound separately. First, by Lemma \ref{lemm:price_perturbation}, we have:
\begin{equation*} 
\begin{split} 
 C  \mathbb E_{\pi} \left [\, \frac{1}{n} \sum_{i = 1}^n \Norm{P_n(w_i = 1; \, \bm W_{-i}) - P_n(w_i = 0; \, \bm W_{-i})}_2  \right ]  = o_p(1).
\end{split} 
\end{equation*} 
Next, we invoke Lemma \ref{lemm:price_perturbation} as well as approximate monotonicity as in Assumption \ref{as:monotone}. 
Specifically, for any $\varepsilon > 0$, then Lemma \ref{lemm:price_perturbation} implies that $\Norm{P_n(w_i; \, \bm W_{-i}) - P_n(\bm W)}_2 \leq \varepsilon / (C\sqrt{n})$ with probability tending to 1. By Assumption \ref{as:monotone}, where $C$ is the constant from the assumption, this means that $$ Z_{ij}(P_n(\bm W) +  \varepsilon/\sqrt{n} e_j + U_{i}) \leq Z_{ij}(w, \, P_n(w_i; \, \bm W_{-i}) + U_{i}) \leq Z_{ij}(P_n(\bm W) -  \varepsilon/\sqrt{n} e_j + U_{i}) $$
simultaneously for all $i$, with probability tending to 1. Plugging this into the above, we can now write 
\begin{align*}
\mathbb E_{\pi} [ \abs{R_n(w)} ] & \leq  \mathbb E_{\pi} \left [ C \, \frac{1}{n} \sum_{i = 1}^n  \sum_{j = 1}^J \Big( Z_{ij}(w, \, P_n(\bm W) - \varepsilon/\sqrt{n} e_j + U_{i}) - Z_{ij}(w, \, P_n(\bm W) + \varepsilon/\sqrt{n} e_j + U_{i}) \Big)   \right ] \\  & \qquad \qquad \qquad  \qquad + o_p(1/\sqrt{n}), \\
&\leq - \frac{2C\varepsilon}{\sqrt{n}}   \sum_{j = 1}^J  \frac{\partial}{\partial p_j} z_{n, j}(w, \ p^*_\pi)  + o_p(1/\sqrt{n}),
\end{align*}
where the second line followed from asymptotic equicontinuity of $Z_{i}(w, p + U_i)$, from Lemma \ref{lemm:asymp_equicontinuity}. 
Because this bound holds for any $\varepsilon > 0$, we in fact have $\mathbb E_{\pi} [|R_n(w)| ] = o_p(1/\sqrt{n})$. 
\end{proof}

\begin{lemma}
\label{lemm:price_perturbation}
Under the conditions of Lemma \ref{lem:ptrate}, changing the treatment given to any one unit
cannot change the equilibrium price much:
\begin{equation}
\sup_{i = 1, \ldots, \ n} \mathbb E_{\pi} [ \Norm{P_n(w_i = 1; \bm W_{-i}) - P_n(w_i = 0; \bm W_{-i})}_2] = o_p(1/\sqrt{n}).
\end{equation}

\proof
By Assumption \ref{as:interference} and a union bound, we know that with probability at least $1 - 2ne^{-c_1n}$,
\begin{equation}
\Norm{\frac{1}{n} \p{Z_i\p{w, \, P_n(w_i = w, \, \bm W_{-i}) + U_i} +  \sum_{j \neq i} Z_j\p{W_j, \, P_n(w_i = w, \, \bm W_{-i}) + U_j}}}_2 \leq a_n
\end{equation}
for all $i = 1, \, \ldots, \, n$ and $w \in \cb{0, \, 1}$.
By boundedness as in Assumption \ref{as:regularity}, this also implies that, with high probability for all $i$ and $w$,
\begin{equation}
\label{eq:Zj_unif_delta}
\Norm{\frac{1}{n} \sum_{j = 1}^n Z_j\p{W_j, \, P_n(w_i = w, \, \bm W_{-i}) + U_j}}_2 \leq a_n + \frac{2M}{n} = o(1/\sqrt{n}).
\end{equation}
Since we have now shown that $\sup_{i, w}  ||\bar Z_n(P_n(w_i =w; \, \bm W_{-i})) ||_2 = o(1/\sqrt n)$ with probability $1 - 2ne^{-c_1n}$, we can now follow \eqref{eq:ztb} to \eqref{eq:Pn_tail_bound} to show the following tail bound. For $n$ large enough, there is a finite constant $C$ such that 
\begin{equation*} 
\mathbb P\left [ \sup_{i, w} ||P_n(w_i =w; \, \bm W_{-i}) - p^*_{\pi} || > \frac{1}{\sqrt{n}} \frac{4 + 2t}{\lambda_{\min}\p{\xi_z}} \right] \leq C t^{2J} e^{-2t^2}, 
\end{equation*} 
and following the same argument as \eqref{eq:expz1} to \eqref{eq:expz2}, we can show: 
\begin{align}
\sup_{i,w}  \left | \left |  P_n(w_i = w, \, \bm W_{-i}) - p^*_{\pi} + \xi_z^{-1}\, \frac{1}{n}  \sum \limits_{i=1}^n Z_i(W_i, p^*_{\pi} + U_i) \right | \right |_2 = o_p(1/\sqrt{n}),
\end{align}
which implies our desired result as the asymptotic linear representation is the same regardless of $i$ and $w$. We also note that as in Lemma \ref{lem:ptrate}, the $o_p(n^{-1/2})$ remainder term in this expansion,  $R_n$, also has the property $\mathbb E_{\pi} [R_n] = o_p(n^{-1/2})$. 
\endproof

\end{lemma}

\begin{proof}[Proof of Lemma \ref{lem:inf_ap}]
Recall that  $\tau_{\text{MPE}}$ is defined as
$$ \tau_{\text{MPE}} =   \sum_{k=1}^n  \frac{\partial}{\partial \pi_k}  \frac1n \sum_{i=1}^n \EE[\pi]{ Y_i(W_i, P_n(\bm W) + U_{i}}. $$
Writing $f_\pi(\bm w)$ for the distribution function of $\bm w$, we can write
this more explicitly as
\[ \tau_{\text{MPE}} =  \sum \limits_{k=1}^n \frac{\partial}{\partial \pi_k} \sum \limits_{\bm w} f(\bm w) \,  \frac1n \sum \limits_{i=1}^n  Y_i(w_i, P_n(\bm W)+ U_{i}), \]
Now, under our design, $W_i \sim \mbox{Bernoulli}(\pi_i)$ independently for each sample, and so
$f_\pi(\bm w) = \prod_{k=1}^n \pi_k^{w_k } ( 1- \pi_k)^{1 - w_k}$.
Taking the derivative with respect to $\pi_k$ yields
\[ \frac{\partial }{\partial \pi_k} f_\pi(\bm w) = \frac{ w_k - \pi_k }{\pi_k (1- \pi_k)} f(\bm w), \]
and so
\begin{equation}
\begin{split}
\tau_{\text{MPE}} &= \sum \limits_{k=1}^n  \sum \limits_{\bm w} \frac{\partial}{\partial \pi_k} f(\bm w)  \frac{1}{n} \sum \limits_{i=1}^n Y_i(w_i, P_n(\bm W)+ U_{i})   \\
&   = \EE[\pi]{ \sum \limits_{k=1}^n \frac{ W_k - \pi_k}{\pi_k(1- \pi_k)}  \frac1n \sum \limits_{i=1}^n Y_i(W_i, P_n(\bm W)+ U_{i}) }.
\label{eqn:total1}
\end{split}
\end{equation}
At the end of this proof, we will show that
\begin{equation}
\label{eq:Bn}
\frac1n \sum \limits_{i=1}^n Y_i(W_i, P_n(\bm W)+ U_{i}) = \frac1n \sum \limits_{i=1}^n Y_i(W_i, p^*_\pi + U_{i})
- \xi_y^{\top} \xi_z^{-1} \frac1n \sum \limits_{i=1}^n Z_i(W_i, p^*_\pi + U_{i}) + B_n,
\end{equation}
where the remainder term $B_n$ satisfies $\limn n \EE{B_n^2}$ = 0. By Cauchy-Schwarz
\begin{align*}
\EE{\abs{\EE[\pi]{ \sum \limits_{k=1}^n \frac{ W_k - \pi_k}{\pi_k(1- \pi_k)} B_n}}}
&\leq \EE{\EE[\pi]{ \sum \limits_{k=1}^n \frac{ W_k - \pi_k}{\pi_k(1- \pi_k)} B_n}^2} \\
&\leq \EE{n \EE[\pi]{B_n^2}} = n \EE{B_n^2} = o(1),
\end{align*}
and so this remainder term in fact has a lower-order effect on $\tau_{\text{MPE}}$.

Finally, $W_k \indep Y_i(W_i, p^*_\pi + U_{i}), \, Z_i(W_i, p^*_\pi + U_{i})$ for $i \neq k$, and so
\begin{align*}
&\EE[\pi]{ \sum \limits_{k=1}^n \frac{ W_k - \pi_k}{\pi_k(1- \pi_k)}  \frac1n \sum \limits_{i=1}^n Y_i(W_i, p^*_\pi + U_{i}) } = \frac1n \sum \limits_{i=1}^n \p{Y_i(1, p^*_\pi + U_{i}) - Y_i(0, p^*_\pi + U_{i}) } \\
&\EE[\pi]{ \sum \limits_{k=1}^n \frac{ W_k - \pi_k}{\pi_k(1- \pi_k)}  \xi_y \xi_z^{-1} \frac1n \sum \limits_{i=1}^n Z_i(W_i, p^*_\pi + U_{i}) } = \xi_y^{\top} \xi_z^{-1} \frac1n \sum \limits_{i=1}^n \p{Z_i(1, p^*_\pi + U_{i}) - Z_i(0, p^*_\pi + U_{i})},
\end{align*}
thus immediately implying the desired result (as these quantities are IID averages that
concentrate around the claimed terms in $\tau^*_{\text{MPE}}$).

We now return to verifying that $\EE{B_n^2} = o(1/n)$. We can expand out this residual as
\begin{equation} \label{eq:b_n} 
\begin{split} 
B_n &= \frac1n \sum \limits_{i=1}^n Y_i(W_i, P_n(\bm W)+ U_{i}) - \frac1n \sum \limits_{i=1}^n Y_i(W_i, p^*_\pi + U_{i}) - \frac1n \sum \limits_{i=1}^n\p{y_n\p{P_n(\bm W)} -y_n\p{p^*_\pi}} \\
&\quad\quad + \frac1n \sum \limits_{i=1}^n\p{y_n\p{P_n(\bm W)} -y_n\p{p^*_\pi}} - \xi_y^{\top} \xi_z^{-1} \frac1n \sum \limits_{i=1}^n Z_i(W_i, p^*_\pi + U_{i}).
\end{split} 
\end{equation}
For the first line, let 
\[ R_n =  \frac1n \sum \limits_{i=1}^n Y_i(W_i, P_n(\bm W)+ U_{i}) - \frac1n \sum \limits_{i=1}^n Y_i(W_i, p^*_\pi + U_{i}) - \frac1n \sum \limits_{i=1}^n\p{y_n\p{P_n(\bm W)} -y_n\p{p^*_\pi}}. \] 
Let $\delta_n = n^{-1/3}$. By Lemma \ref{lemm:asymp_equicontinuity}, $\mathbb E[ R^2_n \mathbbm{1}( || P_n(\bm W) - p^*_{\pi} ||_2 \leq \delta_n)] = o(1/n)$. 
\begin{equation} \label{eq:formalize_residual_bound}  
\begin{split}
\EE{R_n^2} &\leq \EE{R_n^2 1\p{\cb{\Norm{P_n - p^*_\pi}_2 \leq \delta_n}}} + \EE{R_n^2 1\p{\cb{\Norm{P_n - p^*_\pi}_2 > \delta_n}}} \\
&\leq o(1/n) + \PP{\Norm{P_n - p^*_\pi}_2 > \delta_n} M^2, \\
& = o(1/n). 
\end{split} 
\end{equation} 
where the last step is by the tail bound in \eqref{eq:Pn_tail_bound}, which indicates that $\PP{\Norm{P_n - p^*_\pi}_2 > \delta_n}  = o(1/n)$. And, $M$ is  finite constant since $R_n^2$ is uniformly bounded. For the second line in \eqref{eq:b_n}, 
\begin{align*}
& \frac1n \sum \limits_{i=1}^n\p{y_n\p{P_n(\bm W)} -y_n\p{p^*_\pi}} - \xi_y^{\top} \xi_z^{-1} \frac1n \sum \limits_{i=1}^n Z_i(W_i, p^*_\pi + U_{i}) \\
&\quad\quad\quad\quad=  \frac1n \sum \limits_{i=1}^n\p{y\p{P_n(\bm W)} -y\p{p^*_\pi}} - \xi^{\top}_y \xi_z^{-1} \frac1n \sum \limits_{i=1}^n Z_i(W_i, p^*_\pi + U_{i}) + o(1/\sqrt{n}),
\end{align*}
and
$$  \frac1n \sum \limits_{i=1}^n  \xi_y^{\top} \p{P_n(\bm W)- p^*_\pi} =  \xi^{\top}_y \xi_z^{-1} \frac1n \sum \limits_{i=1}^n Z_i(W_i, p^*_\pi + U_{i}) + o_p(1/\sqrt{n}). $$
Thus
\begin{align*}
&\EE{\p{ \frac1n \sum \limits_{i=1}^n\p{y_n\p{P_n(\bm W)} -y_n\p{p^*_\pi}} - \xi_y^{\top} \xi_z^{-1} \frac1n \sum \limits_{i=1}^n Z_i(W_i, p^*_\pi + U_{i})}^2} \\
&\quad\quad\quad\quad \leq \EE{\p{\frac1n \sum \limits_{i=1}^n \p{y\p{P_n(\bm W)} - y\p{p^*_\pi}  - \xi_y^{\top} \p{P_n(\bm W) - p^*_\pi} }}^2} + o(1/n) \\
&\quad\quad\quad\quad= \oo\p{\EE{\Norm{P_n(\bm W) - p^*_\pi}_2^2}} + o(1/n) \\ & \quad\quad\quad\quad= o(1/n).
\end{align*}
On the second last line, we used the bound on the remainder term from Lemma \ref{lem:ptrate}.  On the last line we used continuous twice differentiability of $y(p)$ and the bound \eqref{eq:Pn_tail_bound}
to control the second moment of market price fluctuations.
\end{proof}

\subsection{Proof of Theorem \ref{theo:adeinf}}
\label{ap:adeinf}

Thanks to the self-normalized form of the difference-in-means estimator, we have
\begin{equation}
\htau_{\text{ADE}} - \tau^*_{\text{ADE}} = \frac{1}{n} \sum_{i = 1}^n \frac{W_i\p{Y_i - y_n(1, \, p^*_\pi)}}{\hpi} - \frac{(1 - W_i) \p{Y_i - y_n(0, \, p^*_\pi)}}{1 - \hpi}.
\end{equation}
By asymptotic equicontinuity as shown in Lemma \ref{lemm:asymp_equicontinuity} 
followed by an application of Theorem \ref{theo:prate},
\begin{align*}
\frac{1}{n} \sum_{i = 1}^n \frac{W_i\p{Y_i - y_n(1, p^*_\pi)}}{\pi}
&= \frac{1}{n} \sum_{i = 1}^n \frac{W_i \, \varepsilon_{in}(1)}{\pi} + y_n(1, \, P_n) - y_n(1, \, p^*_\pi) + o_p\p{1/\sqrt{n}} \\
&= \frac{1}{n} \sum_{i = 1}^n \frac{W_i \, \varepsilon_{in}(1)}{\pi} - \nabla_p^\top y(1, \, p^*_\pi) \xi_z^{-1} Z_i(W_i, \, p^*_\pi + U_i) + o_p\p{1/\sqrt{n}}, \\
\frac{1}{n} \sum_{i = 1}^n \frac{(1 - W_i)\p{Y_i - y_n(0, p^*_\pi)}}{1 - \pi}
&= \frac{1}{n} \sum_{i = 1}^n \frac{(1 - W_i) \, \varepsilon_{in}(0)}{1 - \pi} + y_n(0, \, P_n) - y_n(0, \, p^*_\pi) + o_p\p{1/\sqrt{n}} \\
&= \frac{1}{n} \sum_{i = 1}^n \frac{(1 - W_i) \, \varepsilon_{in}(0)}{1 - \pi} - \nabla_p^\top y(0, \, p^*_\pi) \xi_z^{-1}Z_i(W_i, \, p^*_\pi + U_i) + o_p\p{1/\sqrt{n}},
\end{align*}
Then, taking the difference, we have 
\begin{equation*} 
\htau_{\text{ADE}} - \tau^*_{\text{ADE}} = \frac{1}{n} \sum_{i = 1}^n \frac{W_i \, \varepsilon_{in}(1)}{\pi}  - \frac{(1 - W_i) \, \varepsilon_{in}(0)}{1 - \pi} - \Delta_{in}(W_i, p^*_{\pi})  + o_p\p{1/\sqrt{n}}, \\
\end{equation*} 
where $ \Delta_{in}(w, p) = \nabla_p^{\top}[  y(1, p) - y(0, p) ] \xi_z^{-1}  Z_i(w, \, p + U_i).$ The central limit theorem follows by standard Lindeberg arguments and Slutsky's theorem, since $\hat \pi \rightarrow_p \pi$. Note that the variance of the limiting distribution does not depend on $U_i$: 
\[ \mathbb E \left [ \left (  \frac{W_i}{\pi} \varepsilon_{in}(1) + \frac{ ( 1- W_i) }{ 1- \pi} \varepsilon_{in}(0) - \Delta_{in}(W_i, p^*_{\pi}  )\right)^2 \right] = \mathbb E \left [ \left (  \frac{W_i}{\pi} \varepsilon_{i}(1) + \frac{ ( 1- W_i) }{ 1- \pi} \varepsilon_{i}(0) - \Delta_{i}(W_i, p^*_{\pi}) \right)^2 \right] + o(1) \]  by weak continuity in Assumption \ref{as:regularity}. 

\subsection{Proof of Proposition \ref{prop:neyman}}

First, introduce the notation $a_i(1) = \epsilon_i(1) - \pi \Delta_i(1, p^*_{\pi})$, and $a_i(0) = \epsilon_i(0) + (1 - \pi) \Delta_i(0, p^*_{\pi}).$ 
In an extension of the classical results in \citet{neyman1923applications}, we can show that $ \sigma^2_D$ is conservative for $\bsigma^2_D$. 
\begin{equation*} 
\begin{split} 
 \sigma^2_D  & =  \EE{\p{ \frac{W_i \varepsilon_i(1)}{ \pi}  - \frac{( 1- W_i)  \varepsilon_i(0)}{  1- \pi }  - \Delta_i(W_i, p^*_{\pi})}^2},  \\
	 & = \EE{\p{ \frac{W_i a_i(1)}{ \pi}  - \frac{( 1- W_i)  a_i(0)}{  1- \pi } }^2}, \\ 
	 & = \EE{ \frac{ a_i(1)^2 }{\pi} }  + \EE{ \frac{ a_i(0)^2 }{1 - \pi} }, \\ 
	 & =   \EE{\p{ \frac{  ( 1 -  \pi)  a_i^2(1)}{ \pi}  + \frac{  \pi  a_i^2(0)}{  1- \pi }  + a^2_i(1) + a^2_i(0) }}, \\
	 & \geq \EE{\p{ \frac{ ( 1- \pi) a_i^2(1)}{ \pi}  + \frac{ \pi a_i^2(0)}{  1- \pi }  + 2  \mathbb E[a_i(1)^2]^{1/2} \mathbb E[a_i(0)^2]^{1/2} }}, \\  & 
	 \geq \EE{  \frac{  ( 1 -  \pi)  a_i^2(1)}{ \pi}  + \frac{  \pi  a_i^2(0)}{  1- \pi }  + 2 a_i(1) a_i(0) }, \\ & 
	 = \pi \p{1 - \pi} \EE{ \left (  \frac{a_i(1)}{\pi} + \frac{a_i(0)}{1 - \pi}  \right)^2}, \\ & 
	 =  \pi \p{1 - \pi} \EE{ \left (\frac{ \varepsilon_i(1) }{\pi} + \frac{ \varepsilon_i(0)}{1 - \pi}  - \p{\Delta_i(1, p^*_{\pi}) - \Delta_i(0, p^*_{\pi})} \right )^2}, \\ &
	 = \bsigma_D^2 
\end{split} 
\end{equation*} 
where the first inequality is from the AM-GM inequality and the second inequality is from Cauchy-Schwarz. 



\subsection{Proof of Theorem \ref{theo:aieinf}}
\label{ap:aieinf}
As usual we write unscaled perturbations as $S_i = U_{i} / h_n$,
and recall that $S_i \in \cb{\pm 1}^J$ uniformly at random.
Our estimator of the average indirect effect can equivalently be written as
\begin{equation}
\label{eq:aie_pf_decomp}
\begin{split}
&\htau_{\text{AIE}} = -(\Yvec^\top \Svec) \p{\Zvec^\top \Svec}^{-1} \htau_{\text{ADE}}^z = -E_Y E_Z^{-1} \htau_{\text{ADE}}^z, \\
&E_Y = \Yvec^\top \Svec (\Svec^\top \Svec)^{-1}, \ \ \ \ E_Z = \Zvec^\top \Svec (\Svec^\top \Svec)^{-1}.
\end{split}
\end{equation}
One can replicate the proof of Theorem \ref{theo:adeinf} to verify that $\htau_{\text{ADE}}^z = z(1, \, p^*_\pi) - z(0, \, p^*_\pi) + \oo_P(1/\sqrt{n})$.
Meanwhile, as shown below, $E_Y$ and $E_Z$ will prove to have slower convergence rates. Thus, to capture the asymptotic behavior of
$\htau_{\text{AIE}}$ it will suffice to characterize $E_Y$ and $E_Z$; noise in $\htau_{\text{ADE}}^z$ will make an asymptotically negligible contribution to the error.

We start by studying the term $E_Y$ below; the same argument can also be applied to $E_Z$.
By asymptotic equicontinuity as in Lemma \ref{lemm:asymp_equicontinuity},
\begin{align*}
\frac{1}{n} \Yvec^\top \Svec
&=\frac1n \sum \limits_{i=1}^n Y_i(W_i, p^*_{\pi} + h_n S_i) S_i + (P_n(\bm W) - p^*_{\pi})^{\top} \mathbb E[ y'(p^*_{\pi} + h_n S_i) S_i] + o_p(1/\sqrt{n}) \\ 
& = \frac1n \sum \limits_{i=1}^n Y_i(W_i, p^*_{\pi} + h_n S_i) S_i + (P_n(\bm W) - p^*_{\pi})^{\top} \mathbb E[S_i] y'(p^*_{\pi}) + O (h_n  || P _n(\bm W) - p^*_{\pi} ||_2)  +  o_p(1/\sqrt{n}) \\
&= \frac{1}{n} \sum_{i = 1}^n Y_i(W_i, \, p^*_\pi + h_n S_i)S_i + o_p(1/\sqrt{n})
\end{align*}
The second step above is by a Taylor expansion of $y'(p^* + h_n S_i)$, using the assumption that the second derivatives of $y(p)$ are bounded and that $\mathbb E[S_i] = 0$.
Clearly $n^{-1} \Svec^\top \Svec \rightarrow_p I_{J \times J}$, and so the above implies that
$$ E_Y =  \text{vec}\p{Y_i(W_i, \, p^*_\pi + h_n S_i)}^\top \Svec \p{\Svec^\top\Svec}^{-1} + o_p(1/\sqrt{n}). $$
Furthermore, again by twice differentiability of $y(\cdot)$, we can invoke Taylor expansion to verify that
\begin{align*}
E_Y &= y(p^*_\pi) \, \mathbf{1}^\top \Svec \p{\Svec^\top\Svec}^{-1} + h_n \xi^{\top}_y \, \Svec^\top \Svec \p{\Svec^\top\Svec}^{-1} + \oo\p{h_n^2} \\
&\quad+ \text{vec}\p{Y_i(W_i, \, p^*_\pi + h_n S_i) - y(p^*_\pi + h_n S_i)}^\top \Svec
\p{\Svec^\top\Svec}^{-1} + o_p(1/\sqrt{n}), \\
&= h_n \xi^{\top}_y  + \text{vec}\p{Y_i(W_i, \, p^*_\pi)}^\top \Svec \p{\Svec^\top\Svec}^{-1} \\
&\quad + \text{vec}\p{Y_i(W_i, \, p^*_\pi + h_n S_i) - y(p^*_\pi + h_n S_i) - \p{Y_i(W_i, \, p^*_\pi) - y(p^*_\pi)}}^\top \Svec
\p{\Svec^\top\Svec}^{-1} + o_p(1/\sqrt{n}).
\end{align*}
Finally, by randomness of the $S_i$ and
by weak continuity of net demand as given in Assumptions \ref{as:regularity} (and as translated into outcomes by
Assumption \ref{as:lips}),
\begin{align*}
&\EE{\p{Y_i(W_i, \, p^*_\pi + h_n S_i) - y(p^*_\pi + h_n S_i) - \p{Y_i(W_i, \, p^*_\pi) - y(p^*_\pi)}} S_i} = 0, \\
&\EE{\Norm{\p{Y_i(W_i, \, p^*_\pi + h_n S_i) - y(p^*_\pi + h_n S_i) - \p{Y_i(W_i, \, p^*_\pi) - y(p^*_\pi)}} S_i}_2^2} \\
&\quad\leq\EE{\Norm{Y_i(W_i, \, p^*_\pi + h_n S_i) - y(p^*_\pi + h_n S_i) - \p{Y_i(W_i, \, p^*_\pi) - y(p^*_\pi)}}_\infty^2 \cond S_i} = \oo\p{h_n}.
\end{align*}
These terms are all independent in $i$ so
$$ \EE{\Norm{\frac{1}{n} \sum_{i = 1}^n \p{Y_i(W_i, \, p^*_\pi + h_n S_i) - y(p^*_\pi + h_n S_i) -
\p{Y_i(W_i, \, p^*_\pi) - y(p^*_\pi)}}S_i}_2^2} = \oo\p{h_n/n}, $$
and so, by Chebyshev's inequality and again because $n^{-1} \Svec^\top \Svec \rightarrow_p I_{J \times J}$,
the last term in our previous expression for $E_Y$ is $\oo_P(\sqrt{h_n/n})$.
We have thus shown that
\begin{align*}
E_Y = h_n \xi^{\top}_y  + \text{vec}\p{Y_i(W_i, \, p^*_\pi)}^\top \Svec
\p{\Svec^\top\Svec}^{-1} + o_p(1/\sqrt{n}),
\end{align*}
and one can derive a similar expression for $E_Z$.

Putting everything together, and writing $\Yvec^*$ for the vector with
$i$-th entry $Y_i(W_i, \, p^*_\pi)$ and
$\Zvec^*$ for the matrix with $i$-th row
$Z_i(W_i, \, p^*_\pi)$,
we then get
\begin{align*}
E_Y E_Z^{-1} = \p{ h_n \xi^{\top}_y + \Yvec^{*\top} \Svec \p{\Svec^\top\Svec}^{-1}} \p{h_n \xi_z + \Zvec^{*\top} \Svec \p{\Svec^\top\Svec}^{-1}}^{-1} +  o_p\p{\frac{1}{ h_n \sqrt{n}}}.
\end{align*}

Now, notice that $\Yvec^*$ and $\Zvec^*$ are independent of $\Svec$ so
$$\Yvec^{*\top} \Svec \p{\Svec^\top\Svec}^{-1}, \ \ \Yvec^{*\top} \Svec \p{\Svec^\top\Svec}^{-1} = \oo\p{1/\sqrt{n}}, $$
whereas $h_n$ goes to zero slower than $1/\sqrt{n}$ and $\xi_z$ is full rank by Assumption \ref{as:market}. Thus,
we can Taylor-expand the matrix inverse to get
\begin{align*}
E_Y E_Z^{-1} = \xi^{\top}_y\xi_z^{-1} + h_n^{-1}\Yvec^{*\top} \Svec \p{\Svec^\top\Svec}^{-1}\xi_z^{-1} - h_n^{-1} \xi^{\top}_y \xi_z^{-1} \Zvec^{*\top} \Svec \p{\Svec^\top\Svec}^{-1} \xi_z^{-1}  + o_p\p{\frac{1}{h_n \sqrt{n}}},
\end{align*}
where the last step is by Taylor expansion.
Then, because $n \p{\Svec^\top\Svec}^{-1} \xi_z^{-1} \rightarrow_p \xi_z^{-1}$ and
by Slutsky's lemma,
\begin{align*}
E_Y E_Z^{-1} = \xi^{\top}_y\xi_z^{-1} + \frac{1}{n h_n} \sum_{i = 1}^n \p{Y_i(W_i, \, p^*_\pi) - \xi_y^{\top} \xi_z^{-1} Z_i(W_i, \, p^*_\pi)}  S_i^\top \xi_z^{-1} + o_p\p{1/ (h_n \sqrt{n})}.
\end{align*}
Finally, recalling the decomposition \eqref{eq:aie_pf_decomp} for $\htau_{\text{AIE}}$, the fact that
$\tau^*_{\text{AIE}} = -\xi^{\top}_y\xi_z^{-1}\tau^{*,z}_{\text{ADE}}$ and that
estimation errors in $\htau_{\text{ADE}}^z$ are lower order, we see that
\begin{align*}
\htau_{\text{AIE}} - \tau^*_{\text{AIE}} = - \frac{1}{n h_n} \sum_{i = 1}^n \p{Y_i(W_i, \, p^*_\pi) - \xi_y^{\top} \xi_z^{-1} Z_i(W_i, \, p^*_\pi)}  S_i^\top \xi_z^{-1} \tau^{*,z}_{\text{ADE}} + o_p\p{1/ (h_n \sqrt{n})}.
\end{align*}
Plugging in $U_i = h_n S_i$ then yields the desired asymptotic linear expansion.
The central limit theorem again follows by a standard Lindeberg argument.

\subsection{Proof of Theorem \ref{theo:varest}}
\label{app:varest}

All parameter estimates plugged into \eqref{eq:sigmaD} are consistent, and so
$\hat \sigma^2_D = \hat \sigma^{*2}_D+ o_p(1)$ with
\begin{equation}
\hat \sigma^{*2}_D = \frac{1}{n} \sum_{i=1}^n \p{\frac{W_i \p{Y_i - y_n(1, \, P_n)} }{ \pi} - \frac{(1 - W_i )\p{Y_i - y_n(0, \, P_n)}} {1 - \pi} -   \Delta_i(W_i, \, P_n+ U_i)}^2 .
\end{equation}
If $\hat \sigma^{*2}_D$ were an IID average, the claim from the theorem
statement would be immediate; however, we still need to
address dependence due to the shared price shocks from $P_n$.

To this end, we introduce an auxiliary random variable $T_i \in \cb{0, 2}$ uniformly at random.
It is well known from the literature on resampling-based variance estimation
\citep[e.g.,][]{praestgaard1993exchangeably} that \begin{equation}
\label{eq:02boot}
\begin{split}
&\hat \sigma^{*2}_D = n\Var[T]{\frac{1}{n} \sum_{i=1}^n T_i \p{\frac{W_i \p{Y_i - y_n(1, \, P_n)} }{ \pi} - \frac{(1 - W_i )\p{Y_i - y_n(0, \, P_n)}} {1 - \pi} -   \Delta_i(W_i, \, P_n + U_i)}},
\end{split}
\end{equation}
where $\Var[T]{.}$ denotes variance induced by only the random weights $T_i$. Let \[ F_{in}(p) =  T_i \p{\frac{W_i \p{Y_i - y_n(1, \, P_n)} }{ \pi} - \frac{(1 - W_i )\p{Y_i - y_n(0, \, P_n)}} {1 - \pi} -   \Delta_i(W_i, \, P_n + U_i)}. \]  This is the sum of bounded functions that meet the conditions of Lemma \ref{lemm:asymp_equicontinuity}, multiplied by a bounded and independent random variable, so the conclusions of Lemma \ref{lemm:asymp_equicontinuity} apply also to sample averages of $F_{in}(p)$. So, we can write 
\begin{align*}
&\frac{1}{n} \sum_{i=1}^n T_i \p{\frac{W_i \p{Y_i - y_n(1, \, P_n)} }{ \pi} - \frac{(1 - W_i )\p{Y_i - y_n(0, \, P_n)}} {1 - \pi} -   \Delta_i(W_i, \, P_n)} \\
&\quad\quad=\frac{1}{n} \sum_{i=1}^n T_i \p{\frac{W_i \p{Y_i(1, \, p^*_\pi + U_i) - y_n(1, \, p^*_\pi)} }{ \pi} - \frac{(1 - W_i )\p{Y_i(0, \, p^*_\pi + U_i) - y_n(0, \, p^*_\pi)}} {1 - \pi} -   \Delta_i(W_i, \, p^*_\pi + U_i)} \\
&\quad\quad\quad\quad\quad\quad +\nabla_p^{\top}[  y(1, \, p^*_\pi) - y(0, \, p^*_\pi) ] \xi_z^{-1}  \p{z_n\p{P_n} - z_n\p{p^*_\pi}} + B_n(\Tvec),
\end{align*}
where $B_n(\Tvec)$ is the residual term $R_n$ from Lemma \ref{lemm:asymp_equicontinuity}, making clear the dependence on $T_i$, so $\limn n\EE{B_n(\Tvec)^2} = 0$. Thus, from \eqref{eq:02boot}, we immediately
see that
\begin{equation}
\label{eq:02boot_result}
\begin{split}
\hat \sigma^{*2}_D &= \frac{1}{n} \sum_{i=1}^n \p{\frac{W_i\varepsilon_{in}(1) }{ \pi} - \frac{(1 - W_i )\varepsilon_{in}(0)} {1 - \pi} -   \Delta_i(W_i, \, p^*_\pi + U_i)}^2 + n\Var[T]{B_n(\Tvec)} \\
&\quad\quad\quad\quad + 2 \Cov[T]{\frac{1}{n}\sum_{i=1}^n T_i \p{\frac{W_i\varepsilon_{in}(1) }{ \pi} - \frac{(1 - W_i )\varepsilon_{in}(0)} {1 - \pi} -   \Delta_i(W_i, \, p^*_\pi + U_i)}, \, B_n(\Tvec)},
\end{split}
\end{equation}
with notation $\varepsilon_{in}(w) =  Y_i(w, p^*_\pi + U_i) - y_n(w, p^*_\pi)$, augmented from the definition of $\varepsilon_{i}(w)$ in the statement of Theorem \ref{theo:adeinf}. Now,
\begin{equation}
\label{eq:nBnbound}
n\EE{\Var[T]{B_n(\Tvec)}} \leq n\Var{B_n(\Tvec)} \leq n\EE{B_n(\Tvec)^2} = o(1),
\end{equation}
and so the terms depending on $B_n(\Tvec)$ are asymptotically negligible; the covariance
term needs to be bounded by Cauchy-Schwarz before applying \eqref{eq:nBnbound}, as below:

\begin{align*}
&n \Cov[T]{\frac{1}{n}\sum_{i=1}^n T_i \p{\frac{W_i\varepsilon_i(1) }{ \pi} - \frac{(1 - W_i )\varepsilon_i(0)} {1 - \pi} -   \Delta_i(W_i, \, p^*_\pi)}, \, B_n(\Tvec)} \\
&\quad\quad\leq \sqrt{n\Var[T]{\frac{1}{n}\sum_{i=1}^n T_i \p{\frac{W_i\varepsilon_i(1) }{ \pi} - \frac{(1 - W_i )\varepsilon_i(0)} {1 - \pi} -   \Delta_i(W_i, \, p^*_\pi)}}} \sqrt{n\EE{\Var[T]{B_n(\Tvec)}}} \\
&\quad\quad=\sqrt{\frac{1}{n}\sum_{i=1}^n\p{\frac{W_i\varepsilon_i(1) }{ \pi} - \frac{(1 - W_i )\varepsilon_i(0)} {1 - \pi} -   \Delta_i(W_i, \, p^*_\pi)}^2} \sqrt{o_p(1)} \\
&\quad\quad= o_p(1).
\end{align*}

Finally, the dominant
term in \eqref{eq:02boot_result} is an IID average whose mean is 

\[ \mathbb E\left[\p{\frac{W_i\varepsilon_{in}(1) }{ \pi} - \frac{(1 - W_i )\varepsilon_{in}(0)} {1 - \pi} -   \Delta_i(W_i, \, p^*_\pi + U_i)}^2 \right] = \sigma^2_D + o(1). \] 
The consistency of $\hsigma_I^2$ can be established via an analogous argument, which we omit for brevity. 



\subsection{Proof of Proposition \ref{prop:hte} }

The proof of Theorem 1 in Appendix B.1 of \citet{hu2021average} shows that under a general interference pattern that $\frac{\partial}{\partial \pi'_k} \mathbb E_{\pi'}[ Y_i(\bm W) ]_{\pi'= \pi} = \mathbb E_{\pi} \left [ Y_i(w_k = 1; W_{-k} ) - Y_i(w_k = 0; W_{-k} )\right ].$ Plugging this into our conditional expectation of interest,

   \begin{align*}  \mathbb E \left [ \frac{\partial}{\partial \pi_k} \sum \limits_{i=1}^n \mathbb E_{\pi}[Y_i(\bm W)]\Big | X_k = x\right] &= \mathbb E \left [  \sum \limits_{i=1}^n  \mathbb E_{\pi} \left [ Y_i(w_k = 1; W_{-k} ) - Y_i(w_k = 0; W_{-k} )\right ] \Big | X_k = x\right ] \\
   & = \sum \limits_{i=1}^n  \mathbb E \left [ Y_i(w_k = 1; W_{-k} ) - Y_i(w_k = 0; W_{-k} ) | X_k = x\right ] \\
   & =\mathbb E \left [ Y_i(w_i = 1; W_{-i} ) - Y_i(w_i = 0; W_{-i} ) | X_i = x\right ] \\ & \qquad + \sum \limits_{j \neq i} \mathbb E \left [ Y_j(w_i = 1; W_{-i} ) - Y_j(w_i = 0; W_{-i} ) | X_i = x\right ] \\
   & = \btau_{\text{CADE}}(x) + \btau_{\text{CAIE}}(x)
   \end{align*}

\subsection{Proof of Theorem \ref{theo:hte}}

By Assumption \ref{as:lips} the outcomes $Y_i(w, \, p)$
can be written as the sum of a Lipschitz component and a Lipschitz-in-net-demand
component, and that net demand is weakly monotone by Assumption \ref{as:monotone}.
Thus, there is a constant $M$ such that, for $w = 0, \, 1$,
\begin{align*}
&\abs{Y_i\p{w, \, P(w_i = w, \, W_{-i})} - Y_i\p{w, \, p^*_\pi}} \\
&\quad\quad\quad\quad\leq M \p{\delta_P(w) + \sum_{j = 1}^J \p{Z_{ij}\p{w, \, p^*_\pi - C \delta_P(w) e_j} -Z_{ij}\p{w, \, p^*_\pi + C \delta_P(w) e_j}}},
\end{align*}
where $\delta_P(w) = \Norm{ P(w_i = w, \, W_{-i}) - p^*_\pi}_2$.
Furthermore, we can adapt the proof leading to \eqref{eq:Pn_tail_bound} in the proof
of Lemma \ref{lem:ptrate} to verify that, for $w \in \cb{0, \, 1}$,
$$ \PP{\Norm{ P(w_i = w, \, W_{-i}) - p^*_\pi}_2 > n^{-1/3} \cond X_i = x} = \oo\p{e^{-n^{1/3}}}. $$
Thus, thanks to weak continuity of net demand as in Assumption \ref{as:regularity},
$$ \limn \EE{\abs{\EE[\pi]{Y_i\p{w, \, P(w_i = w, \, W_{-i})}} - Y_i\p{w, \, p^*_\pi}} \cond X_i = x} = 0. $$
We can then conclude that
$$ \limn \bar \tau_{\text{CADE}}(x) - \EE{Y_i\p{1, \, p^*_\pi} - Y_i\p{0, \, p^*_\pi} \cond X_i = x} = 0, $$
which implies the first part of the claimed result.

For the second claim, we follow the structure from the proof of Theorem \ref{theo:pop} and
note that, by Proposition \ref{prop:hte}, it suffices to show that, for any set $S$ such that $\PP{X_k \in S} > 0$,
\begin{equation}
\label{eq:HTE_pf_goal}
\limn \EE{\frac{\partial}{\partial \pi_k} \EE[\pi]{\sum_{i = 1}^n Y_i} \cond X_k \in S} = \EE{\tau^*_{\text{CADE}}(X_k) + \tau^*_{\text{CAIE}}(X_k) \cond X_k \in S}.
\end{equation}
Now, fix such an $S$. Given that $\PP{X_k \in S} > 0$, there is a $c > 0$
such that with probability $1 - e^{cn}$ there is at least one observation $X_k \in S$,
and so, using a convention that $0/0=0$, 
\begin{equation}
\label{eq:HTE_pf_not_empty}
\EE{\frac{\partial}{\partial \pi_k} \EE[\pi]{\sum_{i = 1}^n Y_i} \cond X_k \in S}
= \EE{\frac{1}{\abs{\cb{k : X_k \in S}}}\sum_{X_k \in S} \frac{\partial}{\partial \pi_k} \EE[\pi]{\sum_{i = 1}^n Y_i}} + \oo\p{e^{-cn}}.
\end{equation}
We can then follow the proof of Lemma \ref{lem:ade_ap}, including the
transformation of a derivative into a covariance as in \eqref{eqn:total1} and
the expansion \eqref{eq:Bn}, to verify that
\begin{equation}
\label{eq:HTE_MPE_resid}
\begin{split}
&\EE{\frac{1}{\abs{\cb{k : X_k \in S}}}\sum_{X_k \in S} \frac{\partial}{\partial \pi_k} \EE[\pi]{\sum_{i = 1}^n Y_i}} \\
&\quad\quad\quad = \EE{\tau^*_{\text{CADE}}(X_k) + \tau^*_{\text{CAIE}}(X_k) \cond X_k \in S} \PP{\abs{\cb{k : X_k \in S}} > 0}  \\
&\quad\quad\quad\quad\quad\quad + \EE{\frac{n}{\abs{\cb{k : X_k \in S}}}\sum_{X_k \in S} \frac{W_k - \pi_k}{\pi_k(1 - \pi_k)} B_n},
\end{split}
\end{equation}
where the residual term $B_n$ is as defined in \eqref{eq:Bn}. Now, we already know from the proof of
Lemma \ref{lem:ade_ap} that $\limn n \EE{B_n^2} = 0$. By Cauchy-Schwarz,
\begin{align*}
\EE{\abs{\frac{n}{\abs{\cb{k : X_k \in S}}}\sum_{X_k \in S} \frac{W_k - \pi_k}{\pi_k(1 - \pi_k)} B_n}}
\leq \sqrt{n\EE{B_n^2}} \sqrt{n\EE{\p{\frac{1}{\abs{\cb{k : X_k \in S}}}\sum_{X_k \in S} \frac{W_k - \pi_k}{\pi_k(1 - \pi_k)}}^2}}.
\end{align*}
Because $0 < \eta \leq \pi_k \leq 1 - \eta$ in Design \ref{def:rct}, we have
$$ n\EE{\p{\frac{1}{\abs{\cb{k : X_k \in S}}}\sum_{X_k \in S} \frac{W_k - \pi_k}{\pi_k(1 - \pi_k)}}^2} \leq \frac{n}{\eta(1 - \eta)} \, \EE{\frac{1\p{\cb{\abs{\cb{k : X_k \in S}} > 0}}}{\abs{\cb{k : X_k \in S}}}}, $$
and the right-hand side term above converges to $1/\p{\eta(1 - \eta)\PP{X_k \in S}}$ as $n$ gets large. Thus,
$$ \limn \EE{\frac{n}{\abs{\cb{k : X_k \in S}}}\sum_{X_k \in S} \frac{W_k - \pi_k}{\pi_k(1 - \pi_k)} B_n} = 0, $$
and so \eqref{eq:HTE_pf_not_empty} and \eqref{eq:HTE_MPE_resid} together imply \eqref{eq:HTE_pf_goal}.

\subsection{Proof of Theorem \ref{theo:cade_est}}
\label{sec:hatcate}

For simplicity, we will assume throughout this proof that $X_i$ has a continuous
distribution, and so the set of $k$-nearest neighbors is almost surely unique.\footnote{If
$X_i$ has a discrete distribution, we also need to address random tie-breaking; this
introduces some extra notational overhead, but no conceptual difficulties.}
Now, by treatment randomization $\abs{\cb{i \in N_k(x) : W_i = 1}} / k \rightarrow_p \pi$, and so
because the outcomes are bounded
\begin{equation}
\label{eq:CADE_Bernoulli}
\htau_{\text{CADE}}(x) = \frac{1}{k} \sum_{\cb{i \in N_k(x)}} \p{\frac{W_i}{\pi}\,  Y_i - \frac{1 - W_i}{1 - \pi}\, Y_i} + o_p(1).
\end{equation}
Next, let $B_\varepsilon(x) = \cb{x' : \Norm{x - x'}_2 \leq \varepsilon}$, and
define $\varepsilon_n(x) = \inf\cb{\varepsilon : \PP{X_i \in B_\varepsilon(x)} \geq k/n}$.
When $X_i$ has a continuous distribution, we have $\PP{\Norm{X_i - x}_2 = \varepsilon_n(x)} = 0$,
and so
$$\abs{\cb{i : X_i \in B_{\varepsilon_n}(x)}} \sim \text{Binomial}(k/n, \, n). $$
It immediately follows that the sets $\cb{i : X_i \in B_{\varepsilon_n}(x)}$ and
$N_k(x)$ disagree on only $\oo_p(\sqrt{k})$ elements, and so \eqref{eq:CADE_Bernoulli} implies
that (again because the $Y_i$ are bounded)
\begin{equation}
\label{eq:CADE_IID}
\htau_{\text{CADE}}(x) = \frac{1}{\abs{\cb{i : X_i \in B_{\varepsilon_n}(x)}}} \sum_{\cb{i : X_i \in B_{\varepsilon_n}(x)}} \p{\frac{W_i}{\pi}\,  Y_i - \frac{1 - W_i}{1 - \pi}\, Y_i} + o_p(1).
\end{equation}
We note that the leading term in the above expression is an IID average for any given $n$.
Furthermore, because $\PP{X \in B_\varepsilon(x)} > 0$ for any $\varepsilon > 0$ by assumption,
we have $\varepsilon_n \rightarrow 0$;
and, because the conditional distribution of $Y_i(w, \, p)$ given $X_i = x$ varies continuously
in $x$ and $Y_i(w, \, p)$ is bounded, conditional moments of $Y_i(w, \, p)$ also vary continuously
in $x$. Thus, the condition \eqref{eq:covariance_convergence} from Lemma \ref{lemm:asymp_equicontinuity}
holds for the potential outcome functions used to construct \eqref{eq:CADE_IID},
and so for any sequence $p_n \rightarrow_p p$,
\begin{equation}
\begin{split}
&\limn \Bigg| \frac{1}{\abs{\cb{i : X_i \in B_{\varepsilon_n}(x)}}}  \sum_{\cb{i : X_i \in B_{\varepsilon_n}(x)}} \bigg(\frac{W_i}{\pi} \, Y_i(1, p_n)  \\
&\quad\quad\quad\quad -
 \p{\frac{W_i}{\pi}\,  Y_i(1, p) + y(1, p_n; \, X_i) - y(1, p; \, X_i)} \bigg)\Bigg|
  = o_p\p{\frac{1}{\sqrt{k}}},
\end{split}
\end{equation}
and a similar expansion holds for $(1 - W_i) Y_i(0, \, p_n ) / (1 - \pi)$.
Finally, because $P_n \rightarrow_p p^*_\pi$ by Theorem \ref{theo:prate}, $k \to \infty$, and
$y(w, \, p; \, x)$ is continuous in $p$ and $x$ (and thus uniformly continuous
in $p$ in a neighborhood of $x$), we can combine all above results to get
\begin{equation}
\htau_{\text{CADE}}(x)
= \frac{1}{k} \sum_{\cb{i \in N_k(x)}} \p{\frac{W_i}{\pi}\,  Y_i(1, \, p^*_\pi) - \frac{1 - W_i}{1 - \pi}\, Y_i(0, \, p^*_\pi)} + o_p(1).
\end{equation}
We also note that this argument can be extended to obtain rates of convergence
under smoothness assumptions following \citet{stone1980optimal}.

\subsection{Proof of Proposition \ref{prop:opt}}

By uniqueness of equilibrium prices, $p^*_{\nu} = p^*_{\pi}$ if and only if
$$\mathbb E[ \nu(X_i) Z_i(1, p^*_{\pi}) + ( 1- \nu(X_i) )Z_i(0, p^*_{\pi})] = \mathbb E[\pi(X_i) Z_i(1, p^*_{\pi}) +(1 - \pi(X_i)) Z_i(0, p^*_{\pi})] = 0. $$
This means the constraint $p^*_{\nu} = p^*_{\pi}$ in optimization problem in Equation \eqref{eqn:target}
can be replaced with the above moment constraint. We can
then transform the constraint by subtracting $\mathbb E[ Z_i(0, p^*_{\pi})]$ from both sides above
and applying the chain rule to get the constraint claimed in Proposition \ref{prop:opt}.
 Meanwhile, for the objective, we can use the equilibrium stability condition to verify
 \begin{align*}
 \EE{Y_i(W_i, p^*_{\nu})}
 &= \EE{\nu(X_i) Y_i(1, \, p^*_\nu) + (1 - \nu(X_i)) Y_i(0, \, p^*_\nu)} \\
 &= \EE{\nu(X_i) Y_i(1, \, p^*_\pi) + (1 - \nu(X_i)) Y_i(0, \, p^*_\pi)} \\
 &= \EE{\nu(X_i) \p{Y_i(1, \, p^*_\pi) - Y_i(0, \, p^*_\pi)}} + \EE{Y_i(0, \, p^*_\pi)}\\
 &= \EE{\nu(X_i) \EE{Y_i(1, \, p^*_\pi) - Y_i(0, \, p^*_\pi) \cond X_i}} + \EE{Y_i(0, \, p^*_\pi)},
 \end{align*}
which yields the desired expression after dropping the last term that does not depend on $\nu(\cdot)$.

\subsection{Proof of Theorem \ref{theo:ratio}}

This result follows from \citet{dantzig1951fundamental}. Mapping to the notation of that paper, we introduce a supplementary
random variable $U_i \sim \mbox{Uniform}[0, 1]$ independently of everything else. For any (potentially random) policy $\nu : \xx \rightarrow [0, \, 1]$,
define the set $S_\nu = \{ (x, \, u) : \nu(x) \geq u \}$. Writing $A_i = (X_i, \, U_i)$ and $F$ for the induced distribution on $F$,
the value function $V(\nu) = \EE{\nu(X_i) \tau_{\text{CADE}}^*(X_i)}$ can be re-expressed as 
\begin{equation}
\label{eq:pf_maximizer}
V(\nu) = \int \mathbbm{1}\p{\cb{a \in S_\nu}}  \tau_{\text{CADE}}^*(x) dF(a),
\end{equation}
and the constraint functions $G_{j}(\nu) = \EE{\p{\nu(X_i) - \pi(X_i)} \tau_{\text{CADE}, j}^{*,z}(X_i)} $ can be re-expressed as
\begin{equation}
\label{eq:pf_constraints}
G_j(\nu) = \int \mathbbm{1}\p{\cb{a \in S_\nu}}   \p{\nu(x) - \pi(x)} \tau_{\text{CADE}, j}^{*,z}(x) dF(a).
\end{equation}
Now, let
\begin{equation}
\mathcal{B} = \cb{b : G(\nu) = b \text{ for some } 0 \leq \nu(\cdot) \leq 1},
\end{equation}
and let $\operatorname{span}\p{\mathcal{B}} \subseteq \RR^J$ be its linear span.
Then, Theorem 3.1 of \citet{dantzig1951fundamental} tells us that, as long as $0$ is an interior point of $\mathcal{B}$
with respect to $\operatorname{span}\p{\mathcal{B}}$, then
the supremum of \eqref{eq:pf_maximizer} subject to \eqref{eq:pf_constraints} is achieved, and
any $S_\nu$ satisfying the constraints \eqref{eq:pf_constraints} is value-maximizing if and only there exists $c \in \RR^J$ 
such that, almost surely
\begin{equation}
\label{eq:DWset}
\begin{split}
&\tau_{\text{CADE}}^*(X_i) \geq \sum_{j = 1}^J c_j \tau_{\text{CADE}, j}^{*,z}(X_i) \text{ whenever } A \in S_\nu, \\
&\tau_{\text{CADE}}^*(X_i) \leq \sum_{j = 1}^J c_j \tau_{\text{CADE}, j}^{*,z}(X_i) \text{ whenever } A \not\in S_\nu.
\end{split}
\end{equation}
Now, since $U_i$ is drawn uniformly over $[0, 1]$, a policy $\nu$ induces a set $S_\nu$ satisfying \eqref{eq:DWset} if and only if,
for each $x \in \mathcal X$  except on a set of measure zero: 
\begin{enumerate} 
\item  $ \tau_{\text{CADE}}^*(x)  >  c^{\top}  \tau_{\text{CADE}}^{*,z}(x)  $ and $\nu(x) = 1$, 
\item   $ \tau_{\text{CADE}}^*(x)  <  c^{\top}  \tau_{\text{CADE}}^{*,z}(x)  $ and $\nu(x) = 0$, 
\item  $ \tau_{\text{CADE}}^*(x)  = c^{\top}  \tau_{\text{CADE}}^{*,z}(x)  $ and $\nu(x) \in [0, 1]$. 
\end{enumerate}
Thus, any policy is optimal if and only if it satisfies the constraints \eqref{eq:pf_constraints} and is
(almost surely) of the above form.

It remains to determine behavior of $\nu$ in the region where $\tau_{\text{CADE}}^*(x)  = c^{\top}  \tau_{\text{CADE}}^{*,z}(x)$.
Given $c$, for any policy of the above form, satisfying the constraints \eqref{eq:pf_constraints} entails
\begin{equation}
\begin{split}
&c^\top \EE{\p{\nu(X_i) - \pi(X_i)} \tau_{\text{CADE}}^{*,z}(X_i) 1\p{\cb{\tau_{\text{CADE}}^*(x) = c^{\top}  \tau_{\text{CADE}}^{*,z}(x)}}} \\
&\quad\quad + c^\top \EE{\p{1 - \pi(X_i)} \tau_{\text{CADE}}^{*,z}(X_i) 1\p{\cb{\tau_{\text{CADE}}^*(x)  > c^{\top}  \tau_{\text{CADE}}^{*,z}(x)}}} \\
&\quad\quad\quad\quad + c^\top \EE{\p{0 - \pi(X_i)} \tau_{\text{CADE}}^{*,z}(X_i) 1\p{\cb{\tau_{\text{CADE}}^*(x)  < c^{\top}  \tau_{\text{CADE}}^{*,z}(x)}}} = 0.
\end{split}
\end{equation}
Now, in the 3rd region we're considering $\tau_{\text{CADE}}^*(x)  = c^{\top}  \tau_{\text{CADE}}^{*,z}(x)$, and so any policy
satisfying these constraints must in fact have the same value.
Thus, if there exists a constant $b$ such that a policy of the above form with $\nu(x) = b$ for all $x$
$ \tau_{\text{CADE}}^*(x)  = c^{\top}  \tau_{\text{CADE}}^{*,z}(x)  $ satisfies the constraints, then this policy is in fact optimal.

It now remains to check the interior-point condition.  Recall that, in our Design \ref{def:rct}, we have ``overlap'' in the sense that
$\eta \leq \pi(x) \leq 1 - \eta$ for all $x \in \xx$ for some $\eta > 0$. The main step is to show that, under our conditions,
\begin{equation}
\label{eq:partial_symmetry}
-\eta \mathcal{B} \subset \mathcal{B}.
\end{equation}
To verify this, pick any $b \in \mathcal {B}$. There must be a $\nu(\cdot)$ such that $\EE{\p{ \nu(X) - \pi(X)} \tau^{*,z}_{\text{CADE}}(X)}  =b$.
Given such a $\nu$, let $\tilde \nu(\cdot) = \pi(\cdot) - \eta (\nu(\cdot) - \pi(\cdot))$. By overlap, $0 \leq \tilde \nu(\cdot) \leq 1$, i.e.,
$\tilde \nu(\cdot)$ is a valid policy. Furthermore, clearly, $\EE{\p{\tilde \nu(X) - \pi(X)} \tau^{*,z}_{\text{CADE}}(X)} = -\eta b$, i.e.,
$-\eta b \in \mathcal{B}$.
Finally, because $\mathcal{B}$ is also convex, \eqref{eq:partial_symmetry} immediately implies the interior-point condition: if we pick a maximal
set of linearly independent vectors in $\mathcal{B}$, then their mirror images (up to a factor $\eta$) are also
in $\mathcal{B}$, and the convex hull of both the original vectors and their mirror images contains an open set
around 0 and is in turn contained in $\mathcal{B}$.

\subsection{Consistency of Estimated Targeting Rule} 
\label{as:target_con} 

When the optimal rule  is deterministic and unique, it has the form $\nu^*(X_i) = \mathbbm{1}( \tau^*_{\text{CADE}}(X_i) \geq c^*\tau^{*, z}_{\text{CADE}}(X_i) )$. By Theorem \ref{theo:ratio}, we can write the optimal rule as the unique solution to the following $J$-dimensional moment condition, where $\tau^*$ represents the vector of functions that concatenates  $\tau^*_{\text{CADE}}(X_i)$ and $\tau^{*, z}_{\text{CADE}}(X_i)$ and $\tau$ is some bounded function with the same domain and codomain. 
\[ G(c^*; \tau^*) = 0, \mbox{ where } G(c ; \tau) = \mathbb E \left [ \Big ( \mathbbm{1} ( \tau(X_i) \geq c^{\top} \tau^z(X_i)) - \pi(X_i) \Big) \tau^z(X_i) \right]   \] 

$c^*$ is finite, so we can specify a ball in $\mathbb R^J$ such that the distance from $c^*$ to the surface of this ball is at least $M$. Specify such a ball as $\mathcal B_M$.  We can estimate the optimal rule by solving the following $J$-dimensional score condition. The conditional average direct effects are nuisance functions, since they need to be estimated, where $\hat \tau$ is some estimate of the population conditional average treatment effects on outcomes and net demand:

\begin{equation*} 
\begin{split}
& \hat c  \in \{ c \in \mathbb R^J:  \hat G_n(c; \hat \tau) = o_p(1) \}, \\ 
&  \hat G_n(c; \hat \tau) = \frac{1}{n} \sum \limits_{i=1}^n (\mathbbm{1}( \hat \tau(X_i) \geq c^{\top} \hat \tau^z(X_i) - \pi(X_i)) \hat \tau^z(X_i). 
\end{split}
\end{equation*} 

The proof of Proposition \ref{prop:chat} shows that the optimal-equilibrium neutral rule is the unique (and well-separated) solution to the $J$ population constraints defined by $G(\cdot)$. Then, the consistency of the estimated rule follows from uniform convergence of the empirical constraint functions to the population constraint functions.

\begin{proposition} \label{prop:chat} 

 For every $v \in S^{J-1}$ and $b \in \mathbb R^{+}$, where $S^{J-1}$ is the unit sphere in $\mathbb R^J$, assume that the random variable $ \tau^*_{\text{CADE}}(X_i) - b \cdot v^{\top} \tau^{z, *}_{\text{CADE}}(X_i)$ has uniformly bounded density, and for every $v \in S^{J-1}$ that $v^{\top} \tau^{z, *}_{\text{CADE}}(X_i)$ has uniformly bounded density. In addition, assume that the covariance matrix of $ \tau^{*, z}_{\text{CADE}}(X_i)$ is positive definite, and that $\hat G_n(\hat c; \hat \tau) = o_p(1)$.  Finally, assume that $\hat \tau: \mathcal X \to \mathbb R$ and $\hat \tau^z: \mathcal X \to \mathbb R$ are functions that are uniformly bounded and are consistent estimators of the population conditional average treatment effects in the following sense: 

\begin{equation*} 
\begin{split} 
&  \mathbb E_{T}[ (\hat \tau(X_i) - \tau^*_{\text{CADE}}(X_i))^2] = o_p(1), \\ 
&   \mathbb E_{T} [ || \hat \tau^z_j(X_i) - \tau^{*, z}_{j, \text{CADE}}(X_i))||_2^2]   = o_p(1), 
\end{split} 
\end{equation*} 
where $\mathbb E_{T}[\cdot]$ is the expectation over a random sample of test data, conditional on the training data used for estimation. 

Then, the estimated targeting rule is consistent for the optimal equilibrium-stable targeting rule in the population, 
\[ \hat c = c^* + o_p(1). \] 

\end{proposition} 

First, we use a change of variables to make the parameter space compact. For every $c \in \mathbb R^{J}$, we can write $c = f(a) \cdot v$, where  $v = c/||c||_2 \in S^{J-1}$ and $a = f^{-1}( || c||_2) \in [0, 1)$ for some continuous and strictly monotonic function $f: [0, 1] \to [0, +\infty]$ with $f(0) =0$ and $f(1) = \infty$.  For this proof, it will be useful to characterize properties of the population constraint function over the extended (and compact) space of $a \in  [0, 1]$, even though the optimal rule, and all of the estimated rules are found within $a \in [0, 1)$. For a more compact notation define $\theta$ as the vector combining $v$ and $a$, where $\theta \in \Theta$, and $\Theta = S^{J-1} \times [0, 1]$. 
Let \[ \nu (X_i; \theta, \tau)  = \begin{cases} \mathbbm{1}(\tau(X_i) > f(a)v^{\top} \tau^z(X_i)),  & 0 \leq  a < 1, \\ 
							 \mathbbm{1}( 0 > v^{\top} \tau^z(X_i) ), & a =1. \end{cases}\] 
Now we can define the population and empirical constraint at a given treatment rule and for a given set of conditional mean functions: 
\[ G(\theta; \tau)  =   \mathbb E_T \left [ \Big ( \nu (X_i; \theta,  \tau)  - \pi(X_i) \Big) \tau^z(X_i) \right], \qquad G_n(\theta^*;  \tau) = \frac{1}{n} \sum \limits_{i=1}^n  \Big ( \nu (X_i; \theta,  \tau)  - \pi(X_i) \Big) \tau^z(X_i) .  \] 
Any estimated treatment rule, characterized by $\hat \theta$,  approximately satisfies the empirical constraint with conditional average treatment effects estimated by data-splitting, so that $G_n(\hat \theta; \hat \tau) = o_p(1)$. 

Under our assumptions, Theorem \ref{theo:ratio} indicates that there is a unique and deterministic rule in the population characterized by $G(\theta^*; \tau^*) =0$, where $\tau^*$ collects $\tau^*_{\text{CADE}}(X_i)$ and $\tau^{*, z}_{\text{CADE}}(X_i)$ and $\theta^*$ collects $a^*$ and $v^*$. It is deterministic because the set $\{ x \in \mathcal X: \tau^*_{\text{CADE}}(x) = f(a^*) \cdot v^{*\top} \tau^{z, *}_{\text{CADE}}(x)\}$ has measure 0. By Theorem \ref{theo:ratio},  any optimal equilibrum-neutral rule that is deterministic has the structure $\mathbbm{1}(\tau^*_{\text{CADE}}(x)  > f(a^*) v^{*, \top}  \tau^{z, *}_{\text{CADE}}(x))$. Since the covariance matrix of $ \tau^{*, z}_{\text{CADE}}(X_i)$ is positive definite, there can be only one such rule that meets the population constraints (this rules out label swapping between two goods in the market, for example). 

Next, we want to show that each element of $G(\theta; \tau^*)$ is a continuous function in $\theta$ for all $\theta \in \Theta$. Choose some $a_0, v_0$ and arbitrary $j \in \{ 1, \ldots, J \}$. For some sequence $\theta_n$, where $\theta_n \to \theta_0 = (a_0, v_0)$, by the boundedness of $\tau^z(X_i)$, there is some finite $M$ such that  

\begin{equation*} 
\begin{split} 
  |G_j(\theta_n; \tau^*) - G_j(\theta_0; \tau^*) | & \leq   P ( \nu (X_i; \theta_n, \tau^*) \neq  \nu (X_i; \theta_0, \tau^*)) M \\ 
 & = O( || \theta_n - \theta_0 ||_2)    \\
 & = o(1) 
 \end{split}
\end{equation*} 
The first step is by Lemma \ref{lem:supc}. This proves that when $a=1$, then $G(\theta; \tau^*)$ is continuous in $\theta$. Since $G(\theta; \tau)$ is a continuous function in $\theta$,  $\hat \theta$ is also in a compact space, and $\theta^*$ is the unique solution to $G(\theta^*; \tau^*) = 0$, then we know that if $\epsilon >0$, then $\sup \limits_{\theta: || \theta - \theta^* ||_2 > \epsilon} || G(\theta) ||_2 > 0$.  
The final step is to show uniform consistency: 

\[ \sup \limits_{\theta \in \Theta } | \hat G_n( \theta; \hat \tau) - G(\theta; \tau^*) | = o_p(1).    \] It will be useful to make the following decomposition: 
\begin{equation} \label{eq:splitsup} 
  | \hat G_n( \theta; \hat \tau) - g(\theta; \tau) | \leq |  G_n( \theta; \hat \tau) -  G(\theta; \hat \tau) | + | G(\theta ; \hat \tau) - G(\theta; \tau^*) |.
  \end{equation}   For the first term of \eqref{eq:splitsup}, we write the empirical average using data-splitting, where $I_k$ are the indexes of data in split $k$, and $\hat \tau^{k(i)}(X_i)$ are conditional average treatment effects estimated on data that is not in $I_k$. We drop the CADE subscripts here to keep the notation more manageable. Then, by treating $\hat \tau^{k}$ as fixed within each split, we use tail bounds for empirical processes indexed by VC-classes. The details are as follows: 
\begin{equation*} 
\begin{split} 
& \sup_{\theta \in \Theta}  | G_n( \theta; \hat \tau) -  G(\theta; \hat \tau)|  \\ 
&  = \sup \limits_{\theta \in \Theta}   \left | \frac{1}{n} \sum \limits_{i=1}^n ( \nu(X_i; \theta, \hat \tau)  -\pi(X_i)) \hat \tau^z(X_i) -  \mathbb E_T[   ( \nu(X_i; \theta, \hat \tau)  -\pi(X_i)) \hat \tau^{ z}(X_i) ] \right | \\ 
& = \sup \limits_{\theta \in \Theta}   \left | \sum \limits_{k =1 }^K n_k/n \frac{1}{n_k} \sum \limits_{i \in I_k} ( \nu(X_i; \theta, \hat \tau ^{k(i)} )  -\pi(X_i)) \hat \tau^{k(i), z}(X_i) -  \mathbb E_T[   ( \nu(X_i; \theta, \hat \tau^{k(i)})  -\pi(X_i)) \hat \tau^{ k(i), z}(X_i) ] \right | \\
& \leq  \sum \limits_{i=1}^K \frac{n_k}{n}  \sup \limits_{\theta \in \Theta} | A_{nk}(\theta) | \\ 
& \stackrel{(1)}{=} o_p(1)
\end{split} 
\end{equation*} 

Within each split, we can treat $\hat \tau^{k}$ as fixed. $\mathcal F = \{  \mathbbm{1}( \tau(X_i) \geq f(a) v^{\top}  \tau^{ z}(X_i)) \tau^{ z}(X_i) : a \in [0, 1], v \in S^{J-1} \}$ for a fixed and uniformly bounded $\tau(X_i)$ is a VC-class multiplied by a bounded random variable. Then, the tail bound for  empirical processes indexed by VC-classes (e.g. Theorem 2.6.7 of \citet{van1996weak}) gives a tail bound for $ \sup_{\theta \in \Theta} |A_{nk} (\theta)|$  that does not depend on a given realization of $\hat \tau$. This implies $\lim \limits_{n \to \infty} Pr( \sup \limits_{\theta \in \Theta}  |A_{nk} | > t)= 0$ for any $t > 0$. 

To handle the second term in \eqref{eq:splitsup}, 

\begin{equation*} 
\begin{split} 
\sup_{\theta \in \Theta} | G(\theta ; \hat \tau) - G(\theta; \tau^*) |   & = \sup_{\theta \in \Theta} | \mathbb E_T[   ( \nu(X_i; \theta, \hat \tau)  -\pi(X_i)) \hat \tau^{ z}(X_i) ] -  \mathbb E_T[   ( \nu(X_i; \theta,  \tau^*)  -\pi(X_i)) \tau^{*,  z}(X_i) ] | \\ 
& \leq  \sqrt { \mathbb E_{T}[( \hat \tau^{z}(X_i)  - \tau^{*, z}(X_i) )^2] } +  \sup_{\theta \in \Theta} M \sqrt {  \mathbb E_{T} [(\nu(X_i; \theta, \hat \tau) - \nu(X_i; \theta, \tau^*))^2] } \\ 
& \leq o_p(1) +  \sup_{\theta \in \Theta} M \sqrt { P( \nu(X_i; \theta, \hat \tau) \neq \nu(X_i; \theta, \tau^*))} \\ 
& \stackrel{(1)} = o_p(1) +   O( || \hat \tau^z(X_i) - \tau^{*, z}(X_i) ||_2 + | \tau^*(X_i) - \hat \tau(X_i) |)\\ 
& = o_p(1) 
\end{split} 
\end{equation*} 

where $M$ is finite since net demand is bounded,  (1) is from Lemma \ref{lem:supc} and the final step is from the assumption on the mean-square convergence of the nuisance functions. We have now shown that $\sup \limits_{\theta \in \Theta } | \hat G_n( \theta; \hat \tau) - G(\theta; \tau^*) | = o_p(1).$

Now, that we have proved that the unique population treatment rule is well-separated and that the sample constraints converge uniformly to the population constraints, following Theorem 5.9 of \citet{van2000asymptotic}, we have consistency of $\hat \theta$ to $\theta^*$. The details follow. Since $\hat \theta $  satisfies the empirical constraints, then $|| G_n(\hat \theta; \hat \tau) ||_2 \leq  || G_n(\theta^*; \tau^*) ||_2 + o_p(1)$.  By the convergence of the empirical constraints in the RHS, $|| G_n(\hat \theta; \hat \tau) ||_2 \leq  ||G(\theta^*; \tau^*) ||_2 + o_p(1)$. Adding and subtracting $||G(\hat \theta; \tau^*)||_2$, we have: 
\begin{equation*} 
\begin{split} 
  || G(\hat \theta; \tau^*) ||_2 - || G_n(\hat \theta; \hat \tau) ||_2  &  \geq   || G(\hat \theta;  \tau^*) ||_2 -  ||G(\theta^*; \tau^*) ||_2  - o_p(1) \\ 
    || G(\hat \theta;  \tau^*) ||_2 -  ||G(\theta^*; \tau^*) ||_2 &  \leq \sup_{\theta \in \Theta} || G(\theta; \tau^*) ||_2 - || G_n(\theta; \hat \tau) ||_2 \\ 
 || G(\hat \theta;  \tau^*) ||_2    & \leq  \sup_{\theta \in \Theta} || G(\theta; \tau^*) -  G_n(\theta; \hat \tau) ||_2 \\ 
   || G(\hat \theta;  \tau^*) ||_2 & \leq  o_p(1) 
\end{split} 
\end{equation*} 

Since $\theta^*$ is well-separated, then for any $\epsilon > 0$, for any $\theta$ such that $|| \theta - \theta^* || > \epsilon$, then $||G(\theta)||_2 > \eta$. So, for any $\epsilon > 0$, $P(|| \hat \theta_n - \theta^* || > \epsilon ) \leq P(||G(\hat \theta; \tau^*)||_2 > \eta) \rightarrow_p 0$. Since $\hat \theta$ is just a re-parameterization of $\hat c$, then we have shown that $\hat c = c^* + o_p(1)$. 

\begin{lemma} \label{lem:supc} 
Under the Assumptions (and notation)  of Proposition \ref{prop:chat}, 

\begin{equation*} 
\begin{split} 
 & P( \nu (X_i; \theta_n, \tau^*) \neq  \nu (X_i; \theta_0, \tau^*)) = O(  || \theta_n - \theta_0||_2 ), \\ 
 & \sup \limits_{\theta \in \Theta} P( \nu(X_i; \theta, \hat \tau) \neq \nu(X_i; \theta, \tau^*)) = O( || \hat \tau^z(X_i) - \tau^{*, z}(X_i) ||_2 + | \tau^*(X_i) - \hat \tau(X_i) |).  
\end{split} 
\end{equation*} 

\end{lemma} 
\begin{proof} 

For $P ( \nu (X_i; \theta_n, \tau^*) \neq  \nu (X_i; \theta_0, \tau^*)) $ we divide into two cases. In the first case, $a_0  < 1$. We can choose large enough $n$  (to ensure that  $a_n \neq 1$) so that 
\begin{equation*}
\begin{split} 
 &  P ( \nu (X_i; \theta_n, \tau^*) \neq  \nu (X_i; \theta_0;  \tau^*))  \\ & = P ( \min \{  f(a_0) v_0^{\top} \tau^{*, z}(X_i),   f(a_n) v_n^{\top} \tau^{*, z}(X_i) \}  < \tau^*(X_i) < \max \{ f(a_0) v_0^{\top} \tau^{*, z}(X_i),  f(a_n) v_n^{\top} \tau^{*, z}(X_i)\}  )  \\ 
  &  \leq P \Big ( | \tau^*(X_i) - f(a_0) v_0^{\top} \tau^{*, z}(X_i) | < |  (f(a_n) v_n - f(a_0) v_0) \tau^{*, z}(X_i) | \Big) \\ 
 & \leq P \Big( | \tau^*(X_i) - f(a_0) v_0^{\top} \tau^{*, z}(X_i) | < |  f(a_n) v_n - f(a_0) v_0 | M  \Big )\\
 & \leq C  || \theta_n - \theta_0 ||_2, 
\end{split} 
\end{equation*} 

for finite $C$, where $M$ is finite since net demand is bounded. The last step is because $\tau^*(X_i) - a v^{\top} \tau^{*, z}(X_i)$ has bounded  density, so its distribution function is Lipschitz, and $f(\cdot)$ is continuous.  Next, for $a_0 = 1$, we can prove something similar. We can choose large enough (to ensure that $a_n \neq 0$) so that 

\begin{equation*} 
\begin{split} 
&  P ( \nu (X_i; \theta_n, \tau^*) \neq  \nu (X_i; \theta_0;  \tau^*)) \\ 
 & =  P ( \min \{  0,  ( v_0^{\top} - v_n^{\top})  \tau^{*, z}(X_i) + \tau^*(X_i)/a_n \}  < v_0^{\top} \tau^{*, z}(X_i) < \max \{  0,  ( v_0^{\top} - v_n^{\top})  \tau^{*, z}(X_i) + \tau^*(X_i)/a_n \}  )\\ 
 & \leq  P \Big ( | v_0^{\top} \tau^{*, z}(X_i) | <  | (v_0^{\top} - v_n^{\top})  \tau^{*, z}(X_i) + \tau^*(X_i)/f(a_n) | ) \\ 
 & \leq P \Big ( | v_0^{\top} \tau^{*, z}(X_i) | <  | (v_0^{\top} - v_n^{\top})  M + M /f(a_n) | ) \\
 & \leq C || \theta_n - \theta_0 ||_2
\end{split} 
\end{equation*} 

for some finite $C$, where $M$ is finite by boundedness of outcomes and net demand. The last step is because $v^{\top} \tau^{*, z}(X_i)$ has bounded density, so its distribution function is Lipschitz, and $f(\cdot)$ is continuous.

For the second part of the Lemma, also split into two cases. First, when $0 \leq f(a) \leq 2$ (denote the product of this restricted space for $a$ and $S^{J-1}$ as $\Theta^+$), let $b_n = a v^{\top} (\hat \tau^z(X_i) - \tau^{*, z}(X_i)) + \tau^*(X_i) - \hat \tau(X_i)$. 

\begin{equation*} 
\begin{split} 
& \sup_{\theta \in \Theta^{+}} P( \nu(X_i; \theta, \hat \tau) \neq \nu(X_i; \theta, \tau^*)) \\ &  = P( \min \{ b_n , 0 \} < \tau^*(X_i) - f(a)v^{\top} \tau^{*, z}(X_i) < \max \{b_n, 0\} ) \\ 
& \leq  \sup_{\theta \in \Theta^{+}} P(|\tau^*(X_i) - f(a) v^{\top} \tau^{*, z}(X_i) |  \leq | f(a)  v^{\top} (\hat \tau^z(X_i) - \tau^{*, z}(X_i)) + \tau^*(X_i) - \hat \tau(X_i)| ) \\ 
& \leq  \sup_{\theta \in \Theta^{+}}  P(|\tau^*(X_i) - f(a) v^{\top} \tau^{*, z}(X_i) |  \leq  2 || \hat \tau^z(X_i) - \tau^{*, z}(X_i) ||_2 +  | \tau^*(X_i) - \hat \tau(X_i) | ) \\ 
& =  O( || \hat \tau^z(X_i) - \tau^{*, z}(X_i) ||_2 + | \tau^*(X_i) - \hat \tau(X_i) |)  \\ 
\end{split} 
\end{equation*} 

(1) is because $\tau^*(X_i) - a v^{\top} \tau^{*, z}(X_i)$ has a density uniformly bounded over $a$ and $v$. This means that the distribution function of $\tau^*(X_i) - a v^{\top} \tau^{*, z}(X_i)$, which we can call $F(t)$ is Lipschitz in $t$, so

\[ F(2 || \hat \tau^z(X_i) - \tau^{*, z}(X_i) ||_2 +  | \tau^*(X_i) - \hat \tau(X_i) |) - F(0) \leq 2M  || \hat \tau^z(X_i) - \tau^{*, z}(X_i) ||_2 +  M | \tau^*(X_i) - \hat \tau(X_i) | , \]  where $M$ does not depend on $a$ or $v$. When $a \geq 2$, we can argue similarly, where we use $\Theta^{-}$ to denote the product of this restricted space and $S^{J-1}$. Let $c_n = v^{\top} ( \tau^{*, z}(X_i) - \hat \tau^{*, z}(X_i) ) + 1/ f( a) ( \hat \tau^*(X_i) - \tau^*(X_i))$. 

\begin{equation*} 
\begin{split} 
& \sup_{\theta \in \Theta^{-}} P( \nu(X_i; \theta, \hat \tau) \neq \nu(X_i; \theta, \tau^*)) \\ &  = P( \min \{ c_n , 0 \} <  av^{\top} \tau^{*, z}(X_i)  - 1 / a \tau^*(X_i) < \max \{c_n, 0\} ) \\ 
& \leq \sup_{\theta \in \Theta^-} P( v^{\top} \tau^{*, z}(X_i)  - (1 / f(a)) \tau^*(X_i)  \leq || \hat \tau^z(X_i) - \tau^{*, z}(X_i) ||_2 + | \tau^*(X_i) - \hat \tau(X_i) |)  \\ 
& \leq O( || \hat \tau^z(X_i) - \tau^{*, z}(X_i) ||_2 + | \tau^*(X_i) - \hat \tau(X_i) |)\\ 
\end{split} 
\end{equation*} 
Where for the last step this is because under our assumptions, for any $f(a)  \in [2, \infty]$ the density of $v^{\top} \tau^{*, z}(X_i)  - 1 / f(a) \tau^*(X_i) $ is uniformly bounded (the bound does not depend on $v$ or $a$). 
\end{proof}

\section{Concentration Bounds}

Consider a triangular array of random functions $F_{in}(p)$ with $i = 1, \, \ldots, n$,
$n = 1, \, 2, \, \ldots$ and $p \in \RR^J$. Suppose that $F_{in}(\cdot)$ with $i = 1, \, \ldots, n$
are independent and identically distributed, and let $f_n(p) = \EE{F_{in}(p)}$.
We say that the sample average of the $F_{in}(\cdot)$ is asymptotically equicontinuous
at $p^*$ if, for any sequence $\delta_n \rightarrow 0$, we have
\begin{equation}
\label{eq:asymptotic_equicont}
\sup_{\Norm{p - p^*} \leq \delta_n} \abs{\frac{1}{n} \sum_{i = 1}^n \p{F_{in}(p) - F_{in}(p^*) - \p{f_n(p) - f_n(p^*)}}} = o_p\p{\frac{1}{\sqrt{n}}}.
\end{equation}
The motivation behind establishing asymptotic equicontinuity expansions is that we may often be able to assume that
the expected function $f_n(p)$ is differentiable even when the $F_{in}(p)$ themselves are not
(e.g., if the $F_{in}(p)$ capture choices regarding supply and demand that may vary discontinuously
in prices). Asymptotic equicontinuity then allows us to approximate sample averages of $F_{in}(p)$
by Taylor expanding $f_n(p)$.

The goal of this section is to establish asymptotic equicontinuity for a number of function classes
used throughout the proofs. Our argument relies on technical tools from empirical process theory that go
beyond what's used elsewhere in the paper, and here we will only give brief references to the results
needed to establish our desired claims. Section 2 of \citet{van1996weak} provides an excellent introduction
and reference to the tools underlying the arguments made here.

Our approach to establishing asymptotic equicontinuity of relevant quantities
in our marketplace model starts by bounding the bracketing number of approximately monotone functions.
Let $F_{in} : \RR^J \rightarrow \RR$ be random functions drawn i.i.d. from a distribution $Q_n$,
and let $\set \subset \RR^J$ be the set of possible market equilibrium prices. We are interested in a bracket for the function class $\mathcal F_n = \{ F_{in}(\cdot) \to F_{in}(p) : p \in \mathcal S \}$. 
For any pair $p_-, \, p_+$ in $\RR^J$, we define a bracket as
\begin{equation}
\label{eq:bracket_def}
[p_-, \, p_+]= \cb{p \in \set : F_{in}(p_-) \geq F_{in}(p) \geq F_{in}(p_+) \text{ for all }  F_{in}(\cdot) } .
\end{equation}
We define the square of the $L_2(Q_n)$-length of the bracket as $d^2_{Q_n}(p_-, \, p_+) = \EE[Q_n]{(F_{in}(p_-) - F_{in}(p_+))^2}$, and
the $\varepsilon$-bracketing number of $\mathcal F_n$ under $Q_n$ as 
\begin{equation}
N_{[]} (\varepsilon, \, \mathcal F_n, \, L_2(Q_n)) = \inf\cb{ K : \set \subseteq \bigcup_{k = 1}^K [p_-(k), \, p_+(k)] \text{ with } d_{Q_n}\p{p_-(k), \, p_+(k)} \leq \varepsilon \text{ for all } k}.
\end{equation}
The following result bounds the bracketing number of functions that are approximately monotone
and weakly continuous in the sense of Assumption \ref{as:monotone}.

\begin{lemma}
\label{lemm:bracket}
Consider a class of random functions $F_{in} : \RR^J \rightarrow \RR$ with $F_{in} \in \mathcal G_n$
almost surely, and $\set$ is a compact subset of $\RR^J$. Let $\mathcal F_n = \{ F_{in}(\cdot) \to F_{in}(p) : p \in \mathcal S \}$. Suppose that, for all
$p \in \mathcal S$, $\varepsilon \leq 1$ and $\Norm{\delta}_2 \leq C \varepsilon$,
we have
\begin{equation}
\label{eq:approx_mon_ap}
F_{in}(p - \varepsilon e_j) \geq F_{in}(p + \delta) \geq F_{in}(p + \varepsilon e_j).
\end{equation}
Suppose also that the $F_{in}$ have a distribution $Q_n$ under which
\begin{equation}
\label{eq:weak_cont_ap}
\EE[Q_n]{\p{F_{in}(p) - F_{in}(p')}^2} \leq L \Norm{p - p'}_2 \text{ for all } p,\, p' \in \RR^J.
\end{equation}
Then, there exists $\theta(\set, \, C, \,  L)$ depending only on the geometry of $\set$
and the constants given above such that, for all $\varepsilon \leq 1$,
\begin{equation}
N_{[]} (\varepsilon, \, \mathcal F_n, \, L_2(Q_n)) \leq \theta(\set, \, C, \,  L) \, \varepsilon^{-2J}.
\end{equation}
\begin{proof}
Let $B_\delta(p) = \cb{p' \in \set : \Norm{p' - p}_2 \leq \delta}$. By \eqref{eq:approx_mon_ap},
we have
$$ B_{C \zeta}(p) \subseteq [p - \zeta e_j, \, p + \zeta e_j]. $$
Meanwhile, by \eqref{eq:weak_cont_ap} we have $d_{Q_n}^2\p{p - \zeta e_j, \, p + \zeta e_j} \leq 2 L \zeta$,
and so $d_{Q_n}\p{p - \zeta e_j, \, p + \zeta e_j} \leq \varepsilon$ if $\zeta = \varepsilon^2 / (2L)$.
These two facts together imply that
\begin{align*}
N_{[]} (\varepsilon, \, \mathcal F_n, \, L_2(Q_n))
\leq \inf\cb{ K : \set \subseteq \bigcup_{k = 1}^K B_{C (\varepsilon^2/(2L))}(p_k) \text{ with } p_k \in \RR^J},
\end{align*}
where the right-hand side quantity is the $C (\varepsilon^2/(2L))$-covering number of
$\set$ under the usual norm $\Norm{.}_2$. For $\varepsilon \leq 1$ , this quantity can
be further be bounded as \citep[Theorem 27.3]{polyanskiy2024information}
\begin{equation}
\label{eq:covering_geom}
\begin{split}
&\inf\cb{ K : \set \subseteq \bigcup_{k = 1}^K B_{C (\varepsilon^2/(2L))}(p_k) \text{ with } p_k \in \RR^J} \\
&\quad\quad\quad\quad\leq 3^J\p{C\p{\frac{\varepsilon^2}{2L}}}^{-J} \frac{\operatorname{vol}\p{\operatorname{conv}\p{\set * B_{C/(2L)}}}}{\operatorname{vol}\p{B_1}},
\end{split}
\end{equation}
where $B_r$ denotes the radius-$r$ ball in $\RR^J$ centered at 0,
$\acal * \mathcal{B} = \cb{a + b : a \in \acal, \, b \in \mathcal{B}}$, $\operatorname{conv}(.)$
denotes the convex hull of a set and $\operatorname{vol}(.)$ denotes the volume of a set.
The scaling in \eqref{eq:covering_geom} establishes the desired claim. 
\end{proof}
\end{lemma}

\begin{lemma}
\label{lemm:subgaussian}
We say that a random function $F_{in}$ is $\zeta$-subgaussian over $\set$ if there
exists a constant $C$ such that, for all $n \geq 1$ and $t>0$, that 

\begin{equation}
\label{eq:subgaussian}
\PP{\sup_{p \in \set} \abs{\frac{1}{n} \sum_{i = 1}^n F_{in}(p) - f(p)} > \frac{t}{\sqrt{n}}} \leq C t^\zeta e^{-2t^2}.
\end{equation}
Under Assumptions \ref{as:monotone}, \ref{as:lips} and  \ref{as:regularity},
let $U_{i}$ be random price perturbations with $\EE{U_{i}} = 0$
and $\abs{U_{i}} \leq h_n$ almost surely for some sequence $h_n \rightarrow 0$.
Suppose furthermore that $W_i$ is Bernoulli-randomized as in Design \ref{def:rct}.
Then, the following random functions are $\zeta$-subgaussian: 
\begin{itemize}
\item $Z_i(w, \, p + U_{i})$ for $w = 0,\, 1$.
\item $Z_i(W_i, \, p + U_{i})$ and $(W_i / \pi_i - (1 - W_i) / (1 - \pi_i)) Z_i(W_i, \, p + U_{i})$.
\item $Y_i(w, \, p + U_{i})$ for $w = 0,\, 1$.
\item $Y_i(W_i, \, p + U_{i})$ and $(W_i / \pi_i - (1 - W_i) / (1 - \pi_i)) Y_i(W_i, \, p + U_{i})$.
\end{itemize}
\begin{proof}

For $Z_i(w, \, p + U_{i})$ and $Z_i(W_i, \, p + U_{i})$, we can use  the bracketing bounds 
in Lemma \ref{lemm:bracket}, which imply  $N_{[]} (\varepsilon, \mathcal F_n, L_2(Q_n)) \leq   ( {K}/{\varepsilon} )^{\zeta} $, for $\zeta = 2J$, and $\mathcal F_n$ is constructed by varying $Z_i(w, \, p + U_{i})$ or  $Z_i(W_i, \, p + U_{i})$ over $p \in \mathcal S$. The tail bound follows from Theorem 2.14.9 of \citet{van1996weak}, given boundedness of the net demand functions and  the polynomial bound on the $\varepsilon$-bracketing number.  The result for $W_i / \pi_i \, Z_{i}(W_i, \, p + U_{i})$, 
$(1 - W_i) / (1 - \pi_i) \, Z_{i}(W_i, \, p + U_{i})$ and their differences follows by the same argument because $W_i / \pi_i$ and $(1 - W_i) / (1 - \pi_i)$
are positive and bounded by $\eta^{-1}$, and thus these functions satisfy the conditions of
Lemma \ref{lemm:bracket} with a constant $L/\eta$ in \eqref{eq:weak_cont_ap}. 

For outcomes, we have the decomposition $Y_i(w, p) = H_i(w, p) + \psi(\Gamma_i(w), Z_i(w, p), p)$. We first work with $H_i(w, p)$. Note that bracketing bounds for Lipschitz function classes are standard; see, e.g., Section 2.7.4
of \citet{van1996weak}. Adapting these results to our setting, we let $F_i(w, p) = H_i(w, p) + c$ and $\mathcal F = \{ F_i (w, \cdot) \to F_i(w, p) : p \in \mathcal S \}$. 

By the Lipschitz property, for any $p \in \RR^J$,
$$ \cb{(p', \, 0) : \Norm{p - p'}_2 \leq \varepsilon} \leq [(p, \, -M\varepsilon), \, (p, \, +M\varepsilon)], $$
and $d_P((p, \, -M\varepsilon), \, (p, \, M\varepsilon)) \leq  2M\varepsilon$ for any distribution $Q$ over the
functions $H_i(w, p)$. 

Thus, using notation from Lemma \ref{lemm:bracket},
\begin{equation}
\label{eq:lip_bracket}
N_{[]} (\varepsilon, \, \mathcal F, \, L_2(Q))
\leq \inf\cb{ K : \set \subseteq \bigcup_{k = 1}^K B_{\varepsilon/(2M)}(p_k) \text{ with } p_k \in \RR^J},
\end{equation}
which is in turn $O(\varepsilon^{-J})$ by the same argument as used in \eqref{eq:covering_geom}.
This bracketing bound can be applied to $H_i(w, \, p + U_{i})$, $H_i(W_i, \, p + U_{i})$, etc., thus
enabling us to again apply Theorem 2.14.9 of \citet{van1996weak}.

For the second component of the outcome function, it is useful to use a covering number bound rather than a bracketing number bound. For a function class $\mathcal F_n$, we can define the $\varepsilon$-covering number of $\mathcal F_n$ under  a distribution $Q_n$ as 
$N(\varepsilon, \mathcal F_n, L_2(Q_n))$. This is defined as the minimum number of balls of radius $\varepsilon$ in $L_2(Q_n)$ norm required to cover $\mathcal F_n$. $\varepsilon$-covering numbers are bounded by $2 \varepsilon$-bracketing numbers, as shown in  Section 2.1.1 of \citet{van1996weak}. Theorem 2.14.9 of \citet{van1996weak} provides a tail bound as in the Lemma statement for function classes with 
\[ \sup \limits_{Q_n} N(\varepsilon, \mathcal F_n, L_2(Q_n)) \leq  \left ( \frac{K}{\varepsilon} \right )^{\zeta} \]  for constants $K, \zeta >0$. It remains to provide such a bound for the class $\{ Y_i(\cdot) \to Y_i(w, p): p \in \mathcal S \}$. 

Let $G_{in} = \phi(F_{i1n}, \ldots, F_{iMn})$ be a composition of $M$ random functions, drawn from distribution $Q_n$. Let $\mathcal G_{n} = \{ \phi(F_{i1n}(\cdot), \ldots, F_{iMn}(\cdot)) \to \phi( F_{i1n}(p), \ldots, F_{iMn}(p)) : p \in \mathcal S \}.$ and $\mathcal F_{mn} = \{ F_{imn}(\cdot) \to F_{imn}(p): p \in \mathcal S \}$. Lemma A.6 of \citet{chernozhukov2014gaussian} indicates that as long as $\phi(\cdot)$ is Lipschitz in each of its $M$ arguments, then if there is a polynomial bound on the $\varepsilon$-covering number of $\mathcal F_{mn}$ for $m \in \{1, \ldots, M \}$, there is also a polynomial bound on the $\varepsilon$-covering number of $\mathcal G$. Recall that a $\varepsilon$-covering number is bounded by a $2 \varepsilon$-bracketing number and our bracketing numbers do not depend on $Q_n$. We can now apply the composition result from \citet{chernozhukov2014gaussian} to finish the proof for $Y_i(w, p)$, since we already showed that $H_i(w, p)$ has the right kind of bound on the bracketing number of its function class, $\psi(\cdot)$ is Lipschitz in each of its arguments, and each of its arguments are functions of $p$ that have a polynomial bound on their $\varepsilon$-covering numbers. 
\end{proof}
\end{lemma}

\begin{lemma}
\label{lemm:asymp_equicontinuity}
For $n = 1, \, 2, \, \ldots$, let $F_{in}(p)$
be IID random functions that are
$\zeta$-subgaussian and weakly continuous, and whose covariances converge pointwise
to a finite limit $\Sigma$,
\begin{equation}
\label{eq:covariance_convergence}
\limn \Cov{F_{in}(p) ,\ F_{in}(p')} = \Sigma(p, \, p') \text{ for all } p, \, p'.
\end{equation}
Then, averages of that function are asymptotically equicontinuous, i.e., they satisfy \eqref{eq:asymptotic_equicont} at all $p \in \set$. Furthermore, the residuals in the asymptotic equicontinuty expansion, 
\begin{equation}
\label{eq:asymp_equi_resid}
R_n = \sup \cb{\abs{\frac{1}{n} \sum_{i = 1}^n \p{F_{in}(p) - F_{in}(p^*) - \p{f_n(p) - f_n(p^*)}}} : \Norm{p - p^*} \leq \delta_n, \ p \in \set}
\end{equation}
satisfy $\limsup_{n \rightarrow \infty} n \, \EE{R_n^2} = 0$.
In particular, under the assumptions of Lemma \ref{lemm:subgaussian},  averages of the following random functions are asymptotically equicontinuous and have residuals in the asymptotic equicontinuity expansion meeting the above condition: 
\begin{itemize}
\item $Z_i(w, \, p + U_{i})$ for $w = 0,\, 1$.
\item $Z_i(W_i, \, p + U_{i})$ and $(W_i / \pi_i - (1 - W_i) / (1 - \pi_i)) Z_i(W_i, \, p + U_{i})$.
\item $Y_i(w, \, p + U_{i})$ for $w = 0,\, 1$.
\item $Y_i(W_i, \, p + U_{i})$ and $(W_i / \pi_i - (1 - W_i) / (1 - \pi_i)) Y_i(W_i, \, p + U_{i})$.
\end{itemize}

\begin{proof}
Let $Q_n$ define a distribution over random functions $F_{in}(p)$, for any of the random functions listed in the Lemma, and define the function class $\mathcal F_n = \{ F_{in}(\cdot) \to F_{in}(p): p \in \mathcal S\}$. Theorems 2.11.1 and Theorem 2.11.9 \citet{van1996weak} imply weak convergence of the empirical process
$$ \frac{1}{\sqrt{n}} \sum_{i = 1}^n \p{F_{in}(p) - f_n(p) - F_{in}(p') - f_n(p')} $$
to a Gaussian process indexed by $p$ when for each $n$, the $\varepsilon$-bracketing number or $\varepsilon$-covering number of $\mathcal F_n$ under $Q_n$ is bounded by  $(K/\varepsilon)^V$ for all $0 < \varepsilon < K$, for finite constants $K, V >0$.  Then, asymptotic equicontinuity follows from weak continuity, i.e. that for each $p \in \mathcal S$ and $p' \in \mathcal S$, $\mathbb E[(F_{in}(p) - F_{in}(p'))^2]$ is continuous in $p$. Weak continuity implies that the limiting Gaussian process has continuous sample paths. 

To verify the 2nd-moment bounds on the residuals in the expansions, we first note that
\begin{align*}
R_n &= \sup_{\Norm{p - p^*} \leq \delta_n} \abs{\frac{1}{n} \sum_{i = 1}^n \p{F_{in}(p) - F_{in}(p^*) - \p{f_n(p) - f_n(p^*)}}} \\
&\quad\quad\quad\quad\quad\quad\quad\quad\leq 2 \sup_{p \in \set} \abs{\frac{1}{n} \sum_{i = 1}^n \p{F_{in}(p) - f_n(p)}},
\end{align*}
which has a rapidly decaying tail by Lemma \ref{lemm:subgaussian}. Thus, in particular,
$$ \limsup_{n \rightarrow \infty} n^2\, \EE{R_n^4} < \infty. $$
Now, pick any $c > 0$. By asymptotic equicontinuity, $\limsup \limits_{n \rightarrow \infty} \PP{\abs{R_n} > c/\sqrt{n}} = 0$.
Furthermore
\begin{align*}
\limsup_{n \rightarrow \infty} n\,\EE{R_n^2}
&= \limsup_{n \rightarrow \infty} n\, \EE{1\p{\cb{\abs{R_n} \leq c/\sqrt{n}}} R_n^2} + n\, \EE{1\p{\cb{\abs{R_n} > c/\sqrt{n}}} R_n^2} \\
&\leq c^2 + \limsup_{n \rightarrow \infty} \sqrt{\PP{\abs{R_n} > c/\sqrt{n}}} \sqrt{n^2\, \EE{R_n^4}} \\
&\leq c^2 +  \limsup_{n \rightarrow \infty} \sqrt{\PP{\abs{R_n} > c/\sqrt{n}}}  \limsup_{n \rightarrow \infty} \sqrt{n^2\, \EE{R_n^4}} \\
&= c^2,
\end{align*}
where for the second inequality we used Cauchy-Schwarz. Since we have shown that $0 \leq  \limsup_{n \rightarrow \infty} n\,\EE{R_n^2} \leq \epsilon$ for any $\epsilon > 0$, we see that in fact $\limsup_{n \rightarrow \infty} n\,\EE{R_n^2} = 0$.

For all of the random functions listed, weak continuity holds by Assumption \ref{as:lips} and \ref{as:regularity}, and Lemma \ref{lemm:subgaussian} provides the required subgaussian condition. 
A final check is htat 
\begin{equation}
\label{eq:cov_conv}
\lim_{n \rightarrow \infty} \Cov{Z_{ij}(w, \, p + U_{i}),\ Z_{ij}(w, \, p' + U_{i})} = \Cov{Z_{ij}(w, \, p),\ Z_{ij}(w, \, p')}, 
\end{equation}
which is finite, and the same holds for $Y_i(w, p + U_i)$ and $Y_i(W_i, p + U_i)$. 

\end{proof}
\end{lemma}

\section{Additional Results}
\label{app:add}
This appendix is for results that are cited but not included in the main text.

\subsection{Market-Clearing}

\begin{proposition} \label{prop:clear} \textbf{Single-Good Market}. In a market with $J=1$ goods, assume that outcomes and net demand functions are bounded, and for all $p \in \mathcal S$ and $w \in \{0, 1\}$, $\mathbb P(Z_i(w, p) \text{ is continuous at } p) = 1$ and $\mathbb P(Y_i(w, p) \text{ is continuous at } p) = 1$. In addition, assume that
\begin{itemize}
\item $\mathbb E[Z_i(w, 0) ] > 0$  and there exists a finite $b > 0$ such that $\mathbb E[Z_i(w, b) ] < 0 $ for $ w \in \{0, 1\}$. 
\item With probability 1, $Z_i(w, p)$ is monotonically non-increasing in $p$ for $ w \in \{0, 1\}$.
\end{itemize}
Let $Z_n(\bm w, p) = \frac{1}{n} \sum \limits_{i=1}^n Z_i(w_i, p)$ and $P_n(\bm w) = \arg \min \limits_p | Z_n(\bm w, p) |$ for $\bm w \in \{0, 1\}^n$.  In cases where there are multiple minimizers, choose one by some deterministic rule.

Then, Assumption \ref{as:interference} holds, with $U_i =0$. That is, there exist a sequence $a_n$
with $\lim \limits_{n \to \infty} a_n \, \sqrt{n} = 0$ and a constant $c > 0$ such that, given any treatment vector
$\wvec \in \cb{0,\, 1}^n$,  
\begin{equation}
\label{eq:approxzero_appendix}
\set_{\wvec} = \cb{p \in \RR^J : \Norm{\frac{1}{n} \sum \limits_{i=1}^n Z_{i}(w_i, p)}_2 \leq a_n }.
\end{equation}
is non-empty with probability $1 - e^{-c_1n}$. And $P_n(\bm w) \in \set_{\wvec}$ when it is non-empty.
\end{proposition}

\begin{proof}

Denote by $\mathcal{E}_n$ the event that the realized $Z_n(\bm w, p)$ crosses zero during the interval $(0-s,b+s)$:
 \begin{equation}
 	\mathcal{E}_n = \{ Z_n(0) > 0, \mbox{ and }  Z_n(b) < 0  \},
 	\end{equation}
 and denote by $E_n$ the indicator variable for $\mathcal{E}_n$: $E_n = \mathbb{I}(\mathcal{E}_n) $.

 Suppose that  $E_n = 1$. Here, either $Z_n(P_n) = 0$, or $|Z_n(P_n)| \neq 0$, and we will focus on the latter case.  Suppose $|Z_n(P_n)| \neq 0$. Because we condition on $E_n=1$, we have that $Z_n(0) >0$ and $Z_n(b) <0$. Since $P_n$ is chosen to be a minimizer of $|Z_n(p)|$, the only possibility is that the function $Z_n(\cdot)$ crosses zero at a point of discontinuity, $ P'$, that is,   $Z_n(\tilde p ) <0$ for all $\tilde p > P'$, and $Z_n(\bar p) > 0$ for all $\bar p < P'$.

We have assumed that  $\mathbb P ( Z_i(W_i, \cdot)   \mbox{ is discontinuous at $p$} ) = 0$ for any $p >0$, the $Z_i$'s are drawn independently, and by the boundedness assumption they are bounded in magnitude by $M$. These facts further imply that with probability one, any jump in $Z_n(\cdot)$ cannot exceed magnitude $M/n$, for otherwise it would have required at least two separate $Z_i$ to have a point of discontinuity at the same exact location. Formally, conditional on $E_n=1$, we have that with probability one:
\begin{equation}
|Z_n(P_n)| \leq | \lim_{p \uparrow P' } Z_n(p) - \lim_{p \downarrow P' } Z_n(p)  | \leq  M/n,
\end{equation}

Choosing $a_n = M/n$, we note that $ \lim \limits_{n \to \infty} a_n \sqrt n = 0$, and that we have that the specified $P_n(\bm w) \in \set_{\wvec}$ whenever $\set_{\wvec}$ is non-empty. It is non-empty whenever $E_n = 1$, so to finish the proof we now check the probability that $E_n = 1$.

If $E_n = 0$, there are four possible cases. The first two cases are $Z_n(0) = 0$ or $Z_n(b) = 0$, in which case $Z_n(P_n) =0$. The last two are $Z_n(p) > 0$ for all $ p \in \mathcal S$, or $Z_n(p) < 0$ for all $p \in \mathcal S$, in which case $Z_n(P_n) \leq c_1$ for some constant $c_1 >0$, since each net demand function is bounded. Note that $c_1$ is not less than the specified $a_n$, so to verify the proposition we need that $\mathbb P(E_n = 0)$ is exponentially small. To bound $\mathbb P(E_n = 0)$,
  \begin{align*}
 \mathbb P (E_n = 0) & \leq \mathbb P( Z_n(0) \leq 0)  +  \mathbb P( Z_n(b) \geq 0)  \nonumber \\ &
  = \mathbb P(Z_n(0) - \mathbb E[ Z_n(0) ]  \leq -\mathbb E[Z_n(0)] ) + \mathbb P (Z_n(b) - \mathbb E[Z_n(b)] \geq - \mathbb E[Z_n(b) ]) \\
  & \leq  \mathbb P( | Z_n(0) - \mathbb E[ Z_n(0) ]  | \geq \mathbb E[Z_n(0)] ) + \mathbb P ( | Z_n(b) - \mathbb E[Z_n(b)] |  \geq  - \mathbb E[Z_n(b) ])\\
  &  \leq  \mathbb P( | Z_n(0) - \mathbb E[ Z_n(0) ]  | \geq \varepsilon ) + \mathbb P ( | Z_n(b) - \mathbb E[Z_n(b)] | \geq \varepsilon ),
  \end{align*}
  for $\varepsilon = \min \{ \mathbb E[Z_n(0), - \mathbb E[Z_n(c)]  \} > 0$.
 We can now use Hoeffding's inequality, and obtain that for any $n$ and $\varepsilon>0$,
\begin{align}
\mathbb P(E_n = 0) &  \leq 2 \cdot \exp \left ( \frac{- 2 n \varepsilon^2}{4c_1^2} \right), \nonumber \\
& =  2 \cdot \exp \left ( \frac{- n \varepsilon^2}{2c_1^2} \right).
 \label{eqn:pe0b}
\end{align}
We have now verified that $\set_{\wvec}$ is non-empty with probability $1 - e^{-c_1n}$, for some $c_1 >0$, which completes the proof.


\end{proof}

\subsection{Tighter Conservative Bound for $\bsigma^2_D$ }
\label{ap:bound}

First, introduce the notation $a_i(1) = \epsilon_i(1) - \pi \Delta_i(1, p^*_{\pi})$, and $a_i(0) = \epsilon_i(0) + (1 - \pi) \Delta_i(0, p^*_{\pi}).$ Under the augmented randomized experiment, an estimator is available for the following variance. 
\[ \tilde \sigma^2_D = \mathbb E \left [ \frac{( 1- \pi)}{\pi}  a_i^2(1) + \frac{\pi}{1 - \pi} a_i^2(0) + 2 \mathbb E[a_i(1)^2]^{1/2} \mathbb E[a_i(0)^2]^{1/2} \right]. \] 
In an extension of the classical results in \citet{neyman1923applications}, and described in detail in Section 2.1 of \citet{aronow2014sharp}, we can show that this variance $\tilde \sigma^2_D$ is conservative for $\bsigma^2_D$ and is tighter than $\sigma^2_D$. 
\begin{equation*} 
\begin{split} 
 \sigma^2_D  & =  \EE{\p{ \frac{W_i \varepsilon_i(1)}{ \pi}  - \frac{( 1- W_i)  \varepsilon_i(0)}{  1- \pi }  - \Delta_i(W_i, p^*_{\pi})}^2},  \\
	 & = \EE{\p{ \frac{W_i a_i(1)}{ \pi}  - \frac{( 1- W_i)  a_i(0)}{  1- \pi } }^2}, \\ 
	 & = \EE{ \frac{ a_i(1)^2 }{\pi} }  + \EE{ \frac{ a_i(0)^2 }{1 - \pi} }, \\ 
	 & =   \EE{\p{ \frac{  ( 1 -  \pi)  a_i^2(1)}{ \pi}  + \frac{  \pi  a_i^2(0)}{  1- \pi }  + a^2_i(1) + a^2_i(0) }}, \\
	 & \geq \EE{\p{ \frac{ ( 1- \pi) a_i^2(1)}{ \pi}  + \frac{ \pi a_i^2(0)}{  1- \pi }  + 2  \mathbb E[a_i(1)^2]^{1/2} \mathbb E[a_i(0)^2]^{1/2} }}, \\  & = \tilde \sigma^2_D, 
\end{split} 
\end{equation*} 
where the inequality is from the AM-GM inequality. Additionally, to show this is conservative for $\bsigma^2_D$

\begin{equation*} 
\begin{split}
\bsigma_D^2 & = \pi \p{1 - \pi} \EE{ \left (\frac{ \varepsilon_i(1) }{\pi} + \frac{ \varepsilon_i(0)}{1 - \pi}  - \p{\Delta_i(1, p^*_{\pi}) - \Delta_i(0, p^*_{\pi})} \right )^2}, \\ 
& = \pi \p{1 - \pi} \EE{ \left (  \frac{a_i(1)}{\pi} + \frac{a_i(0)}{1 - \pi}  \right)^2},  \\ 
& =   \EE{  \frac{  ( 1 -  \pi)  a_i^2(1)}{ \pi}  + \frac{  \pi  a_i^2(0)}{  1- \pi }  + 2 a_i(1) a_i(0) }, \\ 
& \leq   \EE{  \frac{  ( 1 -  \pi)  a_i^2(1)}{ \pi}  + \frac{  \pi  a_i^2(0)}{  1- \pi }  + 2\sqrt {  \mathbb E[a_i^2(1)] \mathbb E[a_i^2(0)]} }, \\
& = \tilde \sigma^2_D,
\end{split}
\end{equation*} 
where the inequality is from the Cauchy-Schwarz inequality.

\subsection{Convergence Rate of $\tau_{\text{AIE}}$ to $\tau^*_{\text{AIE}}$}
\label{ap:aierate}
In this section, we provide a simple example where $Z_i(w, p)$ and $Y_i(w, p)$ are differentiable in $p$. In this example, the asymptotic representation of $\tau_{\text{AIE}} - \tau^*_{\text{AIE}}$ depends on $\frac{1}{n} \sum \limits_{i=1}^n \nabla_p Y_i(w, p) - y(w, p) $, and it is straightforward to extend this example so that it also depends on the concentration of $\frac{1}{n} \sum \limits_{i=1}^n \nabla_p Z_i(w, p)$ around its expected derivative. This suggests that it is not possible to get $\sqrt n$ converge of $\tau_{\text{AIE}}$ to $\tau^*_{\text{AIE}}$ under Assumption \ref{as:regularity}, which allows for discontinuous individual-level demand and outcome functions.

Define bounded random variables $(Y_i(1), Y_i(0), Z_i(1), \beta_i)$ that are sampled IID from some distribution. $Z_i(0) = 0$ always. This distribution is such that  $\mathbb E[Z_i(1)]= z(1) $,  $\mathbb E[Y_i(1) - Y_i(0) ] = y(1) - y(0) ]$ and $\mathbb E[\beta_i] = \beta$.  $W_i$ is drawn IID from $\mbox{Bernoulli}(\pi)$.  The rest of the data-generating process is:

\begin{equation*}
\begin{split}
Z_i(w,p) = \theta_0 -   p  + Z_i(W_i), \qquad Y_i(w, p) = -  \beta_i W_i p + Y_i(W_i).
\end{split}
\end{equation*}

In this model, $p^*_{\pi} = \theta_0  + \pi \cdot z(1)$. The population AIE is
\[ \tau^*_{\text{AIE}} = -  \pi \cdot \beta \cdot  z(1).  \]
This model implies that $P_n(W_i = 1; \bm W_{-i}) - P_n(W_i = 0; \bm W_{-i}) = \frac{1}{n} Z_i(1)$ and $Y_j(W_j, P_n(W_j = 1; \bm W_{-i}) - Y_j(W_j, P_n(W_j = 0; \bm W_{-i}) = - \beta_j W_j  \cdot \frac{1}{n} Z_i(1) $.

This implies that $\tau_{\text{AIE}}$ in this model is:
\begin{equation*}
\begin{split}
\tau_{\text{AIE}}  & =     \frac{1}{n} \sum \limits_{i=1}^n \sum \limits_{j \neq i} \pi \cdot   \beta_j   \frac{1}{n} (Z_i(1)  - Z_i(0))  \\
 & = \frac{1}{n} \sum \limits_{i=1}^n Z_i(1) \frac{1}{n} \sum \limits_{j=1}^n  \pi \cdot   \beta_j    \\
 & =  \tau^*_{\text{AIE}} +  \frac{1}{n} \sum \limits_{i=1}^n Z_i(1) \frac{1}{n} \sum \limits_{j=1}^n  \pi \cdot   (\beta_j  - \beta) +   \frac{1}{n} \sum \limits_{i=1}^n  (Z_i(1)  - Z_i(0)) - z(1) - z(0)) \pi \cdot \beta \\
 & \stackrel{(1)} {=}  \tau^*_{\text{AIE}} + z(1)  \frac{1}{n} \sum \limits_{j=1}^n  \pi \cdot   (\beta_j  - \beta) +   \frac{1}{n} \sum \limits_{i=1}^n  (Z_i(1)  - z(1) ) \pi \cdot \beta + o_p(n^{-1/2}).
\end{split}
\end{equation*}
The last line is by the application of the LLN and recognizing that the CLT applies to the term $\frac{1}{n} \sum \limits_{j=1}^n  \pi \cdot   (\beta_j  - \beta) $. In this case there is $\sqrt n$ convergence of $\tau_{\text{AIE}}$ to $\tau^*_{\text{AIE}}$, but the distribution depends on the variance of $\nabla_p Y_j(1, p^*_{\pi})$. In a more general derivation with non-differentiable $Y_i$, the analogous term would not converge at a $\sqrt n $ rate, so the scaling of $\tau_{\text{AIE}}$ to $\tau^*_{\text{AIE}}$ will be slower than $\sqrt n$ under the assumptions in the paper (just as the convergence of $\hat \tau_{\text{AIE}}$ to $\tau^*_{\text{AIE}}$ is slower than $\sqrt n$).


\subsection{Observing a Sub-Sample of the Market}
\label{sec:subsample}
The estimators for $\tau^*_{\text{ADE}}$ and $\tau^*_{\text{AIE}}$ are unchanged when a sub-sample of the $n$ agents in the finite-sample market are observed rather than all of them. However, the expansion for $\hat \tau_{\text{ADE}}$ described in Theorem \ref{theo:adeinf} is affected slightly, which changes the asymptotic variance. This appendix provides some detail  on these claims which were made in the main text, under the assumptions of Theorem \ref{theo:adeinf}.

Let $n$ be the size of the finite-sample market. Let $\mathcal M \subset \{ 1, \ldots, n \}$ index the $m$ total number of observations sampled without replacement from the finite-sample market. We assume that the market price clears the finite-sample market, but we only observe $Y_j$ and $Z_j$ for  $j \in \mathcal M$. Let $\hat s = \frac{1}{m} \sum \limits_{j \in M} W_j$ be the fraction of individuals treated in the sub-sample,  $\hat t = \frac{m}{n} $  be the fraction of individuals observed,  and  $\hat \pi = \frac{1}{n} \sum \limits_{i =1}^n W_i$ be the fraction of individuals treated in the entire finite market. We assume that for all agents that are not in $\mathcal M$, then $W_i = 0$. The sub-sample is a fixed fraction of the finite market, so that as $n \to \infty$, $\hat s \to  s >0 $, $\hat \pi \to  \pi >0$ and $\hat t \to t > 0$.

The estimators $\hat \tau_{\text{ADE}}$ and $\hat \tau_{\text{AIE}}$ are exactly the same as in the main text, except the observations used are in $\mathcal M$ rather than in $\{1, \ldots n \}$. Let $h_n$ remain the same as in the main text, where the shrinkage rate of the price perturbations depend on $n$.

\begin{equation*}
\begin{split}
&  \hat \tau_{\text{m, ADE}} = \frac{1}{m}  \sum \limits_{j \in \mathcal M} \left [  \frac{W_j Y_j }{\hat s}- \frac{(1-W_j)Y_j}{ 1 - \hat s} \right] \\
&  \hat \tau_{\text{m, AIE}} =  - \hat {\gamma}^{\top} \frac{1}{m} \sum \limits_{j \in \mathcal M} \left [ \frac{W_j { Z}_j}{\hat s }-  \frac{(1- W_j) {Z}_j}{ 1 - \hat s } \right ].
\end{split}
\end{equation*}

Let $\bm Y$ be the $m$-length vector of observed outcomes, $\bm U$ is the $m \times J$ matrix of observed price perturbations, and $\bm Z$ is the $m \times J$ matrix of observed net demand. Then, $\hat {\gamma} = (\bm U^{\top} \bm Z) (\bm U^{\top} \bm Y)$.

The asymptotic expansion for $\hat \tau_{m, \text{AIE}}$ is unchanged, apart from depending on $m$ rather than $n$, since the variance of $P_n - p^*_{\pi}$ does not impact the limiting distribution.

\[ \hat \tau_{m, \text{AIE}}  = \tau^*_{\text{AIE}}  -  Q_z^{\top}   \frac{1 }{ \sqrt m h^2_n} \sum \limits_{j \in \mathcal M} U_j \nu_j(W_j)  + o_p(1), \]  where $\nu_j(W_j) = Y_j(W_j, p^*_{\pi}) - Z_j(W_j, p^*_{\pi}) ^{\top} [\xi_z^{-1}]^{\top} \xi_y $ and $Q_z =  \xi_z^{-1} \tau^{*,z}_\text{ADE}$.  The asymptotic variance can still be estimated by $\hat \sigma_I$, where each component is estimated using the observed data only.

The asymptotic  expansion for $\hat \tau_{\text{ADE}}$ is affected, since $P_n - p^*_{\pi}$ does impact the asymptotic expansion, and the finite sample market price depends on  $n$ rather than $m$.

\begin{equation*}
\begin{split}
\hat \tau_{m, \text{ADE}}  & \stackrel{(1)}{=}   \tau^*_{\text{ADE}} + \frac{1}{n}  \sum \limits_{i=1}^n    \left [ \frac{\mathbbm{1}(i \in \mathcal M) }{ \hat  t} \left ( \frac{W_i \varepsilon_i(1)}{ \hat s}  - \frac{( 1- W_i) \varepsilon_i(0) }{  1- \hat s }  \right ) - \Delta_i(W_i, p^*_{\pi})     \right] + o_p(n^{-0.5}) \\
&=  \tau^*_{\text{ADE}}  + \frac{1}{n }  \sum \limits_{i=1}^n    \left [ \frac{\mathbbm{1}(i \in \mathcal M) }{ t} \left ( \frac{W_i \varepsilon_i(1)}{ s}  - \frac{( 1- W_i) \varepsilon_i(0) }{  1 - s }  \right ) - \Delta_i(W_i, p^*_{\pi})     \right] + o_p(n^{-0.5}) \\
 & \stackrel{(2)}{=}   \tau^*_{\text{ADE}}  +  \frac{1}{n}  \sum \limits_{i=1}^n    \frac{\mathbbm{1}(i \in \mathcal M) }{ t} \left ( \frac{W_i \varepsilon_i(1)}{ s}  - \frac{( 1- W_i) \varepsilon_i(0) }{  1 - s }   - t \Delta_i(W_i, p^*_{\pi}) \right )      \\ &  \qquad  - \frac{1}{ n}  \sum \limits_{i=1}^n \mathbbm{1}{(i \notin \mathcal M)} \Delta_i(W_i, p^*_{\pi})  + o_p(n^{-0.5}) 
\end{split}
\end{equation*}
(1) comes from following the same steps as in the proof of Theorem \ref{theo:adeinf}. (2) comes from splitting up the sum for observations in $\mathcal M$ and those not in $\mathcal M$. As before $\varepsilon_i(w) = Y_i(w, p^*_{\pi}) - y(w, p^*_{\pi})$.

The CLT now applies to this expansion:
\begin{equation*}
\begin{split}
& \sqrt n (\hat \tau_{m, \text{ADE}} - \tau^*_{\text{ADE}}) \Rightarrow \mathcal{N}(0, \sigma^2_{t, D}) \\
& \sigma^2_{t, D} =  \frac{1}{t} \mathbb E \left [  \left ( \frac{W_i \varepsilon_i(1)}{ s} - \frac{( 1- W_i) \varepsilon_i(0)}{ (1 - s)  } - t\Delta_i(W_i, p^*_{\pi})  \right)^2  \right]  + ( 1- t) \mathbb E \left [  \Delta_i(W_i, p^*_{\pi})^2 \right ]
\end{split}
\end{equation*}

For inference on $\tau^*_{\text{ADE}}$ using the sub-sample only, we can estimate each component of $\sigma^2_{t, D}$ using only observations in $\mathcal M$.

\subsection{Alternative Estimator of AIE Variance}
\label{ap:2ndorder}

In analyzing the asymptotic properties of $\sigma^2_I$ in the proof of Theorem \ref{theo:aieinf} in Appendix \ref{ap:aieinf}, we dropped a term that was asymptotically negligible, which was $\sqrt n h_n ( \xi_y \xi_z^{-1} (\hat  \tau^{z}_{ \text{ADE}} -  \tau^{*,z}_{ \text{ADE}})).$

 $\tilde \sigma^2_I$ adds a plug-in estimator for the second term, including an estimator for the variance of $\hat {\tau}^{z}_{\text{ADE}} $.
$ \tilde{\sigma}^2_I = \hat {\sigma}^2_I + h^2_n  \hat{ \gamma}^{\top} \hat { \sigma}^{2}_{z, D}  \hat {\bm  \gamma}, $
where
$ \hat { \sigma}^{2}_{z, D}  = \frac{1}{n} \sum \limits_{i=1}^n B_i B_i' $
and the $J \times 1$ vector $\hat { B}_i = \frac{W_i \hat{\varepsilon}^z_i(1) }{\hat \pi} - \frac{( 1-W_i) \hat{\varepsilon}^z_i(0)}{1 - \hat \pi} -  [\hat {\xi}_{z1} -  \hat {\xi}_{z0}]^{\top} \hat {\xi}_z^{-1} {Z}_i $. ${\varepsilon}^z_i(1) = Z_i - \frac{1}{n_w} \sum \limits_{i: W_i = w}  Z_i$. Last, the $J \times J$ matrix $\hat {\xi}_{zw}$ for $w \in \{0, 1\}$ are estimated from regressions of $Z_i$ on $U_{i}$ using only observations such that $W_i = w$. We found in simulations that this second-order correction leads to better coverage properties at smaller sample sizes.

\end{document}